\documentclass[fleqn,usenatbib]{mnras}

\usepackage{newtxtext,newtxmath}
\usepackage[T1]{fontenc}

\DeclareRobustCommand{\VAN}[3]{#2}
\let\VANthebibliography\thebibliography
\def\thebibliography{\DeclareRobustCommand{\VAN}[3]{##3}\VANthebibliography}

\usepackage{graphicx}	
\usepackage{amsmath}	
\usepackage[dvipsnames]{xcolor}

\usepackage{float,subfloat,subcaption,xspace,adjustbox}

\newcommand{\ergs}{\rm erg\,s^{-1}}
\newcommand{\Oi}{O\,{\sc i}\xspace}
\newcommand{\Oii}{[O\,{\sc ii}]\xspace}
\newcommand{\Feii}{Fe\,{\sc ii}\xspace}
\newcommand{\Oiii}{[O\,{\sc iii}]\xspace}
\newcommand{\Hei}{He\,{\sc i}\xspace}
\newcommand{\Hi}{H\,{\sc i}\xspace}
\newcommand{\Hii}{H\,{\sc ii}\xspace}
\newcommand{\Nii}{[N\,{\sc ii}]\xspace}
\newcommand{\Neiii}{[Ne\,{\sc iii}]\xspace}
\newcommand{\Caii}{Ca\,{\sc ii}\xspace}
\newcommand{\Hb}{{\rm H}$\beta$\xspace}

\newcommand{\kms}{\,\rm km\,s^{-1}}

\newcommand{\Mbh}{M_{\rm BH}} 

\newcommand{\Ha}{H$\alpha$\xspace}

\newcommand{\Pab}{Pa-$\beta$\xspace}
\newcommand{\Paa}{Pa-$\alpha$\xspace}

\newcommand{\ruby}{\textit{The Cliff}\xspace} 
\newcommand{\zspec}{z_{\rm spec}}

\newcommand{\lpeak}{\lambda_{\rm peak}}

\newcommand{\betaBB}{\beta_{\rm MBB}}
\newcommand{\buv}{\beta_{\rm UV}}
\newcommand{\bopt}{\beta_{\rm opt}}
\newcommand{\Muv}{M_{\rm UV}}

\newcommand\orcid[1]{\href{http://orcid.org/#1}{\adjustbox{trim={-.15\width} {0\height} {-.15\width} {0\height},clip}{\includegraphics[height=10pt]{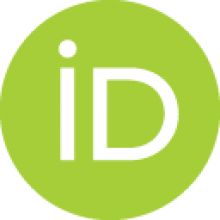}}}}

\title[LRD Demographics]{Little Red Dots host Black Hole Stars: A unified family of gas-reddened AGN revealed by JWST/NIRSpec spectroscopy}

\author[de Graaff et al.]{
Anna de Graaff\orcid{0000-0002-2380-9801}$^{1,2}$\thanks{\url{degraaff@mpia.de}}\thanks{Clay Fellow},
Raphael E. Hviding\orcid{0000-0002-4684-9005]}$^{1}$,
Rohan P. Naidu\orcid{0000-0003-3729-1684}\thanks{NASA Hubble Fellow, Pappalardo Fellow}$^{3}$,
Jenny E. Greene\orcid{0000-0002-5612-3427}$^{4}$,\newauthor 
Tim B. Miller\orcid{0000-0001-8367-6265}$^{5}$,
Joel Leja\orcid{0000-0001-6755-1315}$^{6,7,8}$,
Jorryt Matthee\orcid{0000-0003-2871-127X}$^{9}$,
Gabriel Brammer\orcid{0000-0003-2680-005X}$^{10}$,
Harley Katz\orcid{0000-0003-1561-3814}$^{11,12}$,\newauthor 
Rachel Bezanson\orcid{0000-0001-5063-8254}$^{13}$,
Leindert A. Boogaard\orcid{0000-0002-3952-8588}$^{14}$,
Sownak Bose\orcid{0000-0002-0974-5266}$^{15}$,
John Chisholm\orcid{0000-0002-0302-2577}$^{16,17}$,\newauthor 
Nikko J. Cleri\orcid{0000-0001-7151-009X}$^{6,7,8}$,
Pratika Dayal\orcid{0000-0001-8460-1564}$^{18,19,20}$,
Robert Feldmann\orcid{0000-0002-1109-1919}$^{21}$,
Yoshinobu Fudamoto\orcid{0000-0001-7440-8832}$^{22}$,\newauthor 
Seiji Fujimoto\orcid{0000-0001-7201-5066}$^{19,23}$,
Lukas J. Furtak\orcid{0000-0001-6278-032X}$^{17,16}$,
Karl Glazebrook\orcid{0000-0002-3254-9044}$^{24}$,
Rashmi Gottumukkala\orcid{0000-0003-0205-9826}$^{10}$,\newauthor 
Kasper E. Heintz$^{10,25}$,
Vasily Kokorev\orcid{0000-0002-5588-9156}$^{16,17}$,
Ivo Labbe\orcid{0000-0002-2057-5376}$^{24}$,
Michael V.\ Maseda\orcid{0000-0003-0695-4414}$^{26}$,\newauthor 
Ian McConachie\orcid{0000-0002-2446-8770}$^{26}$,
Themiya Nanayakkara\orcid{0000-0003-2804-0648}$^{24}$,
Erica Nelson\orcid{0000-0002-7524-374X}$^{27}$,
Przemys{\l}aw~Nowaczyk\orcid{0000-0003-0771-8827}$^{28}$,\newauthor 
Pascal~A.~Oesch\orcid{0000-0001-5851-6649}$^{25,10}$,
Hans-Walter Rix\orcid{0000-0003-4996-9069}$^{1}$,
David J. Setton\orcid{0000-0003-4075-7393}$^{4}$, \thanks{Brinson Prize Fellow}
Alberto Torralba\orcid{0000-0001-5586-6950}$^{9}$,\newauthor 
Fabian Walter\orcid{0000-0003-4793-7880}$^{1,29}$,
Bingjie Wang\orcid{0000-0001-9269-5046}$^{4}$,\thanks{NHFP Hubble Fellow}
Andrea Weibel\orcid{0000-0001-8928-4465}$^{25}$,
Arjen van der Wel\orcid{0000-0002-5027-0135}$^{30}$
\\
\\
{\normalsize (Affiliations can be found after the references)}
}

\pubyear{2026}

\begin{document}
\label{firstpage}
\pagerange{\pageref{firstpage}--\pageref{lastpage}}
\maketitle


\begin{abstract}
We construct and characterise a sample of {146} little red dots (LRDs) across {$\mbox{2.0<z<9.3}$}, selecting all sources with v-shaped UV-optical continua from JWST NIRSpec/PRISM spectra and compact morphologies in NIRCam/F444W imaging. We show that LRD continuum spectra are well described by modified blackbodies across $\sim0.4-1.0\,\micron$, with typical $T\sim5000\,$K ($\lpeak\sim0.65\,\micron$) across 2 dex in luminosity, and a tail toward $T\sim2000\,$K. LRDs therefore trace a locus in the Hertzsprung-Russell diagram that is analogous to stars on the Hayashi track, strongly supporting the picture that LRDs are AGN embedded in {optically-thick dense gas envelopes}. Hotter LRDs with $\lpeak<0.65\,\micron$ typically have strong Balmer breaks, redder UV slopes and high optical luminosities; other LRDs show weak or no Balmer breaks, and wide variety in $\buv$ and $L_{5100}$. Crucially, we demonstrate that the UV-optical continua are strongly linked to the \Ha, \Hb, \Oiii and \Oi line properties. There is a tight linear relation between the \Ha and optical continuum luminosities, as well as \Ha and $\textrm{\Oi}_{\lambda8446}$, indicating that Balmer, \Oi and optical emission must primarily be powered by the same source. The Balmer decrement increases strongly toward higher $L_{\rm H\alpha}$, $L_{5100}$ and Balmer break strength, providing key evidence for luminosity-dependent effects of collisional (de-)excitation and resonant scattering in the gaseous envelopes. In contrast, we show that \Oiii emission likely originates from star-forming host galaxies, with strong variation in the AGN-to-host ratio {among the LRD population}. Our work presents an empirical description of the nature and structure of LRDs, defining a new benchmark for ongoing LRD model developments.
\end{abstract}




\section{Introduction} \label{sec:intro}

Understanding the physical nature of Little Red Dots (LRDs), compact high-redshift sources with v-shaped UV-optical continua found by the James Webb Space Telescope (JWST), remains a major open challenge. LRDs were originally proposed to be either dust-reddened active galactic nuclei \citep[AGN; e.g.][]{Matthee2024,Labbe2023b,Kocevski2024}, motivated by their red rest-optical continua and the discovery of broad Balmer lines in a high fraction of LRDs \citep[$>80\%$;][]{Greene2024,Hviding2025}, or massive post-starburst or dusty star-forming galaxies \citep[e.g.][]{Labbe2023,Perez2024,Wang2024b}, due to the frequent detection of Balmer breaks \citep{Setton2024}, weak X-ray emission \citep{Ananna2024,Yue2024}, and lack of hot dust emission in the rest near-infrared \citep[e.g.][]{Williams2024,Wang2024a,Setton2025}. However, recent work has shown that both these models are likely incorrect or at least incomplete, as neither can explain the spectral energy distributions (SEDs) of the most extreme LRDs discovered to date \citep{deGraaff2025,Naidu2025} nor the lack of emission from cold and hot dust in the mid- to far-infrared in a broad range of LRDs \citep{Casey2024,Casey2025,Akins2024,Setton2025,Xiao2025}. 

A new picture has emerged instead, first proposed by \citet{Inayoshi2025}, in which an accreting massive black hole is not reddened by surrounding dust, but by absorption in dense neutral hydrogen gas ($n_{\rm H}\sim 10^{9-11}\,\rm cm^{-3}$, $N_{\rm H}\sim 10^{24-26}\,\rm cm^{-2}$) with high, near-unity covering fraction. Attempts at 1D models of such a ``black hole star'' (BH*) scenario using the spectral synthesis code \texttt{Cloudy} \citep{Ferland2017} have shown that the resulting model spectra can reproduce key features of LRDs, such as very strong Balmer breaks and high equivalent-width Balmer emission lines \citep[e.g.][]{Ji2025,Naidu2025,deGraaff2025,Taylor2025}. The presence of dense gas is also naturally able to explain absorption features observed in the Balmer and \Hei emission lines \citep[][]{Matthee2024,Wang2024a,Juodzbalis2024,DEugenio2025,Naidu2025}. Moreover, radiative transfer effects such as electron scattering \citep{Rusakov2025, Chang2025} or resonant scattering \citep{Naidu2025,Chang2025} in a dense medium could contribute to the broadening of the Balmer lines. Finally, in this picture the dense gas, provided the column density $\gtrsim 10^{25}\,\rm cm^{-2}$, may be Compton thick and absorb emitted X-rays, reconciling the observed X-ray weakness. Alternatively, the low X-ray luminosities could be linked to high black hole accretion rates, as proposed in some super-Eddington models \citep[e.g.][]{Pacucci2024,Inayoshi2024:xray,Lambrides2024}, possibly facilitated by the geometry of the gaseous envelopes. 

Importantly, the interpretation of these various observations has been largely phenomenological, that is, none of the aforementioned studies self-consistently model the AGN properties, the structure of the gaseous envelope, and its sustained accretion onto the black hole. From a theoretical perspective, the existence of black holes embedded in dense gas has been discussed in the context of the direct collapse of the first massive stars and the subsequent rapid growth of black hole seeds \citep[e.g.][]{Begelman2006,Tal2014,Coughlin2024}. The quasistar model proposed by \citet{Begelman2008} is qualitatively most similar to the scenario discussed for LRDs \citep{Inayoshi2025,Ji2025,Naidu2025,deGraaff2025}, but with the key difference that the envelope is massive compared to the black hole mass ($\gtrsim10\times \Mbh$; as opposed to the low-mass envelopes suggested by, e.g., \citealt{Inayoshi2025,Naidu2025}) and consists of a radiative layer and outer photosphere.

\citet{Begelman2025} recently used analytical arguments to propose that LRDs may be the stable (tens of Myr) final stage in the evolution of quasistars, as they suggest that this late phase naturally produces red rest-optical continua and broad emission lines from electron scattering in the photosphere. This is similar to the work of \citet{Kido2025}, who argued that LRDs are black holes residing in massive, optically thick envelopes and demonstrate that this results in photospheres near the Hayashi limit, with temperatures $T\sim5000-7000\,$K that likely resemble the red rest-optical spectra of observed LRDs. To date, only \citet{Liu2025} have made detailed predictions for the expected continuum SEDs (from radiation transport calculations) for super-Eddington accreting black holes embedded in massive envelopes ($M_{\rm env}\sim\Mbh$). The spectra predicted from these models match observed LRD spectra well to first order, as they show red rest-optical continua and strong Balmer breaks.

Although these new observational insights and rapidly developing models are highly promising, key open questions on the structure of the gas and the properties of the central engines remain. A major focus has been to explain the continuum properties of individual LRDs, especially those with strong Balmer breaks \citep[e.g.][]{Ji2025,deGraaff2025,Naidu2025}, as well as a recently discovered local analogue \citep{Lin2025,Ji2025b}. Whether the proposed gaseous envelope models can also describe the rest-optical continua of a broader population of LRDs \citep{Hviding2025} is still unclear, as is the physical connection between the UV and optical continuum properties. Furthermore, the Balmer emission lines have sparked a separate debate. The kinematics of the broad wings may reflect either a broad line region \citep[e.g.][]{Greene2024,Kocevski2024,Brazzini2025} or arise from scattering processes \citep{Rusakov2025,Chang2025,Naidu2025,Torralba2025}, with major implications for the inferred black hole masses and growth rates. In addition, the \Ha equivalent widths exceed that of lower-redshift AGNs by a factor $\gtrsim 2$ \citep[e.g.][]{Lin2024}, raising the question how this can be reconciled with the relatively red accretion disk spectra that have been invoked to explain the observed red rest-optical continua \citep[e.g.][]{Liu2025,Naidu2025,deGraaff2025,Wang2025}. 

To begin to tackle these major questions it is critical to map the large parameter space spanned by LRDs. For example, although all LRDs are selected to have v-shaped UV-optical continua, only $\sim50\%$ have an inflection point at the Balmer limit \citep{Setton2024}. The emission line properties also vary strongly, as some sources show strong nebular emission from a range of metals including ionised Fe \citep[][]{Labbe2024,Lambrides2025,DEugenio2025:iron,Torralba2025}, while in others only hydrogen and helium lines are strongly detected \citep[][]{deGraaff2025,Furtak2024}. The correlation (or lack thereof) between different emission line and continuum properties may thus shed important new light on the physical processes underlying the population of LRDs. 

In this paper, we leverage the complete public archive of observations obtained with the JWST/NIRSpec microshutter array (MSA; \citealt{Ferruit2022}) in order to compile a large sample of 146 LRDs and analyse the population statistics of their key spectral features. The data used and our selection procedure are described in Section~\ref{sec:data}. We then explore the spectral shapes and continuum properties of the LRD sample in Section~\ref{sec:continuum}, revealing that the rest-optical to near-IR continua of LRDs are ubiquitously well described by single-temperature blackbody models. Next, we investigate the properties of strong emission lines (\Ha, \Hb, \Oiii, \Oi) in Section~\ref{sec:elines} and demonstrate, for the first time, a tight link between these lines and the continuum properties. Finally, we discuss our findings in the context of the black hole star models and explore the relative importance of host galaxies to the SEDs in Section~\ref{sec:discussion}, and conclude in Section~\ref{sec:conclusion}. We measure magnitudes using the AB system \citep{abmag}, and, where relevant, adopt a flat $\Lambda$CDM cosmology \citep{Planck2020LCDM} with $\Omega_{\rm m}=0.3$ and $h=0.7$.

\section{Data \& LRD sample}\label{sec:data}

\subsection{NIRSpec spectra}\label{sec:spec_data}

We use all public NIRSpec MSA spectra available in version {4.5} of the DAWN JWST Archive (DJA; \citealt{brammer_djav44})\footnote{\url{https://dawn-cph.github.io/dja/index.html}}, further supplemented with data from the Mirage or Miracle program (GO-5224, PIs Oesch \& Naidu; see e.g. \citealt{Naidu2025z14}). Spectra were reduced with \texttt{msaexp} \citep{msaexp} largely following the methodology described in \citet{Heintz2024} and \citet{deGraaff2024d}, using a local background subtraction from nodded exposures where possible. Briefly, version 4 of the DJA improves upon version 3 by employing an empirical correction to the wavelength calibration based on the source centroid in the shutter, resolving some of the wavelength offsets between the PRISM and higher-resolution spectra reported in \citet{deGraaff2024d}. It also implements new reference files for the flux calibration{, calibrated using observations of standard stars}, which enable spectral extraction beyond the nominal wavelength range \citep[see also][]{Valentino2025,Pollock2025}. The 1D spectra were extracted using an optimal extraction \citep{Horne86}, and the {width of the} extraction kernel was subsequently used together with the source position in the shutter {(based on the nominal pointing)} to correct for {wavelength-dependent point spread function (PSF) and slit losses}. For {compact and} point sources this slit loss correction typically performs well (see, e.g., \citealt{deGraaff2025,Hviding2025}), and we therefore use the spectra without further flux corrections. {We note that the \texttt{msaexp} pipeline does not correct for spatially extended flux that falls outside the shutter. For LRDs with bright offset blue components \citep[e.g.][]{Cloonan2026,Baggen2026} the placement of the slit may therefore significantly alter the observed SED shape; an extreme example demonstrating this effect is shown in Appendix~\ref{apdx:slitloss}.}

All spectra were visually inspected (G. Brammer) to vet the quality of the spectra and automated redshift fitting. We only use spectra with \texttt{grade=3}, which correspond to high-quality data with robustly determined redshifts. This results in a sample of $\sim$17,000 low-resolution PRISM spectra with \texttt{grade=3}, originating from large surveys including, in alphabetical order, CANUCS \citep{canucsdr1}, CAPERS (GO-6368; PI: Dickinson), CEERS \citep{Finkelstein2025}, JADES \citep{Eisenstein2023,CurtisLake2025,Scholtz2025}, NEXUS \citep{Shen2024}, NIRSpec GTO-Wide \citep{Maseda2024}, RUBIES \citep{deGraaff2025}, and UNCOVER \citep{Bezanson2024,Price2025}, as well as a wide range of smaller programs\footnote{for a detailed breakdown, see \url{https://dawn-cph.github.io/dja/blog/2025/05/01/nirspec-merged-table-v4/} and Table~\ref{tab:surveys}}. For sources from the UNCOVER survey we also obtain magnifications from the lens model of the Abell-2744 cluster of \citet{Furtak2023:lensing} and \citet{Price2025}, and use the best-fit values to correct the luminosities measured in Sections~\ref{sec:continuum} and \ref{sec:elines}. For sources in CANUCS clusters, we obtain best-fit lensing magnification factors from the CANUCS first data release \citep{canucsdr1}.

Many of the sources in the DJA were observed not only with the low-resolution ($R\sim 100$) PRISM, but also at higher resolution ($R>1000$). Where available, we use these higher resolution NIRSpec spectra in conjunction with the PRISM data for improved emission line fitting in Section~\ref{sec:elines}. Most relevant to this work are the G395M data (and in some cases also G235M, G235H, or G395H) from the RUBIES, JADES, and NIRSpec GTO-Wide programs.

\subsection{NIRCam imaging}

We obtain JWST/NIRCam image mosaics from the DJA (version 7; at a pixel scale of $0.04\arcsec$), which were reduced with \texttt{grizli} \citep{grizli} as described in detail in \citet{Valentino2023}. We focus only on the F444W filter, which is the most commonly used long-wavelength filter across all JWST fields. Indeed, a very large fraction (85\%) of all \texttt{grade=3} PRISM spectra have available NIRCam/F444W imaging, from a mixture of imaging programs including CANUCS \citep{canucsdr1}, CEERS \citep{Finkelstein2025}, JADES \citep{Eisenstein2023}, NEXUS \citep{Shen2024}, PRIMER (GO-1837; PI: Dunlop), and UNCOVER \citep{Bezanson2024}, as well as pure parallel programs such as PANORAMIC \citep{Williams2024:panoramic}. 

Because some of the spectra were observed prior to NIRCam imaging becoming available (e.g. large parts of the NIRSpec GTO programs were selected based on HST imaging), there may be small mismatches in the astrometric centre used for targeting and the source centroid measured in the F444W image. We therefore cross-match our spectroscopic sample to the photometric catalogues on the DJA to obtain the photometric source centroids. We then extract image stamps and measure fluxes in circular apertures of radii $0.1\arcsec$ and $0.2\arcsec$ centred on these photometric centres, which we use in Section~\ref{sec:selection}.

\subsection{LRD selection}\label{sec:selection}

To select our sample of LRDs we follow a strategy inspired by the work of \citet{Hviding2025}, who showed that the three defining properties of LRDs -- a v-shaped UV-optical continuum, dominant point source morphology in the rest-optical, and broad Balmer lines -- are inextricably linked. Crucially, this implies that selecting sources with v-shaped continua and compact rest-optical morphologies will virtually always yield detection of broad Balmer emission lines, at least for sources brighter than $\rm F444W<26.5$ (100\% of sources where data quality permits detection of broad lines; \citealt{Hviding2025}). 
Because the spectroscopic archive of the DJA primarily consists of PRISM data, for which the low resolution makes it difficult to robustly determine line widths, we only impose v-shape and compactness criteria to select LRDs. In what follows, we assert that these sources also have some broad kinematic component in their Balmer lines. We test this assumption in section~\ref{sec:linefit} using sources for which medium- or high-resolution spectra are available. 

\begin{figure}
    \centering
    \includegraphics[width=\linewidth]{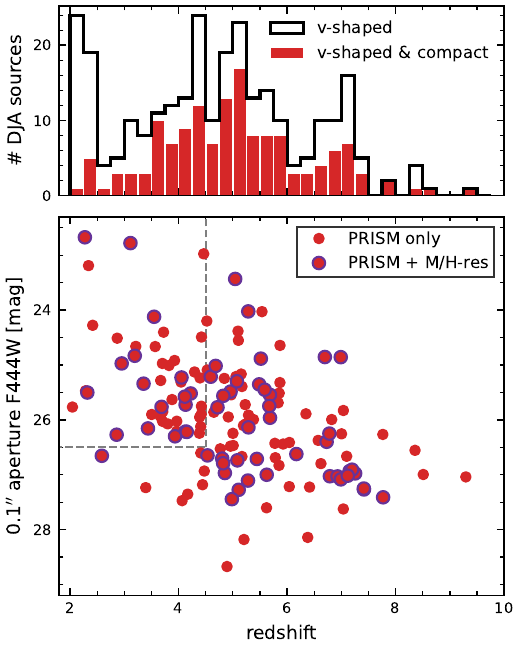}
    \caption{Top: Redshift distribution of all PRISM spectra selected from the public DJA that satisfy the spectroscopic v-shape criteria of \citet{Hviding2025} (black). The sample of v-shaped objects contains both LRDs and a large number of dust-reddened star-forming galaxies at cosmic noon. We define LRDs ({146} unique sources; red) as the subset of sources satisfying strict compactness criteria in NIRCam/F444W imaging. Bottom: NIRCam/F444W magnitudes measured in circular apertures of radius $0.1\arcsec$ (half the MSA slit width) of all LRDs as a function of redshift. Of these sources, {$40\%$ (58/146)} also have medium or high resolution spectra from NIRSpec. Dashed lines indicate the subsample used for the modified blackbody fitting of Section~\ref{sec:MBB}.}
    \label{fig:redshifts}
\end{figure}

We perform broken power-law fits, joint at the Balmer limit ($H_\infty$), of the form $f_\lambda = \alpha (\lambda/H_\infty)^\beta$ to the PRISM spectra across rest-frame $0.12-0.70\,\micron$ in order to determine the UV slope $\buv$ and optical slope $\bopt$. A region of width 100\,\AA\xspace is masked around strong emission lines (\Hi, \Hei, \Neiii, \Oiii, and \Oii) in the fitting. In addition, where photometry at rest-UV wavelengths are available we also fit the photometric UV slopes by integrating the model through the NIRCam bandpasses. We implement the broken power-law model in \texttt{NumPyro} \citep{phan2019} and employ Markov chain Monte Carlo (MCMC) sampling using the No-U-Turn Sampler \citep[NUTS;][]{hoffman2014} comprised of one chain with 250 warmup steps and 2500 posterior samples. We select sources with v-shaped continua by requiring all the following criteria:
\begin{enumerate}
    \item $>95\%$ of all samples satisfy $\bopt>0.0$
    \item $>95\%$ of all samples satisfy $\buv<-0.2$
    \item $>95\%$ of all samples satisfy $\buv - \bopt>0.5$
    \item Criteria 2 \& 3 must be simultaneously satisfied by $\buv$ measured from either photometry or spectroscopy. 
\end{enumerate}
This yields a sample of {333} v-shaped PRISM spectra, the redshift distribution of which is shown in the top panel of Figure~\ref{fig:redshifts}. 

Requiring a v-shape alone selects a large number of dusty star-forming galaxies at cosmic noon. We next impose compactness criteria to select sources for which the rest-optical morphologies are dominated by a point source component. Of the {333} sources, {323} have available NIRCam/F444W imaging. Because the sources are spread across a large number of fields, homogeneous empirical {PSF} models are not immediately available for all objects, which are necessary to perform detailed morphological fitting. We therefore use two different criteria, the first being a circular aperture flux ratio that was calibrated using PSF models for the selection of point-like objects \citep{Labbe2023b}. Second, where possible (the five CANDELS fields and Abell-2744), we also perform two-component point source and S\'ersic profile fitting with \texttt{pysersic} \citep{Pasha2023} using the empirical PSFs of \citet{Weibel2024}, in the exact same manner as described in \citet{Hviding2025}. We hence define compactness as either of
\begin{itemize}\vbox{
    \item $f_{\rm F444W}(0.2\arcsec)/ f_{\rm F444W}(0.1\arcsec) < 1.7$
    \item the point source component dominates ($>50\%$) the F444W model flux in $>95\%$ of all samples. }
\end{itemize}
Of sources that are both v-shaped and compact, {173 (152)} satisfy the first (second) criterion, with large overlap between the two selections. 
In total, our sample consists of {181} v-shaped PRISM spectra with a compact rest-optical morphology, although this still contains some duplicate observations.

We visually inspect the PRISM spectra of the entire sample, removing {7 spectra (6585-59553; a duplicate observation of RUBIES-UDS-40579; RUBIES-UDS-149298; 1208-3108879; 5545-85443; 5545-15494; a duplicate observation of UNCOVER-16594) where the location of the chip gap, data quality issues}, or background subtraction issues prohibit analysis of the key features explored in this paper (i.e. the Balmer break and strong rest-optical lines). Finally, we flag {35} duplicate observations of the same source or of multiple images of the same strongly-lensed source (i.e. the source of \citealt{Furtak2023}). In the following sections, we use the PRISM spectrum with the highest signal-to-noise ratio (S/N).

The redshift and F444W magnitude distribution of the sample of {146} unique LRDs is shown in Figure~\ref{fig:redshifts}. {Just under half of the sample (58/146) benefits from one or more higher resolution observations in the rest-optical, primarily from RUBIES (34) and the NIRSpec GTO programs (JADES and GTO-WIDE, 21), as well as GO-4106 (PI: Nelson, 9) and CEERS (1).} The full LRD sample spans a broad redshift range of {$2.0<z<9.3$ (the extremes being source 1020210 of GO-8018 (PI: Lin) and CAPERS-119334 of \citealt{Taylor2025}), and contains many well-studied LRDs such as A2744-45924 \citep{Labbe2024}, A2744-QSO1 \citep{Furtak2024}, RUBIES-BLAGN-1 \citep{Wang2024a}, \ruby \citep{deGraaff2025}, MoM-BH*-1 \citep{Naidu2025}, and the Rosetta Stone \citep{Juodzbalis2024}}. An overview of the selected sources is provided in Table~\ref{tab:surveys}, and we publicly release a complete table of all spectra and measurements made in this work together with this paper\footnote{\url{https://doi.org/10.5281/zenodo.17665942}}, the column entries of which are described in Table~\ref{tab:data_release}. 

We stress that both our v-shape and compactness selection criteria are strict, and therefore yield a large, but certainly not complete, sample of LRDs. For the purpose of this work we prioritise purity over completeness in order to gain an understanding of the physical properties of and diversity among LRDs. 
The incompleteness can be seen in Figure~\ref{fig:redshifts}, as the sample is skewed toward relatively bright F444W magnitudes. At higher redshifts this is primarily the result of our stringent v-shape selection, which, as discussed in detail in \citet{Hviding2025}, requires sufficient S/N to robustly determine whether $\beta_{\rm opt}>0$. At lower redshifts the cause of the bias is less clear. Our selection is sufficiently sensitive to detect LRDs to fainter magnitudes, and the apparent cutoff at $\rm F444W\sim26.5$ therefore could be physical, translating to a lower luminosity limit of $L\sim10^{43}\,\ergs$. However, the large number of spectroscopic programs among the sample have heterogeneous selection functions, making it difficult to assess whether selection bias could also play a role. 

Finally, we define a subsample of {47} sources that have coverage out to rest-frame $1\,\micron$, i.e. $\zspec<4.5$, and are well-detected, $f_{\rm F444W}(0.1\arcsec) < 26.5$ (where the aperture radius corresponds to half the MSA slit width). This sample is central to our discussion in the following sections, because the broad wavelength coverage allows for the blackbody fitting described in Section~\ref{sec:MBB}.

\subsection{Emission line measurements}\label{sec:linefit}

We use the \texttt{unite} \citep{raphael_erik_hviding_2025_15585035} software package to robustly constrain emission line properties from the NIRSpec MSA spectra, described in detail in  \citet{Hviding2025}. Briefly, \texttt{unite} was designed to fit spectra from multiple NIRSpec dispersers simultaneously, whilst rigorously accounting for the undersampling of the NIRSpec line spread function (LSF) as well as wavelength and flux calibration uncertainties \citep{deGraaff2024d}. For the LSF we assume that of an idealised point source, based on the methods described in Appendix A of \citet{deGraaff2024a}, rescaled by a nuisance parameter to capture calibration uncertainties in the LSF (de Graaff et al. in prep.). 

\begin{figure*}
    \centering
    \begin{subfigure}{0.6\linewidth}
        \includegraphics[width=\linewidth]{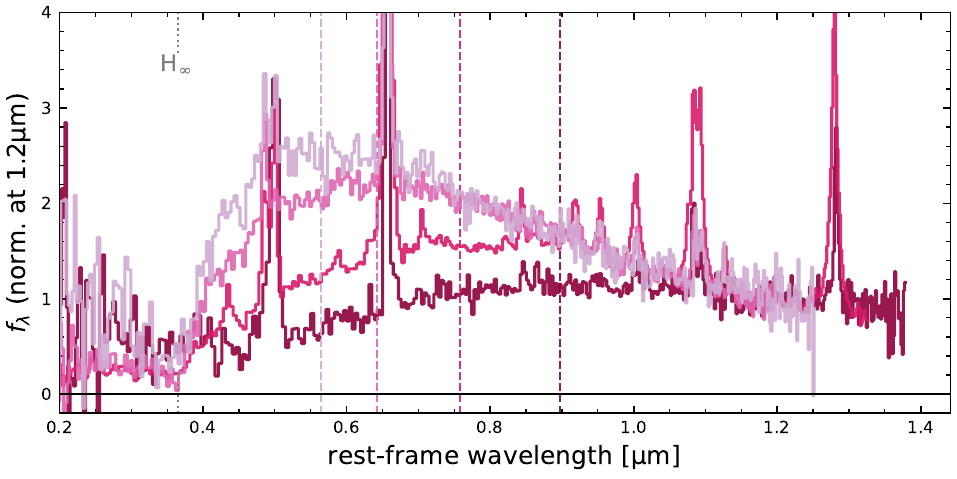}
    \end{subfigure}
    \begin{subfigure}{0.39\linewidth}
        \includegraphics[width=\linewidth]{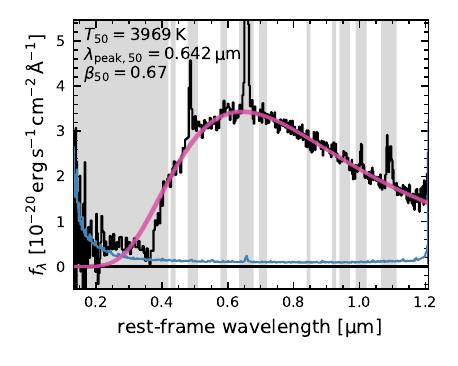}
    \end{subfigure}    
    \caption{Left: NIRSpec/PRISM spectra (normalised at $1.2\,\micron$) of four RUBIES LRDs showcasing their diverse spectral shapes. From light to dark: RUBIES-UDS-144195 \citep{Hviding2025}, \ruby \citep{deGraaff2025}, RUBIES-BLAGN-1 \citep{Wang2024a}, and RUBIES-UDS-175975. The colour coded vertical dashed lines indicate the peak wavelengths of the modified blackbody fits to the spectra. Right: Example of the 3-parameter modified blackbody fitting to the PRISM spectrum of \ruby, with its error spectrum shown in blue. Regions around strong emission lines and the region blueward of $\lambda_{\rm v}$, marked in grey, were masked in the fit. Model lines (pink) represent 100 random draws from the parameter posteriors.}
    \label{fig:shapes}
\end{figure*}

The emission model consists of (i) linear continua, (ii) two-component Gaussians (broad + narrow) for the Balmer lines, (iii) single-component Gaussians (narrow) for the \Oiii$_{\lambda\lambda4959,5007}$ doublet, and (iv) single-component Gaussians for the \Oi$_{\lambda6300}$ and \Oi$_{\lambda8446}$ lines. We do not include the \Nii doublet, as this to date has not been strongly detected in LRDs. The line widths of the broad Balmer components and narrow Balmer and \Oiii lines are tied together respectively, and the flux ratio of the narrow \Oiii doublet is fixed to 1:2.98. The \Oi lines are allowed to vary in width independently of the Balmer and \Oiii lines. In detail, the emission line kinematics may differ for each line, and the broad lines may not be Gaussian \citep[e.g.][]{Rusakov2025}, or there may be absorption features in the Balmer lines \citep[e.g.][]{Matthee2024}. For the majority of the sample we lack either the spectral resolution or S/N to robustly constrain individual profiles of Balmer and metal lines in such detail, but we find that the described setup yields good fits overall and therefore accurately recovers total line fluxes and equivalent widths \citep[see examples in Appendix B of][]{Hviding2025}. We defer a detailed analysis of the line kinematics and presence of absorbers for high S/N spectra to a future paper.

We incorporate all available spectra where possible, provided that the observations were taken in the same slit. That is, we fit simultaneously to PRISM and higher-resolution spectra where available, but do not combine spectra obtained in different masks where the position of the source in the shutter differs. The priors are as described in \citet{Hviding2025}, with the exception of the broad line full width at half maximum (FWHM$_{\rm broad}$), which is set to a uniform distribution in the range $[\rm FWHM_{\rm narrow}+100\,,5000]\,\kms$. The prior for the \Oi lines is set to $[\rm 0,2000]\,\kms$.

In addition, we fit all objects without the broad component for the Balmer lines in order to test our previous assumption that our selected LRDs have broad lines. We perform model comparison using the Widely Applicable Information Criterion \citep[WAIC][]{Watanabe:WAIC} and the detection threshold described in detail in \citet{Hviding2025}. For {53 of the 58} sources with available medium- or high-resolution spectroscopy (in any grating) the spectra cover the \Ha line. We significantly detect broad Balmer lines in {52/53} (98\%), highlighting the high accuracy of our spectro-photometric LRD selection approach. Crucially, these sources with available higher-resolution spectra are broadly distributed in redshift-magnitude space (Figure~\ref{fig:redshifts}), and we can therefore expect similarly high purity for the sources where only PRISM spectra are available. 

Finally, for comparison to the LRD population we also fit the emission lines of the broad line sample of \citet{Hviding2025} that are not LRDs, consisting of 37 sources from RUBIES in the redshift range $3.3<z<7.2$. We use the same model as described above, but include a narrow \Nii doublet with fixed flux ratio 1:2.95.

\section{UV to near-IR continuum shapes}\label{sec:continuum}

The PRISM spectra provide a detailed view of the SED shapes across the rest-UV to optical, as well as the rest near-IR for sources at intermediate redshifts ($z\sim2-4$). Visual inspection of the sample reveals great variety in the spectral shapes of LRDs, demonstrated in Figure~\ref{fig:shapes} (left panel). The sample contains a large number of sources with strong Balmer breaks, as discussed extensively in previous literature (see references in Section~\ref{sec:intro}). Intriguingly, however, their spectral shapes differ significantly, as the rest-optical continua peak at different wavelengths and the rest-UV ranges from flat to blue UV slopes. Other sources (darkest line in Figure~\ref{fig:shapes}) do not show Balmer breaks; their rest-optical continua peak in the near-IR, and the location of the `v' is redward of the Balmer limit. For the reddest source among our sample, RUBIES-EGS-38434, the v is located at $\sim 5700\,$\AA\xspace (shown in Appendix~\ref{apdx:Teff}). 

We also find that, provided the spectra cover rest $>1\,\micron$, the LRDs consistently show a {peak in the rest-optical continuum followed by a} flattening or turnover (in $f_\lambda$) at optical to near-IR wavelengths. In other words, the LRDs appear ubiquitously blue in the near-IR, as was also shown for small samples and individual objects \citep[][]{Williams2024,Wang2024a,Setton2025,deGraaff2025}. We quantify these observations{, i.e. the shapes and peak wavelengths of the rest-optical continua,} in the following subsection by fitting modified blackbody models to the spectra.

\subsection{Modified blackbody fitting}\label{sec:MBB}

Motivated by the shape of the optical to near-IR continua {and the fact that many astrophysical phenomena show blackbody-like spectra}, we fit 3-parameter modified blackbody models of the form
\begin{equation}
    f_\nu = A_{\rm tot} B_\nu(T) \left ( \frac{\nu}{\nu_0} \right )^{\betaBB}\,,
\end{equation}
where $A_{\rm tot}$ is a scale parameter, and $B_\nu$ is the Planck function that depends on temperature $T$. The blackbody function is modified by a power law of slope $\betaBB$ and pivot frequency $\nu_0= c / (5500\,$\AA$) \approx5.45\times10^{14}\,$Hz (where $c$ is the speed of light). The purpose of the pivot frequency is only to reduce the dynamic range in $A_{\rm tot}$, and the specific value chosen does not affect the posterior distribution of $\betaBB$. 

The power-law modification was inspired by other astrophysical contexts where it has been used to account for frequency dependence of the dust absorption cross section \citep[][]{Hildebrand1983}. The mathematical effect of this modification is a broadening or narrowing of the shape of the SED with respect to a true blackbody. Importantly, the deviation from a blackbody can be caused by several effects including, but not limited to, dust attenuation. For example, a distribution in temperatures (i.e. a typical stellar population) would result in a broader SED shape.  In the context of this work, {the 3-parameter modified blackbody primarily provides a flexible empirical framework to quantify the shape (width and peak) and normalisation of the rest-optical continuum, although the power-law modification to the blackbody can also be interpreted as a nuisance parameter to establish how close the SED shapes of LRDs are to single-temperature blackbodies, which we discuss in further detail in Section~\ref{sec:model_discussion}.}

A perfect single-temperature blackbody has $\betaBB=0$. 
Deviations from $\betaBB=0$ primarily affect the Rayleigh-Jeans limit, where $B_\nu \propto \nu^2$ (or $B_\lambda \propto \lambda^{-4}$). That is, SEDs broader than perfect blackbodies will have $\betaBB<0$, and similarly $\betaBB>0$ for SEDs that are narrower in shape. 
There is a well-known strong degeneracy between $\betaBB$ and $T$ \citep{Shetty2009a}, and it is often difficult to precisely constrain the two, especially in the case of low S/N \citep{Shetty2009b}. Instead of temperature, we therefore additionally compute the peak wavelength of the model ($\lpeak$) which we find is more stable against variations in $\betaBB$ and $T$. {The additional benefit of $\lpeak$ is that it is a model-agnostic quantity. The modified blackbody framework used here is not a unique analytical description for the rest-optical SED, but $\lpeak$ may also be obtained from a broad range of alternative models, including fully agnostic polynomial expansions. }

\begin{figure}
    \centering
    \includegraphics[width=\linewidth]{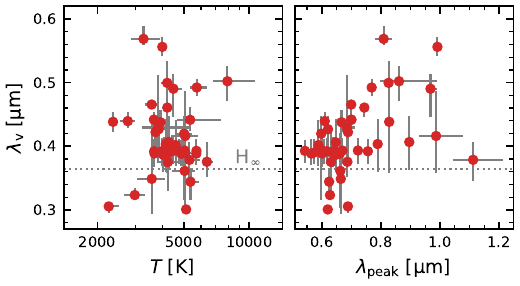}
    \caption{Location of the v-shape turnover wavelength $\lambda_{\rm v}$ versus modified blackbody temperature and $\lpeak$. For most LRDs $\lambda_{\rm v}$ is near the Balmer limit (there is a small systematic bias, see main text), but cooler LRDs with long peak wavelengths can have substantially larger $\lambda_{\rm v}$ than this. }
    \label{fig:vloc}
\end{figure}

\begin{figure*}
    \centering
    \includegraphics[width=0.9\linewidth]{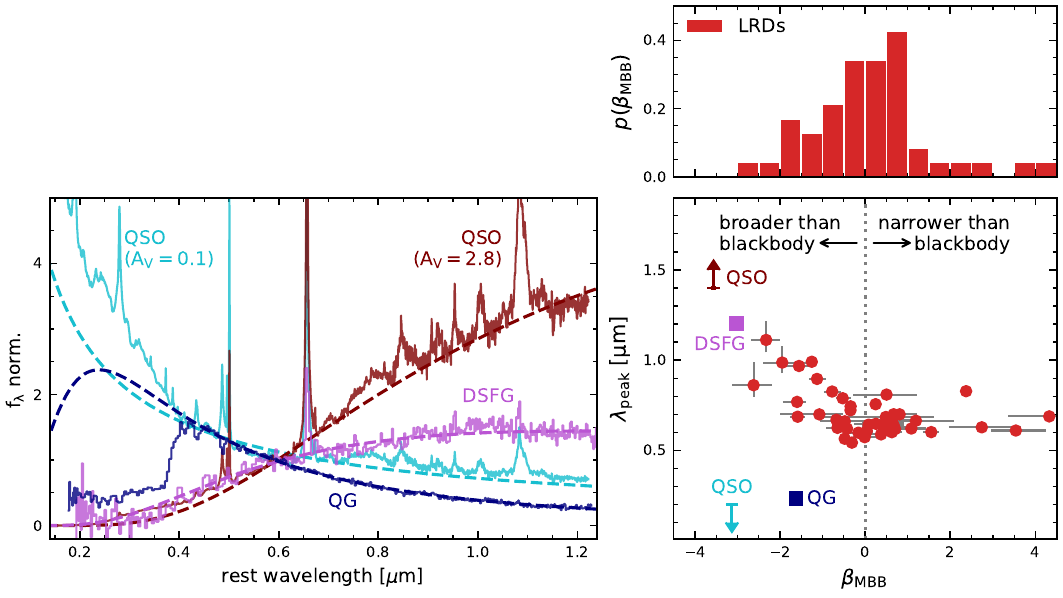}
    \caption{Distribution of the power law slopes and peak wavelengths from modified blackbody fits to the LRD sample restricted to $z<4.5$ (red). For reference, {the results of fits to the spectra of a massive quiescent galaxy, dusty star-forming galaxy, and a quasar composite (for different dust reddening) are also shown.} A slope of $\beta_{\rm MBB}=0$ corresponds to a perfect single-temperature blackbody. Whereas normal galaxies and quasars typically have broad SED shapes, {as their spectra consist of a mixture of stellar types or are dominated by a distribution in gas temperatures}, many LRDs are well approximated by single-temperature blackbodies. The peak wavelengths of galaxies {and quasars} span a broad range due to variation in recent star formation activity and dust attenuation, but LRDs have a characteristic continuum $\lpeak\sim0.6-0.7\,\micron$. }
    \label{fig:BB_beta}
\end{figure*}

Finally, we caution that in order to robustly constrain the three parameters of the model, spectral coverage both blueward and redward of the peak wavelength are crucial. 
For practical purposes we restrict the analysis of our fits to the $z<4.5$ sample highlighted in Figure~\ref{fig:redshifts}, the spectra of which (for all but {four} sources) reach redward of $\lpeak$. Strong emission lines are masked in the fit (the Balmer and Paschen series, \Oiii doublet, \Oi$_{\lambda8446}$, and optical to near-IR \Hei lines). 

\begin{figure*}
    \centering
    \includegraphics[width=\linewidth]{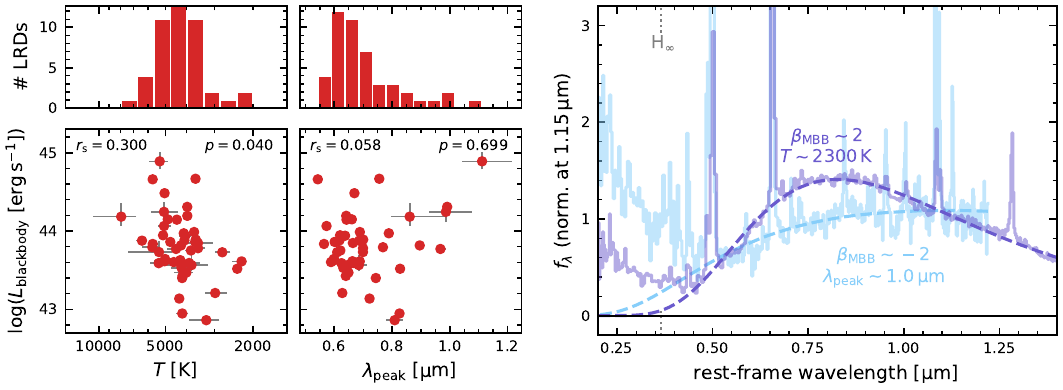}
    \caption{HR diagram of LRDs at $z<4.5$ for the modified blackbody temperature (left) and peak wavelength (middle). LRDs span a wide range in luminosity, with a lower luminosity limit that may be part physical and part driven by the imposed magnitude limit and depth of the PRISM spectra. The rest-optical continua of LRDs typically peak at a wavelength of $\sim 0.65\,\micron$, corresponding to a temperature of $\sim 5000\,$K for a perfect blackbody ($\beta=0$), although some systems are as cold as $\sim2000\,$K. The right panel shows examples of extremes in the HR diagram: a remarkably cold source ($T\sim2300\,$K, $\betaBB\sim2$, $\lpeak\sim0.83\,\mu$m; UNCOVER-20698), and a source that peaks in the near-IR ($\lpeak\sim1.05\,\mu$m, $\betaBB\sim-2$; CAPERS-COSMOS-30440). }
    \label{fig:HR}
\end{figure*}

We also mask the blue side of the v-shaped continuum. To do so, we first reperform our v-shape fitting to the PRISM spectra of Section~\ref{sec:selection}, but instead of fixing the transition wavelength to $H_\infty$, we fit for $\lambda_{\rm v}$ as a free parameter with a uniform prior in the range $[2500-6000]\,$\AA. The distribution in $\lambda_{\rm v}$ is shown in Figure~\ref{fig:vloc}. We find some sources with $\lambda_{\rm v}<3500\,$\AA\ as well as an overall slight offset with respect to the Balmer limit. As also discussed by \citet{Setton2024}, this is a result of the fact that SEDs with (strong) Balmer breaks and associated damping wings are not well described by the simplistic broken power-law model. 
However, for $\sim30\%$ of the sample $\lambda_{\rm v}\gtrsim4000\,$\AA\xspace and we find that this is not an artefact, but an accurate description of the SED (a gallery of SEDs and median $\lambda_{\rm v}$ values can be found in appendix Figure~\ref{fig:all_BBs}). 
We therefore mask all wavelengths blueward of rest 4200\,\AA\xspace or the 84th percentile of $\lambda_{\rm v}$, whichever is greater, for the rest-optical blackbody fitting{, although we note that we have tested a variety of masking strategies (e.g. 4000\,\AA\ or 4500\,\AA, as well as using fixed wavelength ranges rather than $\lambda_{\rm v}$) and find no significant difference in the inferred parameters for 90\% of sources}. We perform MCMC fitting with \texttt{emcee} \citep{emcee}, using a uniform prior for $\betaBB$ ($[-5,5]$) and log-uniform priors for $T$ ($[10^0,10^7]\,$K) and $A_{\rm tot}$. The 3-parameter modified blackbody models fit the LRD sample very well {(we find typically $\chi^2_\nu \sim1$)}, an example of which is shown in Figure~\ref{fig:shapes} (figures for the complete sample can be found in Appendix~\ref{apdx:Teff}). 

\subsection{HR diagram of LRDs}\label{sec:HR}

Figure~\ref{fig:BB_beta} shows the distribution in $\betaBB$ and peak wavelength (a corresponding figure with temperature can be found in Appendix~\ref{apdx:Teff}). {For comparison and to provide intuition for the expected values of $\betaBB$ and $\lpeak$ in non-LRD systems, we also show the NIRSpec/PRISM spectra of a dusty star-forming galaxy (DSFG) and massive quiescent galaxy (QG) obtained from RUBIES and show the result of modified blackbody fits to their rest-optical continua. In this case $\betaBB\lesssim -1$, because typical stellar populations are comprised of a distribution of stellar temperatures, resulting in a broadening of the SED compared to a single-temperature true blackbody. The short peak wavelength of the QG reflects an A-star dominated spectrum, while the long peak wavelength of the DSFG reflects the strong dust-reddening of the stellar continuum. We further show the composite quasar spectrum constructed from the optical-NIR spectra of \citet{VandenBerk2001} and \citet{Glikman2006} provided by the STScI\footnote{\url{https://www.stsci.edu/hst/instrumentation/reference-data-for-calibration-and-tools/astronomical-catalogs/composite-qso-spectra-for-nir}} redshifted to $z=3.5$ and truncated to the wavelength range of the NIRSpec PRISM, with minimal ($\rm A_V=0.1$) and strong ($\rm A_V=2.8$) reddening by an SMC extinction curve applied \citep{Gordon2003}. The rest-optical quasar spectra are more challenging to fit with a 3-parameter blackbody, in part due to broad and prevalent optical \Feii emission. With $\betaBB\sim-3$, the spectra are substantially broader than a true blackbody, as may be expected from quasars with an accretion disc and broad-line region spanning a range in temperature. The peak wavelength is intrinsically very short, but can extend to $>1.5\,\micron$ depending on dust reddening.} 

The LRDs clearly stand out from {this set of ``normal'' sources}, as the average $\betaBB\sim 0$ and the distribution in peak wavelength is strongly clustered around a value of $0.6-0.7\,\micron$. In other words, {the rest-optical spectra of many LRDs are closely approximated by} true single-temperature blackbodies, with near-IR continua that are near the Rayleigh-Jeans limit. {Importantly, this clustering is not a selection effect, because our LRD SED shape selection (Section~\ref{sec:selection}) only requires a rising rest-optical slope over the wavelength range $H_\infty - 0.70\,\micron$. The observed turnover at longer wavelengths and clustering at $\lpeak\sim0.65\,\micron$ therefore forms a real physical characteristic of the LRD population.} The scatter about $\betaBB$ is partially physical, but we also note that in our modelling we have not accounted for wavelength-dependent slit losses beyond the nominal correction made by the reduction pipeline, and any imperfections in this correction are therefore absorbed by $\betaBB$. {We quantify the impact of this in two ways. First, a subset of sources were also in version 3 of the DJA, which used a fundamentally different flux calibration method. By comparing fits between versions 3 and 4 of the DJA, we find that the values of $\betaBB$ can change by up to $\Delta\betaBB<1.0$. Second, for 8 sources we have identified duplicate observations from the same or other programs. These observations also have a different spatial aperture, which can cause real physical variation in the SED (e.g., in the case of a spatially extended UV component{, see Appendix~\ref{apdx:slitloss} for an extreme example}) on top of revealing possible systematic uncertainties in the spectral extraction. Comparison between the fits for sources with duplicate observations reveals changes in $\Delta\betaBB\sim0.1-1.0$, but no systematic bias. We therefore conclude that the value of $\betaBB$ can only be interpreted physically for large increments (i.e. $\Delta\betaBB>1$). }

We identify two types of outliers in Figure~\ref{fig:BB_beta}. First, there are three sources with $\lpeak\approx1.0\,\micron$ and $\betaBB\lesssim-2$, which would imply that their SEDs are red in the rest near-IR. However, these sources (one of which is shown in blue in the right panel of Figure~\ref{fig:HR}) lack coverage redward of $\lpeak$ in the near-IR. As a result, {we find that for these few sources small changes to, for instance, the mask used (e.g. the choice of the blue wavelength cutoff, the specific emission lines and their chosen kinematic widths) can have a significant effect on the inferred $\betaBB$}. Further constraints on the IR SED {from JWST/MIRI imaging or spectroscopy} are therefore needed to robustly constrain their spectral shapes and temperatures. 

The second group of outliers at $\betaBB>1$ is intriguing, as these sources have remarkably narrow SED shapes with v-shapes that lie redward of the Balmer limit. An example is shown in purple in Figure~\ref{fig:HR}. These objects also correspond to some of the coldest temperatures in our sample, reaching $\sim 2000\,$K or $\lpeak\sim0.8\,\micron$. Such a subpopulation of very cool LRDs has, to our knowledge, not been reported thus far, and provides insight into the physical properties of the gaseous structures of LRDs (Section~\ref{sec:model_discussion}).

{Finally, we explore the relation between the inferred luminosity and peak wavelength or temperature in Figure~\ref{fig:HR}, which we will refer to as an LRD version of the Hertzsprung-Russell (HR) diagram.} We use the integrated luminosity of the modified blackbody fits, which is close to the bolometric luminosity of LRDs \citep{Greene2025}. Wien's law approximately holds, as $\betaBB\sim0$, such that the characteristic peak wavelengths of $\sim0.6-0.7\,\micron$ correspond to typical temperatures $T\sim4000-6000\,$K, which agrees well with values estimated for the {LRD} photosphere models of \citet{Liu2025}, \citet{Kido2025}, and \citet{Begelman2025} (see Section~\ref{sec:model_discussion} for further discussion){, as well as the self-gravitating AGN disc models for LRDs of \citet{Chen2026}}. Interestingly, despite the wide range in luminosity spanned by the sample (2 dex), the temperature and $\lpeak$ do not appear to depend strongly on the luminosity, although the very coldest sources ($T<3000\,$K) are all less luminous than the average (i.e., $L<10^{44}\,\ergs$). {Crucially, we stress that the finding that LRDs cluster in a narrow range in rest-optical SED shapes over $>2$ dex in luminosity holds true even when using the model-agnostic quantities $L_{5100}$ and $\lpeak$ (the latter of which can be obtained in principle be measured using a variety of methods), as shown in Figure~\ref{fig:HR5100}.}%

\subsection{Linking the UV and optical }\label{sec:shapes}

\begin{figure}
    \centering
    \includegraphics[width=\linewidth]{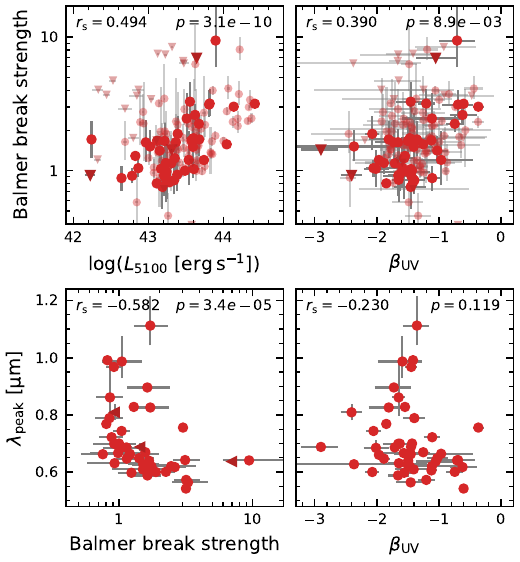}
    \caption{Distribution of parameters describing the continuum SEDs of LRDs: Balmer break strengths, optical luminosities, peak wavelengths, and UV slopes. Results from modified blackbody fitting (bottom) are shown only for the subset of LRDs with $\zspec<4.5$; the upper panels also include the higher redshift subsample. Spearman correlation coefficients and p-values are listed in each panel. Upper limits (triangles) on the Balmer break strength, corresponding to the 84th percentile, are shown for spectra with low S/N at rest $\sim0.36\,\rm \micron$. The Balmer break strength correlates significantly with the luminosity and shape of the UV-optical SED, reflecting a systematic variation in the SEDs across the LRD population. }
    \label{fig:cont_props}
\end{figure}

We have demonstrated that the SEDs of LRDs are well approximated by modified black bodies across $\sim0.4-1.0\,\micron$. Next, we explore whether the optical properties are correlated with the spectral shapes at $<0.4\,\micron$. The UV emission from LRDs has been ascribed to star-forming host galaxies in some studies, as many LRDs show spatially extended UV emission \citep[e.g.][]{Rinaldi2024,Torralba2025a}, while others have found strong UV emission lines in LRDs that can only reasonably be attributed to AGN emission \citep{Labbe2024,Akins2024b}. {In this section, we consider only the UV emission in the aperture of the slit, extracted with the same Gaussian kernel used for the rest-optical and near-IR. This mitigates emission from spatially extended host galaxies, although the observed UV light remains a mixture of possible sources (discussed in further detail in Section~\ref{sec:host_discussion}). }

Following \citet{deGraaff2025}, we calculate the Balmer break strength across $[0.362,0.372]\,\micron$ and $[0.40,0.41]\,\micron$ from the PRISM spectra, estimating uncertainties using random draws of the error spectra. 
Figure~\ref{fig:cont_props} shows that the Balmer break correlates positively with the optical continuum luminosity ($L_{5100}$, measured from the blackbody models). Moreover, there is a weak positive correlation between the UV slope (as measured in Section~\ref{sec:MBB}) and Balmer break, such that sources with redder UV slopes have stronger Balmer breaks. 
In contrast, the peak wavelength obtained from the modified blackbody fits correlates negatively with Balmer break strength, the only exception with a strong Balmer break yet long $\lpeak$ being RUBIES-BLAGN-1 \citep{Wang2024a}. There is great scatter in $\lpeak$ for sources with weak or no Balmer breaks. For a large fraction of our LRD sample $\lambda_{\rm v}>H_\infty$ and hence these sources do not show a Balmer break, resulting in break strengths $\lesssim 1$.

\begin{figure*}
    \centering
    \includegraphics[width=\linewidth]{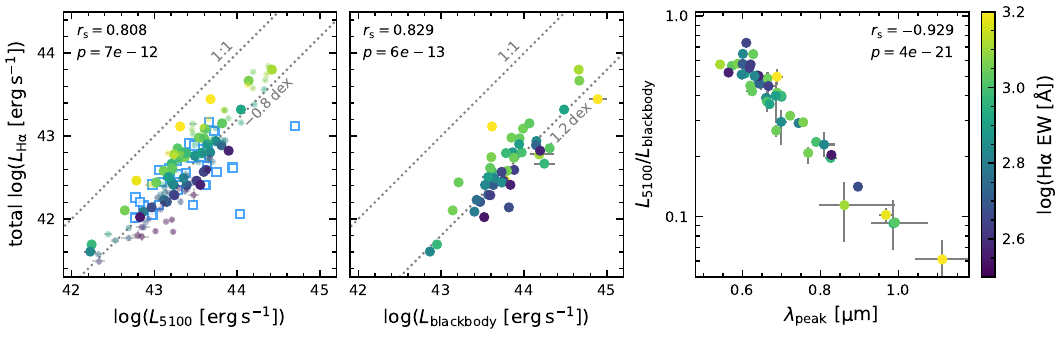}
    \caption{Total (broad + narrow) \Ha line luminosity vs. the luminosity at rest-frame 5100\,\AA\ (left) and integrated blackbody luminosity (middle), colour-coded by total \Ha EW. Large (small) circles show the $z<4.5$ (full) LRD sample, and blue squares the RUBIES broad line sources of \citet{Hviding2025} that are not LRDs. Grey dotted lines indicate linear correlations, offset by a constant factor. The scatter about the unit slope is equally small ($0.2\,$dex) for both $L_{5100}$ and $L_{\rm blackbody}$, despite the broad range in $L_{5100}/L_{\rm blackbody}$ spanned by the sample (right) as the two continuum luminosities are related via the peak wavelength. These tight relations point to a common origin for the production of the \Ha lines and rest-optical continua of LRDs.  }
    \label{fig:LHa_L5100}
\end{figure*}

Crucially, {this points to a coupling between the UV and optical emission, possibly indicating direct a physical association between the two components, although it is also possible that variation in, for instance, the host-AGN ratio drives the observed correlation between Balmer break strength, $L_{5100}$ and $\lpeak$. In either case, it is interesting to note that} the reddest sources in the UV do not correspond to the reddest sources at longer wavelengths, as would {at face value} be expected in case of dust reddening. Sources with the reddest UV slopes have comparatively short $\lpeak$ (bottom right panel of Figure~\ref{fig:cont_props}), whereas sources with bluer UV slopes have a broad range in $\lpeak$. Coupled with the {frequent detection of strong Balmer breaks, a feature only produced in dense neutral hydrogen gas}, this lends further support for the interpretation that the SEDs of LRDs are not reddened by dust, but by the effects of dense, optically-thick gas. First, a large population of $n=2$ hydrogen in such dense gas absorbs UV light; second, at high optical depth high-energy photons are scattered to redder wavelengths. Finally, we mention, but do not show here, that we do not find any correlation between the other parameters of the modified blackbody ($\betaBB$, $T$) and UV slope, although the lack of correlation with temperature may simply be due to the limited dynamic range. 

Altogether these correlations {demonstrate} systematic variations in the spectral shapes of LRDs across the broad population, and as a function of luminosity. Notably, although the scatter in some regimes is large, one subset stands out among the different panels of Figures~\ref{fig:cont_props} and \ref{fig:HR5100}. {Sources with the shortest peak wavelengths} (i.e. $\lpeak<0.65\,\micron$) all tend to have high $L_{5100}$, strong Balmer breaks, and comparatively red UV continua. The most extreme examples of this are the sources reported by \citet{Labbe2024}, \citet{Naidu2025}, and \citet{deGraaff2025}. We discuss the luminosity dependence as well as the origin of the UV emission and possible host galaxy contribution in further detail in Section~\ref{sec:discussion}.

\section{Relation between line and continuum properties}\label{sec:elines}

The diversity in continuum properties found in Section~\ref{sec:continuum} also extends to the emission lines. In this section, we focus on characterising a set of key strong optical emission lines (\Hb, \Oiii, \Ha, \Oi), and show that their properties can be linked to the shapes of the UV to near-IR continuum shapes.

\subsection{\Ha emission}\label{sec:Halpha}

We first explore the properties of the \Ha line, focusing on the total luminosity and equivalent width (EW). We do not consider the broad and narrow components separately, due to the limited spectral resolution for a large fraction of the sample, and the fact that individual components of the double Gaussian model might not be physically meaningful if the true line profile is Lorentzian or exponential. However, observations with the medium-resolution ($R\sim1500$) NIRCam grism have shown that the \Ha lines of LRDs are consistently dominated by the broad component \citep[forming $\sim 70\%$ of the total luminosity;][]{Lin2024}. Emission from potential star-forming host galaxies is thus subdominant to the emission from the central engine and unlikely to affect population trends -- our primary interest -- but may introduce scatter.

\begin{figure*}
    \centering
    \includegraphics[width=\linewidth]{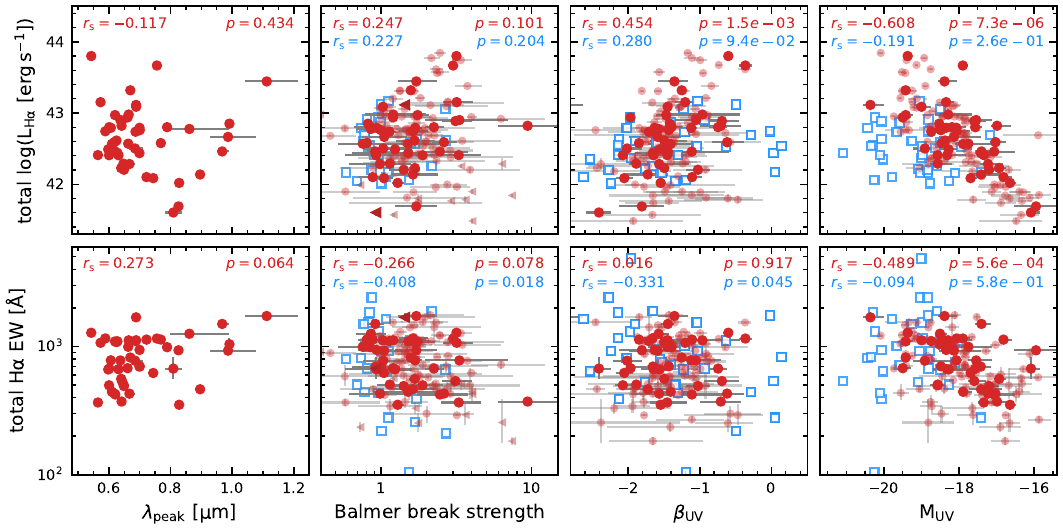}
    \caption{\Ha line properties as a function of the optical continuum shape parameters ($\lpeak$, Balmer break) and UV continuum properties ($\buv$, $\Muv$). Large (small) circles show the $z<4.5$ (full) LRD sample, and blue squares the RUBIES broad line sources of \citet{Hviding2025} that are not LRDs. Spearman correlation coefficients and p-values are shown in each panel for the LRD sample. The \Ha emission is independent of the rest-optical continuum shape, but is correlated with the rest UV, as LRDs with the reddest UV slopes are most luminous in \Ha. The UV luminosities of LRDs correlate with both the \Ha luminosities and EWs, and suggests that a substantial fraction of the UV can be attributed to emission from AGN.  }
    \label{fig:LHa_corr}
\end{figure*}

Figure~\ref{fig:LHa_L5100} shows there is a tight relation between the \Ha luminosity and optical continuum luminosity, $L_{5100}$. Such a relation had long been established for lower redshift AGN \citep[e.g.][]{Greene2005}, but we here demonstrate this correlation holds also for LRDs (both the full sample and $z<4.5$ subsample), as well as the sample of high-redshift broad-line sources that are not LRDs of \citet{Hviding2025}. We note that outliers, i.e. non-LRD broad-line sources at high $L_{5100}$, are AGN residing in massive host galaxies and therefore explained by a significant contribution from the host to the continuum. The slope of the relation (from linear regression to the $z<4.5$ sample\footnote{When fitting the full LRD sample these coefficients are {$43.30\pm0.03$ and $1.04\pm0.06$}, respectively.}, with uncertainties estimated from bootstrapping) is close to unity,
\begin{equation}
    L_{\rm H\alpha} = {10^{43.24\pm0.06}} {\ergs} \left(\frac{L_{5100}}{10^{44}\,\ergs} \right)^{{0.91\pm0.07}}\,,
\end{equation}
and the offset is approximately equal to that of the non-LRD sample. The scatter is modest at 0.23\,dex and, as expected given the proximity in wavelength, correlates strongly with \Ha EW.

Interestingly, the relation remains equally tight (0.23\,dex scatter about a unit slope) when considering the integrated luminosity of the blackbody fits instead of $L_{5100}$. The sample spans a wide range in $\lpeak$ and hence also $L_{5100}/L_{\rm blackbody}$,
\begin{equation}
    \frac{L_{\rm 5100}}{ L_{\rm blackbody}} = 10^{{-0.646\pm0.010}} {\ergs} \left(\frac{\lpeak}{0.8\,\micron} \right)^{{-1.94\pm0.07}}\,.
\end{equation}
The fact that the scatter is unchanged in spite of this wide variation may be due to a weak correlation between $\lpeak$ and the \Ha EW (right panel of Figure~\ref{fig:LHa_L5100}). 
The slope of the relation between $\log L_{\rm H\alpha}$ and $\log L_{\rm blackbody}$ is similarly close to unity and only the offset differs,
\begin{equation}
    L_{\rm H\alpha} = 10^{{42.82\pm0.04}} {\ergs} \left(\frac{L_{\rm blackbody}}{10^{44}\,\ergs} \right)^{{0.94\pm0.08}}\,.
\end{equation}

The implication of these trends is that the \Ha and continuum emission must primarily originate from the same source. Combined with the fact that the non-LRD broad-line sources follow the same scaling relation, this at face value points to photoionisation by AGN, although we will discuss the possibility of collisional excitation and the effects of high optical depth in and near the Balmer transitions in Section~\ref{sec:Balmer_discussion}. Moreover, it implies that $L_{\rm H\alpha}$, $L_{5100}$ are good tracers of the bolometric luminosity of LRDs, in addition to $L_{\rm blackbody}$ that was already presented in \citet{Greene2025}.

Next, we consider correlations between \Ha and the shape of the rest-optical and UV continuum. We do not find any significant trends between the \Ha luminosity or EW and parameters describing the shape of the rest-optical, such as $\lpeak$ and the Balmer break strength shown in Figure~\ref{fig:LHa_corr}, or $T$ and $\betaBB$ (not shown). 
The \Ha emission does depend weakly on the shape of the UV, as sources with redder UV slopes tend to have greater \Ha luminosities, but not necessarily higher EWs. {This appears in contrast with the broad-line sample that are not LRDs for which the \Ha luminosity is uncorrelated with $\buv$, but sources with bluer UV slopes have higher EWs, although a larger sample size of non-LRD broad-line sources is required to robustly confirm this trend.} 

\begin{figure*}
    \centering
    \includegraphics[width=0.75\linewidth]{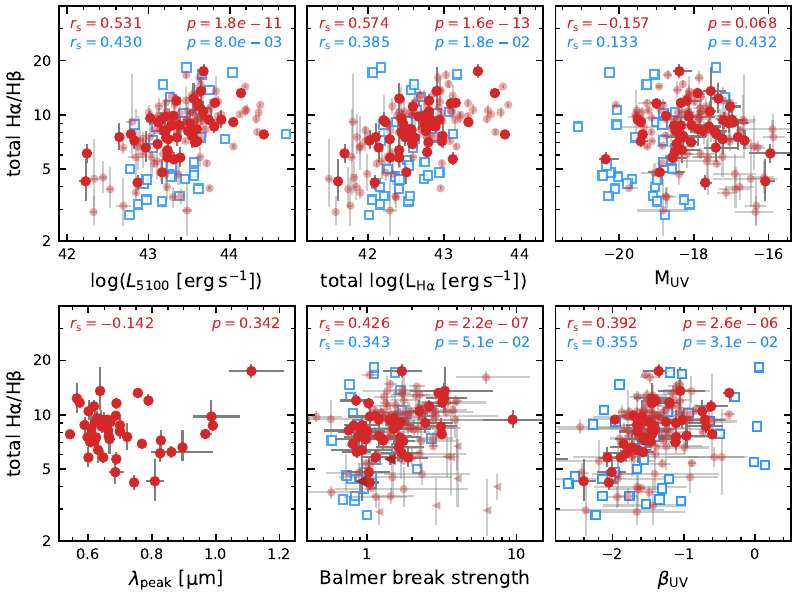}
    \caption{Balmer decrement of the total (broad+narrow) emission lines vs. different luminosities (top) and spectral shape parameters (bottom). Symbols indicate the same as in Figure~\ref{fig:LHa_corr}. LRDs typically have high Balmer decrements ($\rm H\alpha/H\beta\sim10$), which increase for sources with redder UV slopes. However, a lack of correlation with $\lpeak$ argues against dust reddening as the cause. The Balmer decrement correlates strongly with both the rest-optical continuum and \Ha luminosity, and weakly with Balmer break strength. This likely points to a luminosity-dependent population of hydrogen in the n=2 state in the gaseous envelopes of LRDs, which in turns affects the balance between the collisional (de-)excitation and optical depths of the \Ha and \Hb lines.  }
    \label{fig:Balmer_dec}
\end{figure*}

We also obtain UV magnitudes, $\Muv$, by evaluating the power-law slopes of Section~\ref{sec:continuum} at $1500\,$\AA\ (noting this is an extrapolation for sources at $z<3$, and that we consider only UV emission within the NIRSpec shutter). Figure~\ref{fig:LHa_corr} shows that the \Ha luminosities of LRDs are strongly correlated with $\Muv$, but systematically offset from the non-LRD population (as was already {demonstrated in the more homogeneous sample analyses of \citealt{Hviding2025} and \citealt{Lin2024}}),
\begin{equation}
    \log(L_{\rm H\alpha} / {\ergs}) = {(-0.39\pm0.06) \Muv + ({43.39\pm0.09}) }\,.
\end{equation}
Under the assumption that the \Ha luminosity primarily reflects emission from AGN, the moderate scatter of 0.34\,dex about the best-fit relation suggests a causal connection between the \Ha and UV emission, therefore implying a substantial fraction of the UV light in LRDs may originate from AGN (in addition to likely emission from host galaxies). 

In summary, despite following the same tight relation between $L_{\rm H\alpha}$ and $L_{5100}$, the \Ha lines of LRDs {seem to differ in some properties from other broad-line AGN. This possibly reflects a difference in the physical mechanisms producing Balmer line emission, which we discuss in Section~\ref{sec:Balmer_discussion}.}

\subsection{Balmer decrement}\label{sec:Balmer_dec}

The correlations between \Ha and the UV-optical continuum properties do not hold in the same way for higher order Balmer lines. We demonstrate this through correlations between the Balmer decrement ($\rm H\alpha/H\beta$) and continuum SED properties. Here, the Balmer decrement is defined as the ratio between the total (i.e. broad+narrow) \Ha and \Hb line flux, again assuming that emission from an underlying host galaxy is subdominant. 

The Balmer decrements of LRDs are typically very high, as the sample average and median of $\rm H\alpha/H\beta\approx8.7$ are three times greater than expected from case B recombination, with some sources as extreme as $\rm H\alpha/H\beta\approx17$ (NEXUS-2314). 
Surprisingly, we find that the Balmer decrement depends strongly on both the continuum luminosity $L_{5100}$ and \Ha line luminosity (Figure~\ref{fig:Balmer_dec}). {A similar trend is weaker among the broad-line sources that are not LRDs, which appear to span a broader range in $\rm H\alpha/H\beta$ at fixed luminosity.}

\begin{figure*}
    \centering
    \includegraphics[width=\linewidth]{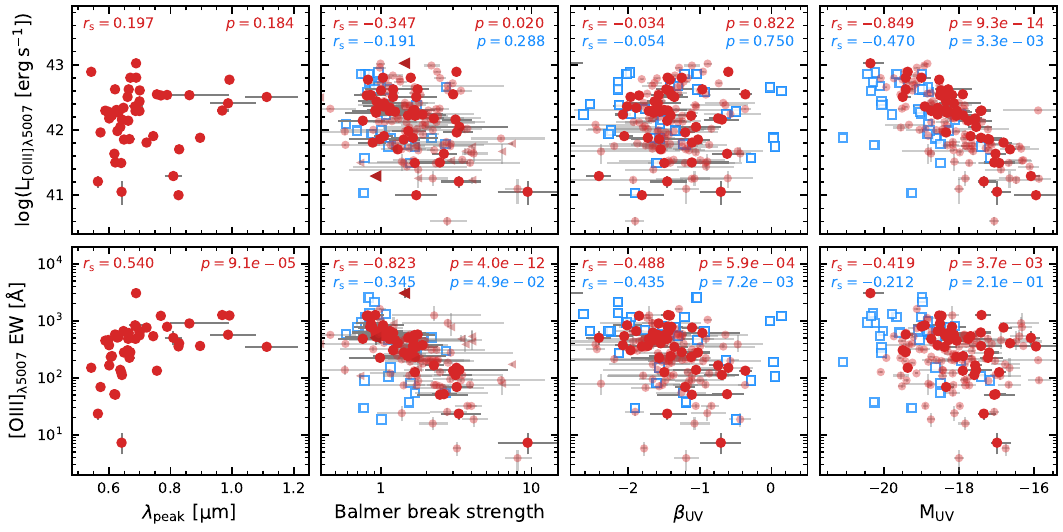}
    \caption{\Oiii$_{\lambda5007}$ line properties as a function of the optical continuum shape parameters ($\lpeak$, Balmer break) and UV continuum properties ($\buv$, $\Muv$). Symbols are as described in Figure~\ref{fig:LHa_corr}. The \Oiii luminosity depends only on $\Muv$, but the EW is correlated strongly with the UV and optical SED shape, especially the Balmer break strength. For LRDs these trends are opposite to those found for \Ha (Figure~\ref{fig:LHa_corr}), and indicate the \Oiii emission has a different origin, likely from host galaxies.   }
    \label{fig:O3_corr}
\end{figure*}

In contrast, there is no dependence on UV luminosity, which may be related to the fact that sources with high Balmer decrements also tend to have redder UV slopes. The reddening of both UV continua and Balmer lines is often interpreted as arising from dust extinction. However, as also discussed in Section~\ref{sec:shapes}, LRDs with redder UV slopes typically do not have the reddest optical continua. Figure~\ref{fig:Balmer_dec} shows that the Balmer decrement indeed does not depend on $\lpeak$ obtained from the modified blackbody fitting, and suggests different mechanisms than dust extinction are at play.

In this context, the weak correlation between Balmer decrement and strength of the Balmer break is noteworthy. A strong Balmer break points to a large population of dense hydrogen gas in the n=2 state, which has formed a key piece of evidence for the presence of gaseous envelopes in LRDs \citep[e.g.][]{Ji2025,Naidu2025,deGraaff2025}. The correlation with Balmer decrement therefore suggests it is driven by the physical conditions of the gas, that is, the effects of scattering and collisional (de-)excitation in dense gas. We further discuss this scenario in Section~\ref{sec:Balmer_discussion}.

\subsection{\Oiii emission}\label{sec:O3}

Next, we explore the properties of the rest-optical \Oiii emission. The kinematic properties of this forbidden transition have already been shown to differ greatly from the Balmer lines, as the \Oiii lines do not have a substantial broad component \citep[e.g.][]{Wang2024b,Juodzbalis2024,Hviding2025}. This suggests that the \Oiii emission has a different origin, and has been attributed to star-forming regions in host galaxies.

Figure~\ref{fig:O3_corr} shows the distribution of the \Oiii$_{\lambda5007}$ luminosity and EW as a function of UV and optical SED properties (these are the same panels as in Figure~\ref{fig:LHa_corr}). The \Oiii luminosity depends strongly on $\Muv$ but not the spectral shape itself, different from \Ha in Section~\ref{sec:Halpha}. Moreover, in this case there is no systematic offset in the line luminosity at fixed $\Muv$ between the LRD and non-LRD broad-line populations. 

The \Oiii EWs of LRDs do depend on the shape of the UV-optical SED, as there is a modest negative trend with $\buv$ such that sources with bluer UV slopes show stronger \Oiii emission. Most remarkable is the fact that the wide spread in EW, spanning nearly three orders of magnitude, is strongly anti-correlated with the strength of the Balmer break. A similar, but positive, correlation is seen with $\lpeak$, although this may simply reflect our previous finding that sources with strong Balmer breaks have short $\lpeak$ (see Section~\ref{sec:continuum}). 

These trends between the line EW and continuum shape are opposite to the results for \Ha of Figure~\ref{fig:LHa_corr}, and point to a different physical origin or geometry. For the \Oiii emission, the increase in line strength and luminosity for sources with bluer UV slopes and brighter $\Muv$ is consistent with expectations from photoionisation in star-forming regions. The fact that the weakest \Oiii lines are found in LRDs with the strongest Balmer breaks provides further insight into the physical origin (i.e. photoionisation by hot young stars or AGN). Because LRDs with strong Balmer breaks are also more luminous (Section~\ref{sec:shapes}), the AGN in these sources may outshine the host galaxies in which they reside. We investigate whether variations in the host galaxy-to-AGN ratio can reproduce the observed trend between \Oiii EW and Balmer break strength in Section~\ref{sec:host_discussion}.

\subsection{\Oi emission}\label{sec:O1}

\begin{figure}
    \centering
    \includegraphics[width=\linewidth]{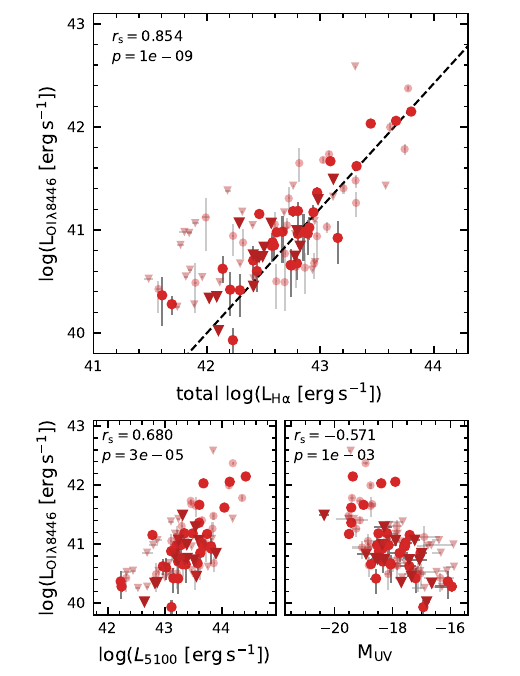}
    \caption{Correlation between the \Oi$_{\lambda8446}$ line luminosity and the total \Ha luminosity (top), and the continuum luminosity $L_{5100}$ and $\Muv$ (bottom).  Large (small) symbols show the $z<4.5$ (full) LRD sample, where triangles represent upper limits (95th percentile) in case of non-detections. The correlation with $L_{\rm H\alpha}$ is particularly strong; the best-fit relation to the full sample (dashed line) is nearly linear, as the power-law slope $\gamma=1.17\pm0.07$.   }
    \label{fig:OI}
\end{figure}

Finally, we also examine the properties of the \Oi$_{\lambda 8446}$ emission line. Different from the Balmer and \Oiii lines, \Oi traces neutral rather than ionised gas, although, as we will discuss further below, the ionisation state of \Oi is coupled to \Hii via charge-exchange reactions and \Oi emission is therefore expected to correlate with the hydrogen recombination lines. The \Oi emission has been of interest in the context of LRDs as it independently probes the physical conditions of the gas near the black hole and interstellar medium \citep{Tripodi2025,Juodzbalis2024,Labbe2024,Kokorev2025}. The 8446\AA\xspace transition in particular stands out as one of the brightest near-IR lines after hydrogen, and is accessible with JWST/NIRSpec up to $z\approx5.5$. Among our LRD sample, spanning a range of surveys with different depths, we detect \Oi$_{\lambda 8446}$ in approximately 60\% of sources (where detection is defined as the integrated line flux $>0$ for $>95\%$ of MCMC samples).

As in the previous sections, we test whether the line luminosity and EW are correlated with UV or optical continuum properties. We find that neither depend significantly on the spectral shape (e.g. $\buv$, $\lpeak$, Balmer break strength), but the luminosity of \Oi$_{\lambda 8446}$ does depend strongly on the luminosity of both the UV and optical, shown in Figure~\ref{fig:OI}. Above all, however, the \Oi$_{\lambda 8446}$ luminosity is strongly correlated with the \Ha luminosity. We fit a power-law relation, accounting for non-detections, and obtain uncertainties from bootstrapping:
\begin{equation}
    L_{\rm OI\,{\lambda8446}} = 10^{{41.13\pm0.04}} {\ergs} \left(\frac{L_{\rm H\alpha}}{10^{43}\,\ergs} \right)^{{1.27\pm0.10}}\,.
\end{equation}

The fact that the relation between $L_{\rm OI\,{\lambda8446}}$ and $L_{\rm H\alpha}$ is close to linear (this even holds true for the line fluxes rather than luminosities, with power-law slope $1.15\pm0.05$) provides direct physical insight. There are various excitation mechanisms for \Oi$_{\lambda 8446}$, including Ly$\beta$ fluorescence, collisional excitation, recombination or continuum fluorescence. The first two mechanisms, and in particular Ly$\beta$ pumping, have been shown to dominate in nearby AGN \citep[e.g.][]{Rodriguez2002}. Because the upper level of the cascade in which \Oi$_{\lambda 8446}$ is produced has the same energy as that of \Ha (i.e. $n=3$ state of hydrogen), the near-linear correlation between \Oi$_{\lambda 8446}$ and \Ha argues strongly in favour of Ly$\beta$ fluorescence as the primary excitation mechanism for LRDs. 
Notably, such a tight linear relation has been found before for Herbig Ae/Be stars (i.e. pre-main sequence stars with accretion disks), with identical normalisation of $L_{\rm H\alpha}/L_{\rm OI\,{\lambda8446}}\approx 70$ \citep{Mathew2018}. As discussed in \citet{Mathew2018}, by comparing this normalisation to the theoretical flux ratio under optically thin conditions \citep[$\sim7500$;][]{Strittmatter1977}, this implies \Ha is optically thick ($\tau_{\rm H\alpha}\sim 100$) and therefore independently supports the conclusion that Ly$\beta$ fluorescence dominates.

Importantly, this also suggests that \Ha and \Oi$_{\lambda 8446}$ are produced in approximately the same region, which may be surprising given the very different emission line kinematics. Whereas the Balmer lines have widths of $\gtrsim1000-4000\,\kms$, the \Oi lines are substantially narrower ($\rm FWHM<1000\,\kms$; \citealt{Tripodi2025,Juodzbalis2024,Labbe2024}). Although beyond the scope of this paper, as at the resolution of the PRISM the width of \Oi is difficult to robustly constrain (further exacerbated by blending from possible \Caii emission), we find that \Oi$_{\lambda 8446}$ is systematically narrower ($\rm FWHM\lesssim1000\,\kms$) also in our large sample of LRDs. This may be explained by the fact that the \Oi emission is weaker, and broad components are therefore difficult to detect unless deep medium-resolution spectroscopy is available \citep[e.g.][]{Kokorev2025}. However, if the absence of a broad component or low broad-to-narrow ratios are confirmed in a larger number of high S/N spectra, this may set important constraints on the gaseous structures in LRDs (see Section~\ref{sec:Balmer_discussion}).

\section{Discussion}\label{sec:discussion}

Using a large spectroscopically-selected sample of LRDs, we have shown that the population of LRDs are unified by the fact that the rest-optical continua are well described by blackbodies of typical $T\sim5000\,$K, although there is great variation in other continuum SED properties, such as the UV slope or Balmer break strength. Strikingly, we have found that the properties of the UV-optical continua are strongly correlated with the properties of strong rest-optical emission lines. In this section we interpret these different correlations in the context of the black hole star, quasistar and supermassive star models recently proposed in the literature.

\subsection{LRD photospheres}\label{sec:model_discussion}

The fact that LRDs are generally well described by modified blackbodies with $|\betaBB|<1$ in the rest-optical to near-IR, in contrast with normal galaxies that have broader SED shapes (Section~\ref{sec:HR}), {is consistent with} the presence of a single dominant thermal component. In turn, this strongly supports the recently proposed scenario where emission from a powerful source -- plausibly an AGN due to the high luminosity -- is reprocessed in dense, optically thick hydrogen gas, and fully dominates the rest-optical continuum. Such a physical picture had so far only been considered for individual LRDs, primarily sources with strong Balmer break features \citep[e.g.][]{Ji2025,Naidu2025,deGraaff2025}. Our work now demonstrates that this holds for the LRD population at large, and that LRDs with strong Balmer breaks form a subset of typically more luminous ($L_{\rm blackbody}>10^{44}\,\ergs$) sources. 

Most remarkably, our HR diagram of LRDs (Figure~\ref{fig:HR}) reveals that the majority of LRDs occupy a narrow range in temperature of $T\sim 5000\,$K for a broad range (2 dex) in luminosity, although the full population spans temperatures $T\sim2000-7000\,$K or $\lpeak\sim0.5-1.1\,\micron$. This near-vertical relation between temperature and luminosity is akin to the Hayashi track \citep{Hayashi1961} of pre-main sequence low-mass stars and red giants that are undergoing core contraction and expansion of their outer layers. Being nearly completely convective stars, their luminosity is largely decoupled from the interior temperature-pressure structure, and the vertical temperature-luminosity relation results from the fact that opacity in the cooler outer radiative layer is dominated by H$^-$ which has a steep temperature dependence \citep[e.g.][]{Kippenhahn1994}. The LRDs are at higher temperatures than the Hayashi tracks of low-mass stars ($\sim 3000$\,K), which could simply reflect the fact that the location of the Hayashi line depends in detail on the specific temperature-density structure of the envelope (for detailed explanation, see also \citealt{Nandal2025}, \citealt{Kido2025}); for instance, for pre-main sequence stars structural variations gives rise to a mass-dependence of the Hayashi limit. In this scenario, we also obtain a direct estimate of the physical scale of LRDs through the Stefan-Boltzmann law, resulting in a typical radius of $R\sim1000\,$AU ($\sim 10^{16}$\,cm; or $R\sim500-3000\,$AU for the distribution of $L$ and $T$ found in Figure~\ref{fig:HR}). {If truly analogous to convective stars, these envelopes must also be in approximate hydrostatic equilibrium if they are to maintain their large size over a lifetime of several Myr, although the degree of support provided by rapid rotation or non-equilibrium models are yet to be explored. }

Recent theoretical studies \citep{Kido2025,Liu2025,Coughlin2024,Begelman2025,Nandal2025,Zwick2025,Zhang2025} have proposed a range of different models to explain the observed properties of LRDs, most of which invoke the presence of a photosphere to produce rest-optical to near-IR continua that closely resemble blackbodies. Crucially, several of these models also predict that the photospheres must be close to the Hayashi limit \citep{Kido2025,Begelman2025,Liu2025} or truly following Hayashi tracks \citep{Nandal2025}, albeit with slightly higher characteristic temperatures of $T\sim5000-7000\,$K than found for our observed sample. The instability to convection in these models may, for example, be explained by outward pressure induced by high surface accretion rates \citep{Nandal2025}. Only the proposed accretion disk models of \citet{Zwick2025} and \citet{Zhang2025} do not invoke an optically-thick envelope or arguments related to the Hayashi limit. These models do accurately predict the temperature distribution, although it is at present unclear whether the observed steep luminosity-temperature dependence would also be expected in this scenario.
Interpreting LRDs as having large, convective envelopes therefore may offer a natural explanation for their wide luminosity range as well as their characteristic temperature.

Crucially, our work goes one step further by showing that the overall UV-optical spectral shape can be linked to the temperature, or rather the peak wavelength, of the rest-optical SED. These different correlations may help to distinguish between the range of theoretical models proposed. Specifically, LRDs with strong Balmer breaks and red UV slopes have shorter $\lpeak\lesssim0.65$. The fact that these features are also correlated with the optical continuum luminosity provides physical insight, as it suggests that more luminous sources are {more likely to outshine their host galaxies or are }able to populate a higher fraction of hydrogen gas in the $n=2$ state, and is therefore possibly related to the overall mass scale or accretion rate.

Moreover, a subset of cool LRDs ($T\sim2000-3000\,$K) seems to fall outside many model predictions. Their SEDs do not show Balmer breaks, and the position of the characteristic v-shape for some is substantially redward of the Balmer limit (Figure~\ref{fig:vloc}). Only \citet{Zwick2025} explicitly model the observed SED of such a cooler LRD \citep[the source of][]{Killi2023} of $T\sim3000\,$K, using a model that differs fundamentally from the picture that LRDs are AGN within optically-thick envelopes. Instead, \citet{Zwick2025} propose that LRDs are supermassive stars surrounded by accretion disks, which produce thermal emission at $\sim 2000-5000\,$K. In this context, the great similarity between the \Oi and \Ha correlation of LRDs and Herbig stars found in Section~\ref{sec:O1} is intriguing.

\begin{figure}
    \centering
    \includegraphics[width=\linewidth]{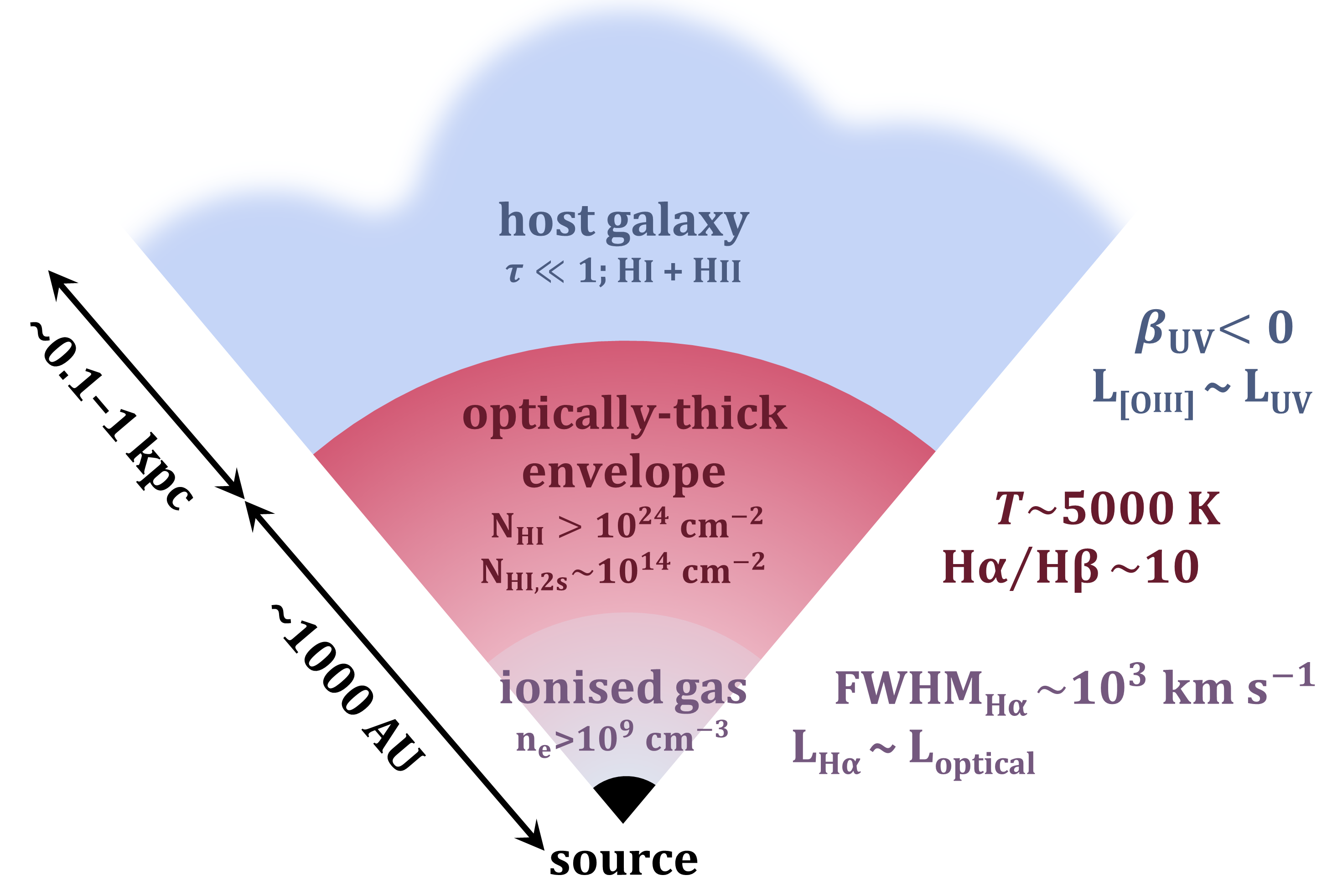}
    \caption{Schematic of the LRD's gaseous structures discussed in Sections~\ref{sec:model_discussion}, \ref{sec:Balmer_discussion} and \ref{sec:host_discussion}.}
    \label{fig:icecream}
\end{figure}

Among the AGN models, the late-stage quasistar model of \citet{Begelman2025} and envelope model of \citet{Kido2025} explicitly predict $T>5000\,$K. However, \citet{Liu2025} show that the observed temperature or peak wavelength of the photosphere depend on gas density, and densities $\rho>10^{-11.5}\,\rm g\,cm^{-3}$ can lead to redder SEDs without Balmer breaks. The existence of cool LRDs therefore does not rule out any of these theoretical models, but indicates that further refinement may be needed, as for instance these objects may have different dominant opacity (H$^-$ rather than bound-free). These sources could possibly even point to evolutionary trends among the LRD population. Although we found only a weak dependence of temperature on the total rest-optical luminosity (Figure~\ref{fig:HR}), which may be due to the small number of such sources in our sample, the coolest sources are all less luminous ($L_{\rm blackbody}<10^{44}\,\ergs$) than LRDs with $T>5000\,$K. This luminosity range is where \citet{Begelman2025} predict the evolutionary stage of normal quasistars, rather than late-stage quasistars (which are $L>10^{44}\,\ergs$), for which the observed temperatures could be different. 

Finally, in this discussion we have largely omitted one key parameter of our modified blackbody fits: the power-law slope $\betaBB$ modifying the Planck function. The sample is distributed around $\betaBB=0$, but with several strong outliers toward positive and negative values, indicating that some of the LRD SEDs are significantly different in shape from a single-temperature blackbody. SEDs for which $\betaBB<0$ are broader in shape, similar to more typical galaxies, and therefore reflects the presence of a broader temperature distribution. We find that the sources for which $\betaBB<-1$ have longer $\lpeak$ and higher UV luminosities, which may imply that the hot (i.e. UV) component of the SED contributes substantially also in the rest-optical. As a result, the sum of the hot and cooler component therefore may be broader, resulting in SEDs similar to, for example, those shown in \citet{Zwick2025}.

In contrast, the small number of sources that are significantly narrower in shape (i.e. $\betaBB>1$) still point to a single thermal component, but with substantial absorption in the Rayleigh-Jeans tail. These steep power-law slopes required for LRDs may be the result of broad molecular absorption features in the gaseous envelopes, analogous to those found in the atmospheres of cool stars. Indeed, we find that several of the sources with high $\betaBB$ also have some of the lowest temperatures, the strongest outlier of which being UNCOVER-20698 (shown in right-hand panel of Figure~\ref{fig:HR}). Detailed follow-up observations of the rest near-IR properties will be crucial to constrain the physical properties of LRDs with narrow SED shapes and cool temperatures, and to establish their connection to the broader LRD population, i.e. whether they represent an evolutionary stage or simply the tail of the distribution.

\subsection{Production and scattering of Balmer lines}\label{sec:Balmer_discussion}

The Balmer lines of LRDs have formed a key topic of debate. As shown in Section~\ref{sec:elines}, LRDs have \Ha luminosities that are higher than typical type-1 AGN of the same $\Muv$ and redshift \citep[see also][]{Hviding2025}, and very strong Balmer decrements ($\rm H\alpha/H\beta \sim 10$). Previously these features were primarily interpreted as strong dust extinction \citep[e.g.][]{Kocevski2023, Greene2024}, but recent follow-up observations of mid- and far-IR wavelengths have ruled out that LRDs are dust-reddened AGN \citep[e.g.][]{Casey2025,Setton2025}. The high Balmer decrements have also been associated with super-Eddington accretion rates \citep{Lambrides2024}, or could arise from resonant scattering of the Balmer lines in dense gas clouds \citep{Chang2025}. 

Moreover, the Balmer line profiles of some LRDs are exponential in shape and show significant absorption features \citep{Rusakov2025,Labbe2024,deGraaff2025}, likely pointing to the effects of electron scattering in dense, and at least partially, ionised gas \citep{Rusakov2025,Chang2025} and possible effects of resonance scattering \citep{Naidu2025}. In this  dense gas picture the gas is optically thick to the Balmer transitions, and the \Ha emission itself may be partially produced in outer layers of the envelope \citep{Begelman2025,Wang2025}, while \Hb emission may be reduced through radiative decay of \Hb photons into \Paa and \Ha photons in resonance scattering \citep{Naidu2025,Nikopoulos2025}. However, \citet{Brazzini2025} argue that scattering processes cannot dominate, due to dissimilarities in the profiles of different hydrogen recombination lines, and that the line widths and luminosities are instead consistent with more typical AGN broad-line regions.

The detailed geometry and physical conditions of LRDs therefore remain major open questions. The measured trends between various emission line and continuum properties among our large LRD sample provide a first, critical step forward to empirically constraining the nature of LRDs. We explore these trends largely agnostic to specific models, but under the assumption that an envelope of dense gas is present. Although we cannot conclusively rule out fundamentally different models \citep[e.g. the accretion disk model of][]{Zwick2025}, such models have not yet been shown to produce Balmer breaks or made predictions for the HR diagram.

First, identical to other high-redshift broad-line AGN, we find tight, near-linear relations between the \Ha and optical continuum luminosities of LRDs (Section~\ref{sec:Halpha}), strongly indicating that \Ha and optical continua are powered by the same source. A positive correlation between the \Ha and UV luminosities with only moderate scatter suggests that this central engine also produces UV continuum emission that can at least partially escape the gaseous envelopes \citep[see e.g.][]{Torralba2025}. Altogether, this points to the presence of a highly ionised region in the vicinity of a central engine (be it accretion onto a black hole or otherwise) where Balmer lines are predominantly produced by hydrogen recombination. 

However, the Balmer transitions in this dense gas are optically thick, corroborated by the consistently low \Ha to \Oi ratio (Section~\ref{sec:O1}), raising the question how these recombination lines can be observed. If the Balmer lines are sufficiently broad, or broadened via electron scattering, the broad wings may have substantially lower optical depth, leading to Balmer line profiles with strong absorption features near the line centres (as shown by \citealt{Chang2025} using radiative transfer simulations). Such absorption features are observed in a subset of LRDs \citep[e.g.][]{Matthee2024,Naidu2025,Juodzbalis2024}, but are not a ubiquitous feature (although this may be an observational bias due to the lack of high-resolution spectroscopy for many LRDs), and are not fully saturated. The emission at the Balmer line centres then likely must be produced, at least partially, in an outer region of lower optical depth. The tight relation with the optical continuum luminosity suggests that this emission is still predominantly physically associated with the central source, rather than for instance star formation in a host galaxy. The observed red optical continua and Balmer break features independently support the presence of such a region or layer of cooler, possibly less dense, gas. We provide a simple schematic overview of this structure in Figure~\ref{fig:icecream}.

Second, we find that the LRD sample shows very high Balmer decrements, which do not appear to be the effect of strong dust attenuation (see Section~\ref{sec:Balmer_dec}). Instead, these findings can be reconciled in the presence of very dense gas, and also suggest that the Balmer emission is not produced solely through photoionisation. Collisional excitation of the $n=2$ to $n=3$ state preferentially produces \Ha emission over that of \Hb, thereby boosting the Balmer decrement. The dependence of the decrement on optical and \Ha luminosity in this picture then may simply reflect the fact that more luminous, hotter sources have a larger population of $n=2$ hydrogen. 
Moreover, at the high electron densities expected in LRDs ($\gtrsim 10^9\,\rm cm^{-3}$; e.g. \citealt{Naidu2025,Ji2025}) collisional de-excitation will also play a role, resulting in reduced \Hb emission and hence also greater Balmer decrements. The third process that can substantially boost $\rm H\alpha/H\beta$ is resonant scattering of the Balmer lines, as \Hb is converted into \Ha and \Paa. As discussed above, the gas surrounding the central engine is optically thick to the Balmer transitions and therefore resonant scattering likely contributes substantially. 

\begin{figure}
    \centering
    \includegraphics[width=\linewidth]{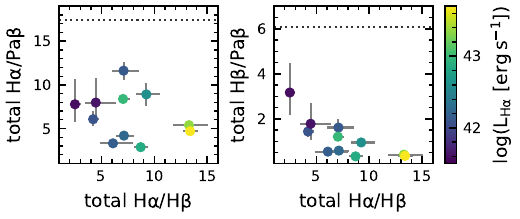}
    \caption{Paschen line ratios (left: $\rm H\alpha/Pa\beta$, right: $\rm H\beta/Pa\beta$) versus Balmer decrement, colour-coded by $L_{\rm H\alpha}$, for LRDs at $z<3.3$. Dotted lines indicate the expected value under the assumption of case B recombination. The very low $\rm H\beta/Pa\beta$ ratios support a scenario in which \Hb is resonantly scattered via \Ha and \Paa. }
    \label{fig:Pa_dec}
\end{figure}

It is difficult to assess which, if any, of these three processes dominates the observed Balmer decrements. The importance of resonant scattering can be quantified by the measurement of \Paa, under the assumption that the gas is optically thin to \Paa, but at $z>2$ this line is redshifted out of the observable wavelength range of JWST/NIRSpec. For a small number ({11}) of sources we are able to measure \Pab instead, and measure the $\rm H\alpha/Pa\beta$ and $\rm H\beta/Pa\beta$ ratios. In Figure~\ref{fig:Pa_dec} we show that these two ratios are substantially lower than expected from case B recombination, and depend on both \Ha luminosity and the Balmer decrement itself. The ratio of $\rm H\beta/Pa\beta$ is especially low, and therefore clearly suggests that resonant scattering contributes to the high Balmer decrements, bolstering previous claims of resonant scattering in LRDs in the literature \citep[e.g.][]{Naidu2025,Nikopoulos2025,Torralba2025}, but in disagreement with the findings of \citet{Juodzbalis2025} and \citet{Brazzini2025}. If the observed Balmer decrement arises predominantly through resonance scattering, the implied hydrogen gas density in the $n=2$ state from recent models of \citet{Chang2025} is $N_{\rm HI,2s}\sim10^{14}\,$cm$^{-2}$.

Scattering and collisional processes may also naturally explain the weaker correlations found between the properties of \Ha and UV-optical continua. In particular, in Section~\ref{sec:Halpha} we found that sources with higher $L_{\rm H\alpha}$ have stronger Balmer breaks and redder UV slopes. This is in contrast with, for example, the \Oiii emission which does not show such trends. Moreover, whereas there is no correlation between $L_{\rm H\alpha}$ and $\lpeak$ in Figure~\ref{fig:LHa_corr} for the entire LRD population, we do find a weak anti-correlation when considering these properties at fixed optical luminosity. In other words, the scatter in the relation between $L_{\rm H\alpha}$ and $L_{\rm blackbody}$ (Figure~\ref{fig:LHa_L5100}) correlates weakly with $\lpeak$, such that hotter LRDs (i.e. shorter $\lpeak$) may be brighter in \Ha. The combination of these weaker correlations then points to the production of \Ha emission from collisional excitation in the outer region of the gaseous envelopes, especially for more luminous sources. This emission should be kinematically distinct from both the broad emission and possible \Ha emission from the host galaxy, and therefore may be confirmed by deep high-resolution spectra.

In summary, in this section we have argued from an empirical standpoint that the Balmer emission in LRDs originates primarily from a central photoionised region and partially from collisional excitation, likely facilitated by the presence of a large population of $n=2$ hydrogen. The correlations between the emission line and continuum properties indicate that scattering processes must be important in altering the emission line ratios. These empirical findings agree well with the recently proposed 2-layer picture of \citet{Chang2025}, of an \Hii and \Hi region in which respectively electron scattering and resonant scattering dominate, and is also qualitatively in line with the quasistar model of \citet{Begelman2025}. To quantify the relative importance of the different physical processes inside LRDs will require further investigation of the Balmer and Paschen lines, by determining their flux ratios and kinematic components in detail.

\subsection{Little Red Dots as BH*s in faint host galaxies }\label{sec:host_discussion}

\begin{figure}
    \centering
    \includegraphics[width=\linewidth]{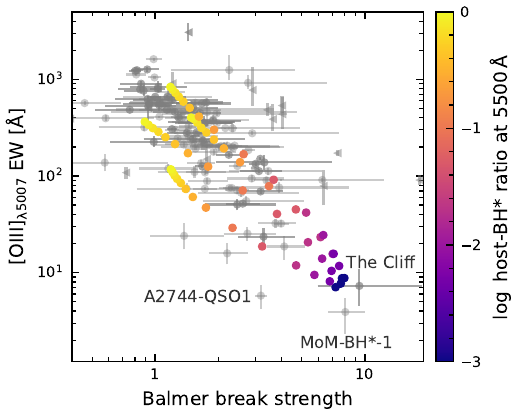}
    \caption{We investigate the strong correlation found between the \Oiii$_{\lambda5007}$ EW and Balmer break strength (Figure~\ref{fig:O3_corr}) empirically, by co-adding the spectra of The Cliff and low-mass ($M_*\sim10^8\,\rm M_\odot$) star-forming galaxies for a wide range in host galaxy-to-BH* ratios and measuring the resulting \Oiii EW and Balmer break strength. Our toy model, evaluated for four star-forming galaxy spectra, traces the slope and scatter of the observed distribution remarkably well. This suggests that LRDs can be described as the simple sum of a host galaxy and BH* component, and that variation in the host contribution alone can explain the weak \Oiii lines observed in LRDs with strong Balmer breaks \citep{Furtak2024,deGraaff2025,Naidu2025}. }
    \label{fig:O3_break}
\end{figure}

The remaining component not yet discussed is the host galaxy in which the BH* or quasistar must reside. That such hosts are present is clear from the fact that many LRDs have extended UV emission \citep[e.g.][]{Rinaldi2024,Torralba2025a}. Clustering analyses have shown that these galaxies are likely to be low in mass \citep{Matthee2024b}, independently supported by the detection of narrow \Oiii emission (implying low dynamical masses; e.g. \citealt{Juodzbalis2024,Wang2024b,Ji2025}), although the brightest LRDs may reside in more massive haloes \citep{Labbe2024,Schindler2025}. 

We find spectroscopic evidence of star-forming host galaxies from the observed UV and \Oiii emission. Our LRD sample shows great diversity in UV luminosity and slope, and for sources with bright UV emission (relative to the rest-optical) this emission contributes to an inferred broadening of the rest-optical SED shape (Section~\ref{sec:model_discussion}). Perhaps most convincing is the \Oiii emission, which shows different correlations than the \Ha emission and, due to its relatively low critical density, likely originates from an interstellar medium. The fact that there is a strong correlation between the \Oiii and UV emission supports this picture. Interestingly, this correlation (when measured within the aperture of the NIRSpec shutter) is approximately equally tight as that between \Ha and $\Muv$, and either suggests that the central UV emission originates from a mixture of AGN and host galaxy, or that the galaxy-scale properties correlate strongly with the properties of the AGN.

Recent work has argued in favour of a physical connection between the properties of the host and AGN. \citet{Maiolino2025} interpret the weak \Oiii emission relative to narrow \Hb emission in A2744-QSO1 (an LRD with a strong Balmer break; \citealt{Furtak2024}) as evidence for a galaxy of extremely low metallicity hosting a primordial black hole. In this context, the strong correlation found between the \Oiii EW and Balmer break strength in Figure~\ref{fig:O3_corr} is striking. 
However, we may also expect there to be natural diversity in the \Oiii EWs among the LRD population as a result of variation in the relative contributions from host and BH* to the total spectrum. 

To test this scenario we construct a toy model, adding the spectra of The Cliff \citep[][representing a ``pure'' AGN or BH*]{deGraaff2025} and low-mass star-forming galaxies at different flux ratios. We then evaluate the \Oiii EW and Balmer break strength as a function of the host:BH* ratio at 5500\,\AA. For this exercise we select four PRISM spectra from the UNCOVER survey \citep{Bezanson2024,Price2025}, chosen to have approximately the same redshift as The Cliff ($3.3<z<3.7$), a stellar mass of $\sim10^{8.0-8.2}\,\rm M_\odot$ (obtained from the DJA catalogues, measured from NIRCam photometry with \texttt{eazy} \citealt{Brammer2008}) and high continuum S/N, as well as diversity in emission line and UV properties. 

Crucially, we find that LRDs can be described as the simple sum of a host galaxy and AGN or BH* component. Figure~\ref{fig:O3_break} shows the resulting tracks of the toy models in the \Oiii EW vs Balmer break strength plane, which trace the observed population of LRDs remarkably well, and therefore indicate that variation in the host contribution dominates. Moreover, although A2744-QSO1 remains an outlier in this plane, our simple test also cautions against a broader interpretation of the \Oiii emission in LRDs as a metallicity indicator.

\section{Conclusions}\label{sec:conclusion}

We have used the complete public DAWN JWST Archive to select a sample of {146} LRDs with v-shaped rest UV-optical NIRSpec/MSA PRISM spectra and compact rest-optical morphologies in NIRCam/F444W imaging. Higher-resolution NIRSpec spectroscopy (available for 47\% of the sample) confirms the high purity of our LRD selection, as we detect broad Balmer lines in 98\% of sources. 

The LRD sample spans a broad range in redshift ($2.0<z<9.3$), luminosity, and SED properties. We perform a systematic characterisation of their SED properties, quantifying the shapes of the UV-optical continua as well as strong optical emission lines, summarised as follows: 
\begin{enumerate}
    \item Most strikingly, we find that the rest-optical to near-IR continua are well described by modified blackbody models with $|\betaBB|\lesssim1$ and are therefore {consistent with} a single thermal component. The LRDs have temperatures in the range $T\sim2000-7000\,$K ($\lpeak\sim0.5-1.2\,\micron$), but the distribution is strongly clustered around $T\sim5000\,$K ($\lpeak\sim0.6\,\micron$) with an asymmetric tail to colder temperatures. The blackbody luminosities span a broad range $\sim10^{43-45}\,\ergs$ and correlate only weakly with temperature in the HR diagram.
    \item The UV continuum properties vary systematically with the rest-optical continuum properties. Sources with strong Balmer breaks typically have redder UV slopes, are hotter (i.e. shorter $\lpeak$) and more luminous. Crucially, sources with the reddest UV slopes are less red in the rest-optical and near-IR than other LRDs, adding further evidence against dust reddening in LRDs and instead favouring reddening by absorption in dense gas.
    \item Balmer line properties are strongly linked to the continuum properties. We find a very tight correlation between the \Ha and rest-optical continuum luminosities, demonstrating that the two must be powered by the same source. We further find a strong correlation between the \Ha and UV emission, indicating that at least some UV emission can be attributed to the AGN. The Balmer decrements are typically very high ($\rm H\alpha/H\beta\sim 9$) and correlate with luminosity and Balmer break strength.
    \item The \Oiii$_{\lambda5007}$ line properties are also connected to the continuum SED, as the \Oiii and UV luminosities are strongly correlated. However, the properties of \Oiii differ in important ways from the Balmer lines, pointing to a different physical origin. Notably, we find a strong correlation between the Balmer break strength and \Oiii$_{\lambda5007}$ EW.
    \item The \Oi$_{\lambda8446}$ line luminosity is strongly correlated with the \Ha, UV and optical continuum luminosities. The relation between the \Oi and \Ha lines is near-linear (power-law slope of $1.21\pm0.09$) and indicates that the \Oi emission is primarily produced by Ly$\beta$ pumping. The normalisation of the relation shows that the gas from which these lines originate must have very high optical depth ($\tau_{\rm H\alpha}\sim100$).
\end{enumerate}

Crucially, our findings show that the LRDs differ systematically from galaxies of similar brightness and redshift, as the SEDs of typical galaxies are substantially broader than single-temperature blackbodies. The LRD properties and correlations thereof also indicate that dust attenuation plays little role in LRDs. Instead, many of the properties found can be naturally reconciled by the effects of dense, optically-thick gas. We therefore interpret our findings in the context of recently proposed black hole star and quasistar models. We hence identify four key components of LRDs:

\begin{enumerate}
    \item The central source produces ionising photons that give rise to Balmer emission. Given the high luminosities and ubiquity of LRDs, we consider accreting massive black holes to be the most likely source, although we cannot rule out categorically different models such as supermassive stars.
    \item A (partially) ionised region that is optically thick to Balmer transitions. Collisional excitation and de-excitation may boost the Balmer decrements. Scattering processes that cause line broadening may explain why Balmer lines are observed in spite of the high optical depth.
    \item An optically-thick envelope and radiative layer produce the observed rest-optical continuum emission. The photosphere (of radius $\sim1000\,$AU) follows a temperature-luminosity track analogous to the Hayashi track observed in pre-main sequence stars and red giants (albeit at slightly higher characteristic $T\sim 5000\,$K), with pressure support from the (potentially convective) envelope balancing gravitational collapse and ram pressure from accretion. 
    A high fraction of $n=2$ hydrogen in the outer layer of the envelope causes the Balmer break and may produce further narrow \Ha emission through collisional excitation. Moreover, resonant scattering likely contributes substantially to the high Balmer decrements. 
    \item The host galaxy is subdominant at rest-optical wavelengths and is primarily observed by its UV and nebular emission, such as \Oiii. There is large variation in the relative contribution of the host to the UV and nebular emission of the overall spectrum. A simple toy model describing LRDs as the sum of a low-mass star-forming galaxy and BH* with a range in host:BH* ratios can explain both the average trend and scatter of observed correlations, such as that between \Oiii$_{\lambda5007}$ EW and Balmer break strength.
\end{enumerate}

In summary, by compiling a large set of spectral diagnostics for a broad sample of LRDs with high-quality NIRSpec spectra, we have sketched an empirical picture of the structures of LRDs. A wide variety of models attempting to explain the observed properties of LRDs have recently been proposed in the literature. Our empirical constraints therefore offer an important means to test whether these models can explain not just individual outliers, but also sample trends. This serves as a critical step toward understanding the true physical nature of the population of LRDs and the detailed properties of their gaseous structures and black holes.

\section*{Acknowledgements}

AdG thanks Devesh Nandal for discussions on the properties of convective envelopes and Hayashi limit, and the students of the I-HOW JWST Workshop (Barbora Adamcova, Biswaraj Palit, Olena Pastoven, Weronika Puchalska) for their help in producing an early version of figures 8 and 10. \\
This work is based on observations made with the NASA/ESA/CSA James Webb Space Telescope. The data were obtained from the Mikulski Archive for Space Telescopes at the Space Telescope Science Institute, which is operated by the Association of Universities for Research in Astronomy, Inc., under NASA contract NAS 5-03127 for JWST. These observations are associated with programs 1180, 1181, 1208, 1212, 1213, 1215, 1286, 1345, 1433, 2198, 2561, 2750, 2767, 4106, 4233, 5105, 5224, 6368, and 6585.
Support for program \#4233 was provided by NASA through a grant from the Space Telescope Science Institute, which is operated by the Association of Universities for Research in Astronomy, Inc., under NASA contract NAS 5-03127. \\
AdG acknowledges support from a Clay Fellowship awarded by the Smithsonian Astrophysical Observatory. REH acknowledges support by the German Aerospace Center (DLR) and the Federal Ministry for Economic Affairs and Energy (BMWi) through program 50OR2403 `RUBIES'.
The Cosmic Dawn Center is funded by the Danish National Research Foundation (DNRF) under grant \#140.
This work has received funding from the Swiss State Secretariat for Education, Research and Innovation (SERI) under contract number MB22.00072, as well as from the Swiss National Science Foundation (SNSF) through project grant 200020\_207349. 
TBM was supported by a CIERA fellowship. PD warmly acknowledges support from an NSERC discovery grant (RGPIN-2025-06182).  KG acknowledges support from Australian Research Council Laureate Fellowship FL180100060. 
LAB acknowledges support from the Dutch Research Council (NWO) under grant VI.Veni.242.055 (https://doi.org/10.61686/LAJVP77714) and the ERC Consolidator grant 101088676 ("VOYAJ"). YF is supported by JSPS KAKENHI Grant Numbers JP22K21349 and JP23K13149.

\section*{Data Availability}

All data used in this work are publicly available through the DAWN JWST Archive (\url{https://dawn-cph.github.io/dja/index.html}). The spectra used in this work are part of version 4.4 of the DJA \citep{brammer_djav44}. We also release our complete LRD catalogue, together with all key measurements made in the modified blackbody fitting and emission line fitting. The table can be found at \url{https://doi.org/10.5281/zenodo.17665942}, and its contents are described in Appendix~\ref{sec:apdx_tables}.


\bibliographystyle{mnras}
\bibliography{lrds} 

@ARTICLE{Ferruit2022,
       author = {{Ferruit}, P. and {Jakobsen}, P. and {Giardino}, G. and {Rawle}, T. and {Alves de Oliveira}, C. and {Arribas}, S. and {Beck}, T.~L. and {Birkmann}, S. and {B{\"o}ker}, T. and {Bunker}, A.~J. and {Charlot}, S. and {de Marchi}, G. and {Franx}, M. and {Henry}, A. and {Karakla}, D. and {Kassin}, S.~A. and {Kumari}, N. and {L{\'o}pez-Caniego}, M. and {L{\"u}tzgendorf}, N. and {Maiolino}, R. and {Manjavacas}, E. and {Marston}, A. and {Moseley}, S.~H. and {Muzerolle}, J. and {Pirzkal}, N. and {Rauscher}, B. and {Rix}, H. -W. and {Sabbi}, E. and {Sirianni}, M. and {te Plate}, M. and {Valenti}, J. and {Willott}, C.~J. and {Zeidler}, P.},
        title = "{The Near-Infrared Spectrograph (NIRSpec) on the James Webb Space Telescope. II. Multi-object spectroscopy (MOS)}",
      journal = {\aap},
         year = 2022,
        month = may,
       volume = {661},
          eid = {A81},
        pages = {A81},
          doi = {10.1051/0004-6361/202142673},
archivePrefix = {arXiv},
       eprint = {2202.03306},
 primaryClass = {astro-ph.IM},
       adsurl = {https://ui.adsabs.harvard.edu/abs/2022A&A...661A..81F}
}

@ARTICLE{Labbe2023,
       author = {{Labb{\'e}}, Ivo and {van Dokkum}, Pieter and {Nelson}, Erica and {Bezanson}, Rachel and {Suess}, Katherine A. and {Leja}, Joel and {Brammer}, Gabriel and {Whitaker}, Katherine and {Mathews}, Elijah and {Stefanon}, Mauro and {Wang}, Bingjie},
        title = "{A population of red candidate massive galaxies  600 Myr after the Big Bang}",
      journal = {\nat},
         year = 2023,
        month = apr,
       volume = {616},
       number = {7956},
        pages = {266-269},
          doi = {10.1038/s41586-023-05786-2},
archivePrefix = {arXiv},
       eprint = {2207.12446},
 primaryClass = {astro-ph.GA},
       adsurl = {https://ui.adsabs.harvard.edu/abs/2023Natur.616..266L}
}

@ARTICLE{Barrufet2025,
       author = {{Barrufet}, L. and {Oesch}, P.~A. and {Marques-Chaves}, R. and {Arellano-Cordova}, K. and {Baggen}, J.~F.~W. and {Carnall}, A.~C. and {Cullen}, F. and {Dunlop}, J.~S. and {Gottumukkala}, R. and {Fudamoto}, Y. and {Illingworth}, G.~D. and {Magee}, D. and {McLure}, R.~J. and {McLeod}, D.~J. and {Micha{\l}owski}, M.~J. and {Stefanon}, M. and {van Dokkum}, P.~G. and {Weibel}, A.},
        title = "{Quiescent or dusty? Unveiling the nature of extremely red galaxies at z > 3}",
      journal = {\mnras},
         year = 2025,
        month = mar,
       volume = {537},
       number = {4},
        pages = {3453-3469},
          doi = {10.1093/mnras/staf013},
archivePrefix = {arXiv},
       eprint = {2404.08052},
 primaryClass = {astro-ph.GA},
       adsurl = {https://ui.adsabs.harvard.edu/abs/2025MNRAS.537.3453B}
}

@ARTICLE{Valentino2023,
       author = {{Valentino}, Francesco and {Brammer}, Gabriel and {Gould}, Katriona M.~L. and {Kokorev}, Vasily and {Fujimoto}, Seiji and {Jespersen}, Christian Kragh and {Vijayan}, Aswin P. and {Weaver}, John R. and {Ito}, Kei and {Tanaka}, Masayuki and {Ilbert}, Olivier and {Magdis}, Georgios E. and {Whitaker}, Katherine E. and {Faisst}, Andreas L. and {Gallazzi}, Anna and {Gillman}, Steven and {Gim{\'e}nez-Arteaga}, Clara and {G{\'o}mez-Guijarro}, Carlos and {Kubo}, Mariko and {Heintz}, Kasper E. and {Hirschmann}, Michaela and {Oesch}, Pascal and {Onodera}, Masato and {Rizzo}, Francesca and {Lee}, Minju and {Strait}, Victoria and {Toft}, Sune},
        title = "{An Atlas of Color-selected Quiescent Galaxies at z > 3 in Public JWST Fields}",
      journal = {\apj},
         year = 2023,
        month = apr,
       volume = {947},
       number = {1},
          eid = {20},
        pages = {20},
          doi = {10.3847/1538-4357/acbefa},
archivePrefix = {arXiv},
       eprint = {2302.10936},
 primaryClass = {astro-ph.GA},
       adsurl = {https://ui.adsabs.harvard.edu/abs/2023ApJ...947...20V}
}

@ARTICLE{Furtak2023,
       author = {{Furtak}, Lukas J. and {Zitrin}, Adi and {Plat}, Ad{\`e}le and {Fujimoto}, Seiji and {Wang}, Bingjie and {Nelson}, Erica J. and {Labb{\'e}}, Ivo and {Bezanson}, Rachel and {Brammer}, Gabriel B. and {van Dokkum}, Pieter and {Endsley}, Ryan and {Glazebrook}, Karl and {Greene}, Jenny E. and {Leja}, Joel and {Price}, Sedona H. and {Smit}, Renske and {Stark}, Daniel P. and {Weaver}, John R. and {Whitaker}, Katherine E. and {Atek}, Hakim and {Chevallard}, Jacopo and {Curtis-Lake}, Emma and {Dayal}, Pratika and {Feltre}, Anna and {Franx}, Marijn and {Fudamoto}, Yoshinobu and {Marchesini}, Danilo and {Mowla}, Lamiya A. and {Pan}, Richard and {Suess}, Katherine A. and {Vidal-Garc{\'\i}a}, Alba and {Williams}, Christina C.},
        title = "{JWST UNCOVER: Extremely Red and Compact Object at z $_{phot}$ ≃ 7.6 Triply Imaged by A2744}",
      journal = {\apj},
         year = 2023,
        month = aug,
       volume = {952},
       number = {2},
          eid = {142},
        pages = {142},
          doi = {10.3847/1538-4357/acdc9d},
archivePrefix = {arXiv},
       eprint = {2212.10531},
 primaryClass = {astro-ph.GA},
       adsurl = {https://ui.adsabs.harvard.edu/abs/2023ApJ...952..142F}
}

@ARTICLE{Labbe2023b,
       author = {{Labbe}, Ivo and {Greene}, Jenny E. and {Bezanson}, Rachel and {Fujimoto}, Seiji and {Furtak}, Lukas J. and {Goulding}, Andy D. and {Matthee}, Jorryt and {Naidu}, Rohan P. and {Oesch}, Pascal A. and {Atek}, Hakim and {Brammer}, Gabriel and {Chemerynska}, Iryna and {Coe}, Dan and {Cutler}, Sam E. and {Dayal}, Pratika and {Feldmann}, Robert and {Franx}, Marijn and {Glazebrook}, Karl and {Leja}, Joel and {Maseda}, Michael and {Marchesini}, Danilo and {Nanayakkara}, Themiya and {Nelson}, Erica J. and {Pan}, Richard and {Papovich}, Casey and {Price}, Sedona H. and {Suess}, Katherine A. and {Wang}, Bingjie and {Weaver}, John R. and {Whitaker}, Katherine E. and {Williams}, Christina C. and {Zitrin}, Adi},
        title = "{UNCOVER: Candidate Red Active Galactic Nuclei at 3 < z < 7 with JWST and ALMA}",
      journal = {\apj},
         year = 2025,
        month = jan,
       volume = {978},
       number = {1},
          eid = {92},
        pages = {92},
          doi = {10.3847/1538-4357/ad3551},
archivePrefix = {arXiv},
       eprint = {2306.07320},
 primaryClass = {astro-ph.GA},
       adsurl = {https://ui.adsabs.harvard.edu/abs/2025ApJ...978...92L}
}

@ARTICLE{Williams2024,
       author = {{Williams}, Christina C. and {Alberts}, Stacey and {Ji}, Zhiyuan and {Hainline}, Kevin N. and {Lyu}, Jianwei and {Rieke}, George and {Endsley}, Ryan and {Suess}, Katherine A. and {Sun}, Fengwu and {Johnson}, Benjamin D. and {Florian}, Michael and {Shivaei}, Irene and {Rujopakarn}, Wiphu and {Baker}, William M. and {Bhatawdekar}, Rachana and {Boyett}, Kristan and {Bunker}, Andrew J. and {Cameron}, Alex J. and {Carniani}, Stefano and {Charlot}, Stephane and {Curtis-Lake}, Emma and {DeCoursey}, Christa and {de Graaff}, Anna and {Egami}, Eiichi and {Eisenstein}, Daniel J. and {Gibson}, Justus L. and {Hausen}, Ryan and {Helton}, Jakob M. and {Maiolino}, Roberto and {Maseda}, Michael V. and {Nelson}, Erica J. and {P{\'e}rez-Gonz{\'a}lez}, Pablo G. and {Rieke}, Marcia J. and {Robertson}, Brant E. and {Saxena}, Aayush and {Tacchella}, Sandro and {Willmer}, Christopher N.~A. and {Willott}, Chris J.},
        title = "{The Galaxies Missed by Hubble and ALMA: The Contribution of Extremely Red Galaxies to the Cosmic Census at 3 < z < 8}",
      journal = {\apj},
         year = 2024,
        month = jun,
       volume = {968},
       number = {1},
          eid = {34},
        pages = {34},
          doi = {10.3847/1538-4357/ad3f17},
archivePrefix = {arXiv},
       eprint = {2311.07483},
 primaryClass = {astro-ph.GA},
       adsurl = {https://ui.adsabs.harvard.edu/abs/2024ApJ...968...34W}
}

@ARTICLE{Kocevski2023,
       author = {{Kocevski}, Dale D. and {Onoue}, Masafusa and {Inayoshi}, Kohei and {Trump}, Jonathan R. and {Arrabal Haro}, Pablo and {Grazian}, Andrea and {Dickinson}, Mark and {Finkelstein}, Steven L. and {Kartaltepe}, Jeyhan S. and {Hirschmann}, Michaela and {Aird}, James and {Holwerda}, Benne W. and {Fujimoto}, Seiji and {Juneau}, St{\'e}phanie and {Amor{\'\i}n}, Ricardo O. and {Backhaus}, Bren E. and {Bagley}, Micaela B. and {Barro}, Guillermo and {Bell}, Eric F. and {Bisigello}, Laura and {Calabr{\`o}}, Antonello and {Cleri}, Nikko J. and {Cooper}, M.~C. and {Ding}, Xuheng and {Grogin}, Norman A. and {Ho}, Luis C. and {Hutchison}, Taylor A. and {Inoue}, Akio K. and {Jiang}, Linhua and {Jones}, Brenda and {Koekemoer}, Anton M. and {Li}, Wenxiu and {Li}, Zhengrong and {McGrath}, Elizabeth J. and {Molina}, Juan and {Papovich}, Casey and {P{\'e}rez-Gonz{\'a}lez}, Pablo G. and {Pirzkal}, Nor and {Wilkins}, Stephen M. and {Yang}, Guang and {Yung}, L.~Y. Aaron},
        title = "{Hidden Little Monsters: Spectroscopic Identification of Low-mass, Broad-line AGNs at z > 5 with CEERS}",
      journal = {\apjl},
         year = 2023,
        month = sep,
       volume = {954},
       number = {1},
          eid = {L4},
        pages = {L4},
          doi = {10.3847/2041-8213/ace5a0},
archivePrefix = {arXiv},
       eprint = {2302.00012},
 primaryClass = {astro-ph.GA},
       adsurl = {https://ui.adsabs.harvard.edu/abs/2023ApJ...954L...4K}
}

@ARTICLE{Matthee2024,
       author = {{Matthee}, Jorryt and {Naidu}, Rohan P. and {Brammer}, Gabriel and {Chisholm}, John and {Eilers}, Anna-Christina and {Goulding}, Andy and {Greene}, Jenny and {Kashino}, Daichi and {Labbe}, Ivo and {Lilly}, Simon J. and {Mackenzie}, Ruari and {Oesch}, Pascal A. and {Weibel}, Andrea and {Wuyts}, Stijn and {Xiao}, Mengyuan and {Bordoloi}, Rongmon and {Bouwens}, Rychard and {van Dokkum}, Pieter and {Illingworth}, Garth and {Kramarenko}, Ivan and {Maseda}, Michael V. and {Mason}, Charlotte and {Meyer}, Romain A. and {Nelson}, Erica J. and {Reddy}, Naveen A. and {Shivaei}, Irene and {Simcoe}, Robert A. and {Yue}, Minghao},
        title = "{Little Red Dots: An Abundant Population of Faint Active Galactic Nuclei at z {\ensuremath{\sim}} 5 Revealed by the EIGER and FRESCO JWST Surveys}",
      journal = {\apj},
         year = 2024,
        month = mar,
       volume = {963},
       number = {2},
          eid = {129},
        pages = {129},
          doi = {10.3847/1538-4357/ad2345},
archivePrefix = {arXiv},
       eprint = {2306.05448},
 primaryClass = {astro-ph.GA},
       adsurl = {https://ui.adsabs.harvard.edu/abs/2024ApJ...963..129M}
}

@ARTICLE{Matthee2024b,
       author = {{Matthee}, Jorryt and {Naidu}, Rohan P. and {Kotiwale}, Gauri and {Furtak}, Lukas J. and {Kramarenko}, Ivan and {Mackenzie}, Ruari and {Greene}, Jenny and {Adamo}, Angela and {Bouwens}, Rychard J. and {Di Cesare}, Claudia and {Eilers}, Anna-Christina and {de Graaff}, Anna and {Heintz}, Kasper E. and {Kashino}, Daichi and {Maseda}, Michael V. and {Tacchella}, Sandro and {Torralba}, Alberto},
        title = "{Environmental Evidence for Overly Massive Black Holes in Low-mass Galaxies and a Black Hole─Halo Mass Relation at z {\ensuremath{\sim}} 5}",
      journal = {\apj},
         year = 2025,
        month = aug,
       volume = {988},
       number = {2},
          eid = {246},
        pages = {246},
          doi = {10.3847/1538-4357/ade886},
archivePrefix = {arXiv},
       eprint = {2412.02846},
 primaryClass = {astro-ph.GA},
       adsurl = {https://ui.adsabs.harvard.edu/abs/2025ApJ...988..246M}
}

@ARTICLE{Greene2024,
       author = {{Greene}, Jenny E. and {Labbe}, Ivo and {Goulding}, Andy D. and {Furtak}, Lukas J. and {Chemerynska}, Iryna and {Kokorev}, Vasily and {Dayal}, Pratika and {Volonteri}, Marta and {Williams}, Christina C. and {Wang}, Bingjie and {Setton}, David J. and {Burgasser}, Adam J. and {Bezanson}, Rachel and {Atek}, Hakim and {Brammer}, Gabriel and {Cutler}, Sam E. and {Feldmann}, Robert and {Fujimoto}, Seiji and {Glazebrook}, Karl and {de Graaff}, Anna and {Khullar}, Gourav and {Leja}, Joel and {Marchesini}, Danilo and {Maseda}, Michael V. and {Matthee}, Jorryt and {Miller}, Tim B. and {Naidu}, Rohan P. and {Nanayakkara}, Themiya and {Oesch}, Pascal A. and {Pan}, Richard and {Papovich}, Casey and {Price}, Sedona H. and {van Dokkum}, Pieter and {Weaver}, John R. and {Whitaker}, Katherine E. and {Zitrin}, Adi},
        title = "{UNCOVER Spectroscopy Confirms the Surprising Ubiquity of Active Galactic Nuclei in Red Sources at z > 5}",
      journal = {\apj},
         year = 2024,
        month = mar,
       volume = {964},
       number = {1},
          eid = {39},
        pages = {39},
          doi = {10.3847/1538-4357/ad1e5f},
archivePrefix = {arXiv},
       eprint = {2309.05714},
 primaryClass = {astro-ph.GA},
       adsurl = {https://ui.adsabs.harvard.edu/abs/2024ApJ...964...39G}
}

@ARTICLE{Greene2025,
       author = {{Greene}, Jenny E. and {Setton}, David J. and {Furtak}, Lukas J. and {Naidu}, Rohan P. and {Volonteri}, Marta and {Dayal}, Pratika and {Labbe}, Ivo and {van Dokkum}, Pieter and {Bezanson}, Rachel and {Brammer}, Gabriel and {Cutler}, Sam E. and {Glazebrook}, Karl and {de Graaff}, Anna and {Hirschmann}, Michaela and {Hviding}, Raphael E. and {Kokorev}, Vasily and {Leja}, Joel and {Liu}, Hanpu and {Ma}, Yilun and {Matthee}, Jorryt and {Nanayakkara}, Themiya and {Oesch}, Pascal A. and {Pan}, Richard and {Price}, Sedona H. and {Spilker}, Justin S. and {Wang}, Bingjie and {Weaver}, John R. and {Whitaker}, Katherine E. and {Williams}, Christina C. and {Zitrin}, Adi},
        title = "{What You See Is What You Get: Empirically Measured Bolometric Luminosities of Little Red Dots}",
      journal = {\apj},
         year = 2026,
        month = jan,
       volume = {996},
       number = {2},
          eid = {129},
        pages = {129},
          doi = {10.3847/1538-4357/ae1836},
archivePrefix = {arXiv},
       eprint = {2509.05434},
 primaryClass = {astro-ph.GA},
       adsurl = {https://ui.adsabs.harvard.edu/abs/2026ApJ...996..129G}
}

@ARTICLE{Eisenstein2023,
       author = {{Eisenstein}, Daniel J. and {Willott}, Chris and {Alberts}, Stacey and {Arribas}, Santiago and {Bonaventura}, Nina and {Bunker}, Andrew J. and {Cameron}, Alex J. and {Carniani}, Stefano and {Charlot}, Stephane and {Curtis-Lake}, Emma and {D'Eugenio}, Francesco and {Ferruit}, Pierre and {Giardino}, Giovanna and {Hainline}, Kevin and {Hausen}, Ryan and {Jakobsen}, Peter and {Johnson}, Benjamin D. and {Maiolino}, Roberto and {Rauscher}, Bernard J. and {Rieke}, Marcia and {Rieke}, George and {Rix}, Hans-Walter and {Robertson}, Brant and {Stark}, Daniel P. and {Tacchella}, Sandro and {Williams}, Christina C. and {Willmer}, Christopher N.~A. and {Baker}, William M. and {Baum}, Stefi and {Bhatawdekar}, Rachana and {Boyett}, Kristan and {Chen}, Zuyi and {Chevallard}, Jacopo and {Circosta}, Chiara and {Curti}, Mirko and {Danhaive}, A. Lola and {DeCoursey}, Christa and {Endsley}, Ryan and {de Graaff}, Anna and {Dressler}, Alan and {Egami}, Eiichi and {Helton}, Jakob M. and {Hviding}, Raphael E. and {Ji}, Zhiyuan and {Jones}, Gareth C. and {Kumari}, Nimisha and {L{\"u}tzgendorf}, Nora and {Laseter}, Isaac and {Looser}, Tobias J. and {Lyu}, Jianwei and {Maseda}, Michael V. and {Nelson}, Erica and {Parlanti}, Eleonora and {Perna}, Michele and {Pusk{\'a}s}, D{\'a}vid and {Rawle}, Tim and {Rodr{\'\i}guez Del Pino}, Bruno and {Rujopakarn}, Wiphu and {Sandles}, Lester and {Saxena}, Aayush and {Scholtz}, Jan and {Sharpe}, Katherine and {Shivaei}, Irene and {Silcock}, Maddie S. and {Simmonds}, Charlotte and {Skarbinski}, Maya and {Smit}, Renske and {Stone}, Meredith and {Suess}, Katherine A. and {Sun}, Fengwu and {Tang}, Mengtao and {Topping}, Michael W. and {{\"U}bler}, Hannah and {Villanueva}, Natalia C. and {Wallace}, Imaan E.~B. and {Whitler}, Lily and {Witstok}, Joris and {Woodrum}, Charity},
        title = "{Overview of the JWST Advanced Deep Extragalactic Survey (JADES)}",
      journal = {\apjs},
         year = 2026,
        month = mar,
       volume = {283},
       number = {1},
          eid = {6},
        pages = {6},
          doi = {10.3847/1538-4365/ae3163},
archivePrefix = {arXiv},
       eprint = {2306.02465},
 primaryClass = {astro-ph.GA},
       adsurl = {https://ui.adsabs.harvard.edu/abs/2026ApJS..283....6E}
}

@ARTICLE{Maseda2024,
       author = {{Maseda}, Michael V. and {de Graaff}, Anna and {Franx}, Marijn and {Rix}, Hans-Walter and {Carniani}, Stefano and {Laseter}, Isaac and {Dudzevi{\v{c}}i{\={u}}t{\.{e}}}, Ugn{\.{e}} and {Rawle}, Tim and {Parlanti}, Eleonora and {Arribas}, Santiago and {Bunker}, Andrew J. and {Cameron}, Alex J. and {Charlot}, Stephane and {Curti}, Mirko and {D'Eugenio}, Francesco and {Jones}, Gareth C. and {Kumari}, Nimisha and {Maiolino}, Roberto and {{\"U}bler}, Hannah and {Saxena}, Aayush and {Smit}, Renske and {Willott}, Chris and {Witstok}, Joris},
        title = "{The NIRSpec Wide GTO Survey}",
      journal = {\aap},
         year = 2024,
        month = sep,
       volume = {689},
          eid = {A73},
        pages = {A73},
          doi = {10.1051/0004-6361/202449914},
archivePrefix = {arXiv},
       eprint = {2403.05506},
 primaryClass = {astro-ph.GA},
       adsurl = {https://ui.adsabs.harvard.edu/abs/2024A&A...689A..73M}
}

@misc{grizli,
       author = {{Brammer}, Gabriel},
        title = "{grizli}",
         year = 2023,
        month = sep,
          eid = {10.5281/zenodo.1146904},
          doi = {10.5281/zenodo.1146904},
      version = {1.9.11},
    publisher = {Zenodo},
       adsurl = {https://ui.adsabs.harvard.edu/abs/2021zndo...1146904B}
}

@ARTICLE{Brammer2008,
       author = {{Brammer}, Gabriel B. and {van Dokkum}, Pieter G. and {Coppi}, Paolo},
        title = "{EAZY: A Fast, Public Photometric Redshift Code}",
      journal = {\apj},
         year = 2008,
        month = oct,
       volume = {686},
       number = {2},
        pages = {1503-1513},
          doi = {10.1086/591786},
archivePrefix = {arXiv},
       eprint = {0807.1533},
 primaryClass = {astro-ph},
       adsurl = {https://ui.adsabs.harvard.edu/abs/2008ApJ...686.1503B}
}

@ARTICLE{Weibel2024,
       author = {{Weibel}, Andrea and {Oesch}, Pascal A. and {Barrufet}, Laia and {Gottumukkala}, Rashmi and {Ellis}, Richard S. and {Santini}, Paola and {Weaver}, John R. and {Allen}, Natalie and {Bouwens}, Rychard and {Bowler}, Rebecca A.~A. and {Brammer}, Gabe and {Carnall}, Adam C. and {Cullen}, Fergus and {Dayal}, Pratika and {Dickinson}, Mark and {Donnan}, Callum T. and {Dunlop}, James S. and {Giavalisco}, Mauro and {Grogin}, Norman A. and {Illingworth}, Garth D. and {Koekemoer}, Anton M. and {Labbe}, Ivo and {Marchesini}, Danilo and {McLeod}, Derek J. and {McLure}, Ross J. and {Naidu}, Rohan P. and {P{\'e}rez-Gonz{\'a}lez}, Pablo G. and {Shuntov}, Marko and {Stefanon}, Mauro and {Toft}, Sune and {Xiao}, Mengyuan},
        title = "{Galaxy build-up in the first 1.5 Gyr of cosmic history: insights from the stellar mass function at z   4-9 from JWST NIRCam observations}",
      journal = {\mnras},
         year = 2024,
        month = sep,
       volume = {533},
       number = {2},
        pages = {1808-1838},
          doi = {10.1093/mnras/stae1891},
archivePrefix = {arXiv},
       eprint = {2403.08872},
 primaryClass = {astro-ph.GA},
       adsurl = {https://ui.adsabs.harvard.edu/abs/2024MNRAS.533.1808W}
}

@software{msaexp,
       author = {{Brammer}, Gabriel},
        title = "{msaexp: NIRSpec analyis tools}",
         year = 2023,
        month = sep,
          eid = {10.5281/zenodo.8319596},
          doi = {10.5281/zenodo.8319596},
      version = {0.6.17},
    publisher = {Zenodo},
       adsurl = {https://ui.adsabs.harvard.edu/abs/2023zndo...8319596B}
}

@ARTICLE{deGraaff2024a,
       author = {{de Graaff}, Anna and {Rix}, Hans-Walter and {Carniani}, Stefano and {Suess}, Katherine A. and {Charlot}, St{\'e}phane and {Curtis-Lake}, Emma and {Arribas}, Santiago and {Baker}, William M. and {Boyett}, Kristan and {Bunker}, Andrew J. and {Cameron}, Alex J. and {Chevallard}, Jacopo and {Curti}, Mirko and {Eisenstein}, Daniel J. and {Franx}, Marijn and {Hainline}, Kevin and {Hausen}, Ryan and {Ji}, Zhiyuan and {Johnson}, Benjamin D. and {Jones}, Gareth C. and {Maiolino}, Roberto and {Maseda}, Michael V. and {Nelson}, Erica and {Parlanti}, Eleonora and {Rawle}, Tim and {Robertson}, Brant and {Tacchella}, Sandro and {{\"U}bler}, Hannah and {Williams}, Christina C. and {Willmer}, Christopher N.~A. and {Willott}, Chris},
        title = "{Ionised gas kinematics and dynamical masses of z {\ensuremath{\gtrsim}} 6 galaxies from JADES/NIRSpec high-resolution spectroscopy}",
      journal = {\aap},
         year = 2024,
        month = apr,
       volume = {684},
          eid = {A87},
        pages = {A87},
          doi = {10.1051/0004-6361/202347755},
archivePrefix = {arXiv},
       eprint = {2308.09742},
 primaryClass = {astro-ph.GA},
       adsurl = {https://ui.adsabs.harvard.edu/abs/2024A&A...684A..87D}
}

@ARTICLE{emcee,
       author = {{Foreman-Mackey}, Daniel and {Hogg}, David W. and {Lang}, Dustin and {Goodman}, Jonathan},
        title = "{emcee: The MCMC Hammer}",
      journal = {\pasp},
         year = 2013,
        month = mar,
       volume = {125},
       number = {925},
        pages = {306},
          doi = {10.1086/670067},
archivePrefix = {arXiv},
       eprint = {1202.3665},
 primaryClass = {astro-ph.IM},
       adsurl = {https://ui.adsabs.harvard.edu/abs/2013PASP..125..306F}
}

@ARTICLE{Wang2024a,
       author = {{Wang}, Bingjie and {de Graaff}, Anna and {Davies}, Rebecca L. and {Greene}, Jenny E. and {Leja}, Joel and {Brammer}, Gabriel B. and {Goulding}, Andy D. and {Miller}, Tim B. and {Suess}, Katherine A. and {Weibel}, Andrea and {Williams}, Christina C. and {Bezanson}, Rachel and {Boogaard}, Leindert A. and {Cleri}, Nikko J. and {Hirschmann}, Michaela and {Katz}, Harley and {Labb{\'e}}, Ivo and {Maseda}, Michael V. and {Matthee}, Jorryt and {McConachie}, Ian and {Naidu}, Rohan P. and {Oesch}, Pascal A. and {Rix}, Hans-Walter and {Setton}, David J. and {Whitaker}, Katherine E.},
        title = "{RUBIES: JWST/NIRSpec Confirmation of an Infrared-luminous, Broad-line Little Red Dot with an Ionized Outflow}",
      journal = {\apj},
         year = 2025,
        month = may,
       volume = {984},
       number = {2},
          eid = {121},
        pages = {121},
          doi = {10.3847/1538-4357/adc1ca},
archivePrefix = {arXiv},
       eprint = {2403.02304},
 primaryClass = {astro-ph.GA},
       adsurl = {https://ui.adsabs.harvard.edu/abs/2025ApJ...984..121W}
}

@ARTICLE{Wang2024b,
       author = {{Wang}, Bingjie and {Leja}, Joel and {de Graaff}, Anna and {Brammer}, Gabriel B. and {Weibel}, Andrea and {van Dokkum}, Pieter and {Baggen}, Josephine F.~W. and {Suess}, Katherine A. and {Greene}, Jenny E. and {Bezanson}, Rachel and {Cleri}, Nikko J. and {Hirschmann}, Michaela and {Labb{\'e}}, Ivo and {Matthee}, Jorryt and {McConachie}, Ian and {Naidu}, Rohan P. and {Nelson}, Erica and {Oesch}, Pascal A. and {Setton}, David J. and {Williams}, Christina C.},
        title = "{RUBIES: Evolved Stellar Populations with Extended Formation Histories at z {\ensuremath{\sim}} 7{\textendash}8 in Candidate Massive Galaxies Identified with JWST/NIRSpec}",
      journal = {\apjl},
         year = 2024,
        month = jul,
       volume = {969},
       number = {1},
          eid = {L13},
        pages = {L13},
          doi = {10.3847/2041-8213/ad55f7},
archivePrefix = {arXiv},
       eprint = {2405.01473},
 primaryClass = {astro-ph.GA},
       adsurl = {https://ui.adsabs.harvard.edu/abs/2024ApJ...969L..13W}
}

@ARTICLE{Killi2023,
       author = {{Killi}, Meghana and {Watson}, Darach and {Brammer}, Gabriel and {McPartland}, Conor and {Antwi-Danso}, Jacqueline and {Newshore}, Rosa and {Coe}, Dan and {Allen}, Natalie and {Fynbo}, Johan P.~U. and {Gould}, Katriona and {Heintz}, Kasper E. and {Rusakov}, Vadim and {Vejlgaard}, Simone},
        title = "{Deciphering the JWST spectrum of a 'little red dot' at z {\ensuremath{\sim}} 4.53: An obscured AGN and its star-forming host}",
      journal = {\aap},
         year = 2024,
        month = nov,
       volume = {691},
          eid = {A52},
        pages = {A52},
          doi = {10.1051/0004-6361/202348857},
archivePrefix = {arXiv},
       eprint = {2312.03065},
 primaryClass = {astro-ph.GA},
       adsurl = {https://ui.adsabs.harvard.edu/abs/2024A&A...691A..52K}
}

@ARTICLE{Rinaldi2024,
       author = {{Rinaldi}, P. and {Navarro-Carrera}, R. and {Caputi}, K.~I. and {Iani}, E. and {{\"O}stlin}, G. and {Colina}, L. and {Alberts}, S. and {{\'A}lvarez-M{\'a}rquez}, J. and {Annunziatella}, M. and {Boogaard}, L. and {Costantin}, L. and {Hjorth}, J. and {Langeroodi}, D. and {Melinder}, J. and {Moutard}, T. and {Walter}, F.},
        title = "{The Emergence of the Star Formation Main Sequence with Redshift Unfolded by JWST}",
      journal = {\apj},
         year = 2025,
        month = mar,
       volume = {981},
       number = {2},
          eid = {161},
        pages = {161},
          doi = {10.3847/1538-4357/adb309},
archivePrefix = {arXiv},
       eprint = {2406.13554},
 primaryClass = {astro-ph.GA},
       adsurl = {https://ui.adsabs.harvard.edu/abs/2025ApJ...981..161R}
}

@ARTICLE{Horne86,
       author = {{Horne}, K.},
        title = "{An optimal extraction algorithm for CCD spectroscopy.}",
      journal = {\pasp},
         year = 1986,
        month = jun,
       volume = {98},
        pages = {609-617},
          doi = {10.1086/131801},
       adsurl = {https://ui.adsabs.harvard.edu/abs/1986PASP...98..609H}
}

@ARTICLE{Heintz2024,
       author = {{Heintz}, K.~E. and {Brammer}, G.~B. and {Watson}, D. and {Oesch}, P.~A. and {Keating}, L.~C. and {Hayes}, M.~J. and {Abdurro'uf} and {Arellano-C{\'o}rdova}, K.~Z. and {Carnall}, A.~C. and {Christiansen}, C.~R. and {Cullen}, F. and {Dav{\'e}}, R. and {Dayal}, P. and {Ferrara}, A. and {Finlator}, K. and {Fynbo}, J.~P.~U. and {Flury}, S.~R. and {Gelli}, V. and {Gillman}, S. and {Gottumukkala}, R. and {Gould}, K. and {Greve}, T.~R. and {Hardin}, S.~E. and {Hsiao}, T.~Y. -Y. and {Hutter}, A. and {Jakobsson}, P. and {Killi}, M. and {Khosravaninezhad}, N. and {Laursen}, P. and {Lee}, M.~M. and {Magdis}, G.~E. and {Matthee}, J. and {Naidu}, R.~P. and {Narayanan}, D. and {Pollock}, C. and {Prescott}, M.~K.~M. and {Rusakov}, V. and {Shuntov}, M. and {Sneppen}, A. and {Smit}, R. and {Tanvir}, N.~R. and {Terp}, C. and {Toft}, S. and {Valentino}, F. and {Vijayan}, A.~P. and {Weaver}, J.~R. and {Wise}, J.~H. and {Witstok}, J.},
        title = "{The JWST-PRIMAL archival survey: A JWST/NIRSpec reference sample for the physical properties and Lyman-{\ensuremath{\alpha}} absorption and emission of {\ensuremath{\sim}}600 galaxies at z = 5.0 ‑ 13.4}",
      journal = {\aap},
         year = 2025,
        month = jan,
       volume = {693},
          eid = {A60},
        pages = {A60},
          doi = {10.1051/0004-6361/202450243},
archivePrefix = {arXiv},
       eprint = {2404.02211},
 primaryClass = {astro-ph.GA},
       adsurl = {https://ui.adsabs.harvard.edu/abs/2025A&A...693A..60H}
}

@ARTICLE{Akins2024,
       author = {{Akins}, Hollis B. and {Casey}, Caitlin M. and {Lambrides}, Erini and {Allen}, Natalie and {Andika}, Irham T. and {Brinch}, Malte and {Champagne}, Jaclyn B. and {Cooper}, Olivia and {Ding}, Xuheng and {Drakos}, Nicole E. and {Faisst}, Andreas and {Finkelstein}, Steven L. and {Franco}, Maximilien and {Fujimoto}, Seiji and {Gentile}, Fabrizio and {Gillman}, Steven and {Gozaliasl}, Ghassem and {Harish}, Santosh and {Hayward}, Christopher C. and {Hirschmann}, Michaela and {Ilbert}, Olivier and {Kartaltepe}, Jeyhan S. and {Kocevski}, Dale D. and {Koekemoer}, Anton M. and {Kokorev}, Vasily and {Liu}, Daizhong and {Long}, Arianna S. and {McCracken}, Henry Joy and {McKinney}, Jed and {Onoue}, Masafusa and {Paquereau}, Louise and {Renzini}, Alvio and {Rhodes}, Jason and {Robertson}, Brant E. and {Shuntov}, Marko and {Silverman}, John D. and {Tanaka}, Takumi S. and {Toft}, Sune and {Trakhtenbrot}, Benny and {Valentino}, Francesco and {Zavala}, Jorge},
        title = "{COSMOS-Web: The Overabundance and Physical Nature of ``Little Red Dots''{\textemdash}Implications for Early Galaxy and SMBH Assembly}",
      journal = {\apj},
         year = 2025,
        month = sep,
       volume = {991},
       number = {1},
          eid = {37},
        pages = {37},
          doi = {10.3847/1538-4357/ade984},
archivePrefix = {arXiv},
       eprint = {2406.10341},
 primaryClass = {astro-ph.GA},
       adsurl = {https://ui.adsabs.harvard.edu/abs/2025ApJ...991...37A}
}

@ARTICLE{Williams2024:panoramic,
       author = {{Williams}, Christina C. and {Oesch}, Pascal A. and {Weibel}, Andrea and {Brammer}, Gabriel and {Cloonan}, Aidan P. and {Whitaker}, Katherine E. and {Barrufet}, Laia and {Bezanson}, Rachel and {Bowler}, Rebecca A.~A. and {Dayal}, Pratika and {Franx}, Marijn and {Greene}, Jenny E. and {Hutter}, Anne and {Ji}, Zhiyuan and {Labb{\'e}}, Ivo and {Manning}, Sinclaire M. and {Maseda}, Michael V. and {Xiao}, Mengyuan},
        title = "{The PANORAMIC Survey: Pure Parallel Wide Area Legacy Imaging with JWST/NIRCam}",
      journal = {\apj},
         year = 2025,
        month = feb,
       volume = {979},
       number = {2},
          eid = {140},
        pages = {140},
          doi = {10.3847/1538-4357/ad97bc},
archivePrefix = {arXiv},
       eprint = {2410.01875},
 primaryClass = {astro-ph.GA},
       adsurl = {https://ui.adsabs.harvard.edu/abs/2025ApJ...979..140W}
}

@ARTICLE{Labbe2024,
       author = {{Labbe}, Ivo and {Greene}, Jenny E. and {Matthee}, Jorryt and {Treiber}, Helena and {Kokorev}, Vasily and {Miller}, Tim B. and {Kramarenko}, Ivan and {Setton}, David J. and {Ma}, Yilun and {Goulding}, Andy D. and {Bezanson}, Rachel and {Naidu}, Rohan P. and {Williams}, Christina C. and {Atek}, Hakim and {Brammer}, Gabriel and {Cutler}, Sam E. and {Chemerynska}, Iryna and {Cloonan}, Aidan P. and {Dayal}, Pratika and {de Graaff}, Anna and {Fudamoto}, Yoshinobu and {Fujimoto}, Seiji and {Furtak}, Lukas J. and {Glazebrook}, Karl and {Heintz}, Kasper E. and {Leja}, Joel and {Marchesini}, Danilo and {Nanayakkara}, Themiya and {Nelson}, Erica J. and {Oesch}, Pascal A. and {Pan}, Richard and {Price}, Sedona H. and {Shivaei}, Irene and {Sobral}, David and {Suess}, Katherine A. and {van Dokkum}, Pieter and {Wang}, Bingjie and {Weaver}, John R. and {Whitaker}, Katherine E. and {Zitrin}, Adi},
        title = "{An unambiguous AGN and a Balmer break in an Ultraluminous Little Red Dot at z=4.47 from Ultradeep UNCOVER and All the Little Things Spectroscopy}",
      journal = {arXiv e-prints},
         year = 2024,
        month = dec,
          eid = {arXiv:2412.04557},
        pages = {arXiv:2412.04557},
          doi = {10.48550/arXiv.2412.04557},
archivePrefix = {arXiv},
       eprint = {2412.04557},
 primaryClass = {astro-ph.GA},
       adsurl = {https://ui.adsabs.harvard.edu/abs/2024arXiv241204557L}
}

@ARTICLE{Setton2024,
       author = {{Setton}, David J. and {Greene}, Jenny E. and {de Graaff}, Anna and {Ma}, Yilun and {Leja}, Joel and {Matthee}, Jorryt and {Bezanson}, Rachel and {Boogaard}, Leindert A. and {Cleri}, Nikko J. and {Katz}, Harley and {Labbe}, Ivo and {Maseda}, Michael V. and {McConachie}, Ian and {Miller}, Tim B. and {Price}, Sedona H. and {Suess}, Katherine A. and {van Dokkum}, Pieter and {Wang}, Bingjie and {Weibel}, Andrea and {Whitaker}, Katherine E. and {Williams}, Christina C.},
        title = "{Little Red Dots at an Inflection Point: Ubiquitous V-shaped Turnover Consistently Occurs at the Balmer Limit}",
      journal = {\apj},
         year = 2025,
        month = dec,
       volume = {995},
       number = {1},
          eid = {118},
        pages = {118},
          doi = {10.3847/1538-4357/ae1500},
archivePrefix = {arXiv},
       eprint = {2411.03424},
 primaryClass = {astro-ph.GA},
       adsurl = {https://ui.adsabs.harvard.edu/abs/2025ApJ...995..118S}
}

@ARTICLE{Furtak2024,
       author = {{Furtak}, Lukas J. and {Labb{\'e}}, Ivo and {Zitrin}, Adi and {Greene}, Jenny E. and {Dayal}, Pratika and {Chemerynska}, Iryna and {Kokorev}, Vasily and {Miller}, Tim B. and {Goulding}, Andy D. and {de Graaff}, Anna and {Bezanson}, Rachel and {Brammer}, Gabriel B. and {Cutler}, Sam E. and {Leja}, Joel and {Pan}, Richard and {Price}, Sedona H. and {Wang}, Bingjie and {Weaver}, John R. and {Whitaker}, Katherine E. and {Atek}, Hakim and {Bogd{\'a}n}, {\'A}kos and {Charlot}, St{\'e}phane and {Curtis-Lake}, Emma and {van Dokkum}, Pieter and {Endsley}, Ryan and {Feldmann}, Robert and {Fudamoto}, Yoshinobu and {Fujimoto}, Seiji and {Glazebrook}, Karl and {Juneau}, St{\'e}phanie and {Marchesini}, Danilo and {Maseda}, Micheal V. and {Nelson}, Erica and {Oesch}, Pascal A. and {Plat}, Ad{\`e}le and {Setton}, David J. and {Stark}, Daniel P. and {Williams}, Christina C.},
        title = "{A high black-hole-to-host mass ratio in a lensed AGN in the early Universe}",
      journal = {\nat},
         year = 2024,
        month = apr,
       volume = {628},
       number = {8006},
        pages = {57-61},
          doi = {10.1038/s41586-024-07184-8},
archivePrefix = {arXiv},
       eprint = {2308.05735},
 primaryClass = {astro-ph.GA},
       adsurl = {https://ui.adsabs.harvard.edu/abs/2024Natur.628...57F}
}

@ARTICLE{Kocevski2024,
       author = {{Kocevski}, Dale D. and {Finkelstein}, Steven L. and {Barro}, Guillermo and {Taylor}, Anthony J. and {Calabr{\`o}}, Antonello and {Laloux}, Brivael and {Buchner}, Johannes and {Trump}, Jonathan R. and {Leung}, Gene C.~K. and {Yang}, Guang and {Dickinson}, Mark and {P{\'e}rez-Gonz{\'a}lez}, Pablo G. and {Pacucci}, Fabio and {Inayoshi}, Kohei and {Somerville}, Rachel S. and {McGrath}, Elizabeth J. and {Akins}, Hollis B. and {Bagley}, Micaela B. and {Bowler}, Rebecca A.~A. and {Bisigello}, Laura and {Carnall}, Adam and {Casey}, Caitlin M. and {Cheng}, Yingjie and {Cleri}, Nikko J. and {Costantin}, Luca and {Cullen}, Fergus and {Davis}, Kelcey and {Donnan}, Callum T. and {Dunlop}, James S. and {Ellis}, Richard S. and {Ferguson}, Henry C. and {Fujimoto}, Seiji and {Fontana}, Adriano and {Giavalisco}, Mauro and {Grazian}, Andrea and {Grogin}, Norman A. and {Hathi}, Nimish P. and {Hirschmann}, Michaela and {Huertas-Company}, Marc and {Holwerda}, Benne W. and {Illingworth}, Garth and {Juneau}, St{\'e}phanie and {Kartaltepe}, Jeyhan S. and {Koekemoer}, Anton M. and {Li}, Wenxiu and {Lucas}, Ray A. and {Magee}, Dan and {Mason}, Charlotte and {McLeod}, Derek J. and {McLure}, Ross J. and {Napolitano}, Lorenzo and {Papovich}, Casey and {Pirzkal}, Nor and {Rodighiero}, Giulia and {Santini}, Paola and {Wilkins}, Stephen M. and {Yung}, L.~Y. Aaron},
        title = "{The Rise of Faint, Red Active Galactic Nuclei at z > 4: A Sample of Little Red Dots in the JWST Extragalactic Legacy Fields}",
      journal = {\apj},
         year = 2025,
        month = jun,
       volume = {986},
       number = {2},
          eid = {126},
        pages = {126},
          doi = {10.3847/1538-4357/adbc7d},
archivePrefix = {arXiv},
       eprint = {2404.03576},
 primaryClass = {astro-ph.GA},
       adsurl = {https://ui.adsabs.harvard.edu/abs/2025ApJ...986..126K}
}

@ARTICLE{Yue2024,
       author = {{Yue}, Minghao and {Eilers}, Anna-Christina and {Ananna}, Tonima Tasnim and {Panagiotou}, Christos and {Kara}, Erin and {Miyaji}, Takamitsu},
        title = "{Stacking X-Ray Observations of ``Little Red Dots'': Implications for Their Active Galactic Nucleus Properties}",
      journal = {\apjl},
         year = 2024,
        month = oct,
       volume = {974},
       number = {2},
          eid = {L26},
        pages = {L26},
          doi = {10.3847/2041-8213/ad7eba},
archivePrefix = {arXiv},
       eprint = {2404.13290},
 primaryClass = {astro-ph.GA},
       adsurl = {https://ui.adsabs.harvard.edu/abs/2024ApJ...974L..26Y}
}

@ARTICLE{Ananna2024,
       author = {{Ananna}, Tonima Tasnim and {Bogd{\'a}n}, {\'A}kos and {Kov{\'a}cs}, Orsolya E. and {Natarajan}, Priyamvada and {Hickox}, Ryan C.},
        title = "{X-Ray View of Little Red Dots: Do They Host Supermassive Black Holes?}",
      journal = {\apjl},
         year = 2024,
        month = jul,
       volume = {969},
       number = {1},
          eid = {L18},
        pages = {L18},
          doi = {10.3847/2041-8213/ad5669},
archivePrefix = {arXiv},
       eprint = {2404.19010},
 primaryClass = {astro-ph.GA},
       adsurl = {https://ui.adsabs.harvard.edu/abs/2024ApJ...969L..18A}
}

@ARTICLE{Perez2024,
       author = {{P{\'e}rez-Gonz{\'a}lez}, Pablo G. and {Barro}, Guillermo and {Rieke}, George H. and {Lyu}, Jianwei and {Rieke}, Marcia and {Alberts}, Stacey and {Williams}, Christina C. and {Hainline}, Kevin and {Sun}, Fengwu and {Pusk{\'a}s}, D{\'a}vid and {Annunziatella}, Marianna and {Baker}, William M. and {Bunker}, Andrew J. and {Egami}, Eiichi and {Ji}, Zhiyuan and {Johnson}, Benjamin D. and {Robertson}, Brant and {Rodr{\'\i}guez Del Pino}, Bruno and {Rujopakarn}, Wiphu and {Shivaei}, Irene and {Tacchella}, Sandro and {Willmer}, Christopher N.~A. and {Willott}, Chris},
        title = "{What Is the Nature of Little Red Dots and what Is Not, MIRI SMILES Edition}",
      journal = {\apj},
         year = 2024,
        month = jun,
       volume = {968},
       number = {1},
          eid = {4},
        pages = {4},
          doi = {10.3847/1538-4357/ad38bb},
archivePrefix = {arXiv},
       eprint = {2401.08782},
 primaryClass = {astro-ph.GA},
       adsurl = {https://ui.adsabs.harvard.edu/abs/2024ApJ...968....4P}
}

@ARTICLE{Akins2024b,
       author = {{Akins}, Hollis B. and {Casey}, Caitlin M. and {Berg}, Danielle A. and {Chisholm}, John and {Cloonan}, Aidan P. and {Franco}, Maximilien and {Finkelstein}, Steven L. and {Fujimoto}, Seiji and {Koekemoer}, Anton M. and {Kokorev}, Vasily and {Lambrides}, Erini and {Robertson}, Brant E. and {Taylor}, Anthony J. and {Coulter}, David A. and {Fox}, Ori and {Karmen}, Mitchell},
        title = "{Strong Rest-UV Emission Lines in a ``Little Red Dot'' Active Galactic Nucleus at z = 7: Early Supermassive Black Hole Growth alongside Compact Massive Star Formation?}",
      journal = {\apjl},
         year = 2025,
        month = feb,
       volume = {980},
       number = {2},
          eid = {L29},
        pages = {L29},
          doi = {10.3847/2041-8213/adab76},
archivePrefix = {arXiv},
       eprint = {2410.00949},
 primaryClass = {astro-ph.GA},
       adsurl = {https://ui.adsabs.harvard.edu/abs/2025ApJ...980L..29A}
}

@ARTICLE{Juodzbalis2024,
       author = {{Juod{\v{z}}balis}, Ignas and {Ji}, Xihan and {Maiolino}, Roberto and {D'Eugenio}, Francesco and {Scholtz}, Jan and {Risaliti}, Guido and {Fabian}, Andrew C. and {Mazzolari}, Giovanni and {Gilli}, Roberto and {Prandoni}, Isabella and {Arribas}, Santiago and {Bunker}, Andrew J. and {Carniani}, Stefano and {Charlot}, St{\'e}phane and {Curtis-Lake}, Emma and {de Graaff}, Anna and {Hainline}, Kevin and {Parlanti}, Eleonora and {Perna}, Michele and {P{\'e}rez-Gonz{\'a}lez}, Pablo G. and {Robertson}, Brant and {Tacchella}, Sandro and {{\"U}bler}, Hannah and {Williams}, Christina C. and {Willott}, Chris and {Witstok}, Joris},
        title = "{JADES - the Rosetta stone of JWST-discovered AGN: deciphering the intriguing nature of early AGN}",
      journal = {\mnras},
         year = 2024,
        month = nov,
       volume = {535},
       number = {1},
        pages = {853-873},
          doi = {10.1093/mnras/stae2367},
archivePrefix = {arXiv},
       eprint = {2407.08643},
 primaryClass = {astro-ph.GA},
       adsurl = {https://ui.adsabs.harvard.edu/abs/2024MNRAS.535..853J}
}

@ARTICLE{Ji2025,
       author = {{Ji}, Xihan and {Maiolino}, Roberto and {{\"U}bler}, Hannah and {Scholtz}, Jan and {D'Eugenio}, Francesco and {Sun}, Fengwu and {Perna}, Michele and {Turner}, Hannah and {Carniani}, Stefano and {Arribas}, Santiago and {Bennett}, Jake S. and {Bunker}, Andrew and {Charlot}, St{\'e}phane and {Cresci}, Giovanni and {Curti}, Mirko and {Egami}, Eiichi and {Fabian}, Andy and {Inayoshi}, Kohei and {Isobe}, Yuki and {Jones}, Gareth and {Juod{\v{z}}balis}, Ignas and {Kumari}, Nimisha and {Lyu}, Jianwei and {Mazzolari}, Giovanni and {Parlanti}, Eleonora and {Robertson}, Brant and {Rodr{\'\i}guez Del Pino}, Bruno and {Schneider}, Raffaella and {Sijacki}, Debora and {Tacchella}, Sandro and {Trinca}, Alessandro and {Valiante}, Rosa and {Venturi}, Giacomo and {Volonteri}, Marta and {Willott}, Chris and {Witten}, Callum and {Witstok}, Joris},
        title = "{BlackTHUNDER ─ A non-stellar Balmer break in a black hole-dominated little red dot at z = 7.04}",
      journal = {\mnras},
         year = 2025,
        month = dec,
       volume = {544},
       number = {4},
        pages = {3900-3935},
          doi = {10.1093/mnras/staf1867},
archivePrefix = {arXiv},
       eprint = {2501.13082},
 primaryClass = {astro-ph.GA},
       adsurl = {https://ui.adsabs.harvard.edu/abs/2025MNRAS.544.3900J}
}

@ARTICLE{Ji2025b,
       author = {{Ji}, Xihan and {D'Eugenio}, Francesco and {Juod{\v{z}}balis}, Ignas and {Walton}, Dominic J. and {Fabian}, Andrew C. and {Maiolino}, Roberto and {Ramos Almeida}, Cristina and {Acosta Pulido}, Jose A. and {Belokurov}, Vasily A. and {Isobe}, Yuki and {Jones}, Gareth and {Maraston}, Claudia and {Scholtz}, Jan and {Simmonds}, Charlotte and {Tacchella}, Sandro and {Terlevich}, Elena and {Terlevich}, Roberto},
        title = "{Lord of LRDs: insights into a 'Little Red Dot' with a low-ionization spectrum at z = 0.1}",
      journal = {\mnras},
         year = 2026,
        month = jan,
       volume = {545},
       number = {3},
          eid = {staf2235},
        pages = {staf2235},
          doi = {10.1093/mnras/staf2235},
archivePrefix = {arXiv},
       eprint = {2507.23774},
 primaryClass = {astro-ph.GA},
       adsurl = {https://ui.adsabs.harvard.edu/abs/2026MNRAS.545f2235J}
}

@ARTICLE{Inayoshi2025,
       author = {{Inayoshi}, Kohei and {Maiolino}, Roberto},
        title = "{Extremely Dense Gas around Little Red Dots and High-redshift Active Galactic Nuclei: A Nonstellar Origin of the Balmer Break and Absorption Features}",
      journal = {\apjl},
         year = 2025,
        month = feb,
       volume = {980},
       number = {2},
          eid = {L27},
        pages = {L27},
          doi = {10.3847/2041-8213/adaebd},
archivePrefix = {arXiv},
       eprint = {2409.07805},
 primaryClass = {astro-ph.GA},
       adsurl = {https://ui.adsabs.harvard.edu/abs/2025ApJ...980L..27I}
}

@ARTICLE{deGraaff2024d,
       author = {{de Graaff}, Anna and {Brammer}, Gabriel and {Weibel}, Andrea and {Lewis}, Zach and {Maseda}, Michael V. and {Oesch}, Pascal A. and {Bezanson}, Rachel and {Boogaard}, Leindert A. and {Cleri}, Nikko J. and {Cooper}, Olivia R. and {Gottumukkala}, Rashmi and {Greene}, Jenny E. and {Hirschmann}, Michaela and {Hviding}, Raphael E. and {Katz}, Harley and {Labb{\'e}}, Ivo and {Leja}, Joel and {Matthee}, Jorryt and {McConachie}, Ian and {Miller}, Tim B. and {Naidu}, Rohan P. and {Price}, Sedona H. and {Rix}, Hans-Walter and {Setton}, David J. and {Suess}, Katherine A. and {Wang}, Bingjie and {Whitaker}, Katherine E. and {Williams}, Christina C.},
        title = "{RUBIES: A complete census of the bright and red distant Universe with JWST/NIRSpec}",
      journal = {\aap},
         year = 2025,
        month = may,
       volume = {697},
          eid = {A189},
        pages = {A189},
          doi = {10.1051/0004-6361/202452186},
archivePrefix = {arXiv},
       eprint = {2409.05948},
 primaryClass = {astro-ph.GA},
       adsurl = {https://ui.adsabs.harvard.edu/abs/2025A&A...697A.189D}
}

@ARTICLE{Pasha2023,
       author = {{Pasha}, Imad and {Miller}, Tim B.},
        title = "{pysersic: A Python package for determining galaxy structural properties via Bayesian inference, accelerated with jax}",
      journal = {The Journal of Open Source Software},
         year = 2023,
        month = sep,
       volume = {8},
       number = {89},
          eid = {5703},
        pages = {5703},
          doi = {10.21105/joss.05703},
archivePrefix = {arXiv},
       eprint = {2306.05454},
 primaryClass = {astro-ph.GA},
       adsurl = {https://ui.adsabs.harvard.edu/abs/2023JOSS....8.5703P}
}

@article{hoffman2014,
  title={The No-U-Turn sampler: adaptively setting path lengths in Hamiltonian Monte Carlo.},
  author={Hoffman, Matthew D and Gelman, Andrew and others},
  journal={J. Mach. Learn. Res.},
  volume={15},
  number={1},
  pages={1593--1623},
  year={2014}
}

@article{phan2019,
  title={Composable Effects for Flexible and Accelerated Probabilistic Programming in NumPyro},
  author={Phan, Du and Pradhan, Neeraj and Jankowiak, Martin},
  journal={arXiv preprint arXiv:1912.11554},
  year={2019}
}

@misc{raphael_erik_hviding_2025_15585035,
  title = {{{TheSkyentist}}/Unite: {{Version}} 0},
  author = {Hviding, Raphael Erik},
  year = {2025},
  month = jun,
  doi = {10.5281/zenodo.15585035},
  howpublished = {Zenodo},
  swhid = {swh:1:dir:a8c79a4538cb89d782d162ccbfc5514d8679b869 ;origin=https://doi.org/10.5281/zenodo.15585034;vi sit=swh:1:snp:8aa315e589f315a109356c7925cf43da4996 38de;anchor=swh:1:rel:8637f0dde82d1d08e9d9e65dee3c 840c80d4aa27;path=TheSkyentist-unite-c790838}
}

@ARTICLE{Gordon2003,
       author = {{Gordon}, Karl D. and {Clayton}, Geoffrey C. and {Misselt}, K.~A. and {Landolt}, Arlo U. and {Wolff}, Michael J.},
        title = "{A Quantitative Comparison of the Small Magellanic Cloud, Large Magellanic Cloud, and Milky Way Ultraviolet to Near-Infrared Extinction Curves}",
      journal = {\apj},
         year = 2003,
        month = sep,
       volume = {594},
       number = {1},
        pages = {279-293},
          doi = {10.1086/376774},
archivePrefix = {arXiv},
       eprint = {astro-ph/0305257},
 primaryClass = {astro-ph},
       adsurl = {https://ui.adsabs.harvard.edu/abs/2003ApJ...594..279G}
}

@ARTICLE{Greene2005,
       author = {{Greene}, Jenny E. and {Ho}, Luis C.},
        title = "{Estimating Black Hole Masses in Active Galaxies Using the H{\ensuremath{\alpha}} Emission Line}",
      journal = {\apj},
         year = 2005,
        month = sep,
       volume = {630},
       number = {1},
        pages = {122-129},
          doi = {10.1086/431897},
archivePrefix = {arXiv},
       eprint = {astro-ph/0508335},
 primaryClass = {astro-ph},
       adsurl = {https://ui.adsabs.harvard.edu/abs/2005ApJ...630..122G}
}

@ARTICLE{Ferland2017,
       author = {{Ferland}, G.~J. and {Chatzikos}, M. and {Guzm{\'a}n}, F. and {Lykins}, M.~L. and {van Hoof}, P.~A.~M. and {Williams}, R.~J.~R. and {Abel}, N.~P. and {Badnell}, N.~R. and {Keenan}, F.~P. and {Porter}, R.~L. and {Stancil}, P.~C.},
        title = "{The 2017 Release Cloudy}",
      journal = {\rmxaa},
         year = 2017,
        month = oct,
       volume = {53},
        pages = {385-438},
          doi = {10.48550/arXiv.1705.10877},
archivePrefix = {arXiv},
       eprint = {1705.10877},
 primaryClass = {astro-ph.GA},
       adsurl = {https://ui.adsabs.harvard.edu/abs/2017RMxAA..53..385F}
}

@ARTICLE{Lambrides2024,
       author = {{Lambrides}, Erini and {Larson}, Rebecca L. and {Garofali}, Kristen and {Ptak}, Andrew and {Chiaberge}, Marco and {Long}, Arianna S. and {Hutchison}, Taylor A. and {Norman}, Colin and {McKinney}, Jed and {Akins}, Hollis B. and {Berg}, Danielle A. and {Chisholm}, John and {Civano}, Francesca and {Cloonan}, Aidan P. and {Endsley}, Ryan and {Faisst}, Andreas L. and {Gilli}, Roberto and {Gillman}, Steven and {Hirschmann}, Michaela and {Kartaltepe}, Jeyhan S. and {Kocevski}, Dale D. and {Kokorev}, Vasily and {Pacucci}, Fabio and {Richardson}, Chris T. and {Stiavelli}, Massimo and {Whalen}, Kelly E.},
        title = "{The case for super-Eddington accretion in JWST broad-line active galactic nuclei during the first billion years}",
      journal = {Nature Astronomy},
         year = 2026,
        month = jun,
       volume = {10},
        pages = {868-879},
          doi = {10.1038/s41550-026-02813-w},
archivePrefix = {arXiv},
       eprint = {2409.13047},
 primaryClass = {astro-ph.HE},
       adsurl = {https://ui.adsabs.harvard.edu/abs/2026NatAs..10..868L}
}

@ARTICLE{Inayoshi2024:xray,
       author = {{Inayoshi}, Kohei and {Kimura}, Shigeo S. and {Noda}, Hirofumi},
        title = "{Weakness of X-rays and variability in high-redshift active galactic nuclei with super-Eddington accretion}",
      journal = {\pasj},
         year = 2025,
        month = aug,
       volume = {77},
       number = {4},
        pages = {811-822},
          doi = {10.1093/pasj/psaf050},
archivePrefix = {arXiv},
       eprint = {2412.03653},
 primaryClass = {astro-ph.HE},
       adsurl = {https://ui.adsabs.harvard.edu/abs/2025PASJ...77..811I}
}

@ARTICLE{Setton2025,
       author = {{Setton}, David J. and {Greene}, Jenny E. and {Spilker}, Justin S. and {Williams}, Christina C. and {Labb{\'e}}, Ivo and {Ma}, Yilun and {Wang}, Bingjie and {Whitaker}, Katherine E. and {Leja}, Joel and {de Graaff}, Anna and {Alberts}, Stacey and {Bezanson}, Rachel and {Boogaard}, Leindert A. and {Brammer}, Gabriel and {Cutler}, Sam E. and {Cleri}, Nikko J. and {Cooper}, Olivia R. and {Dayal}, Pratika and {Fujimoto}, Seiji and {Furtak}, Lukas J. and {Goulding}, Andy D. and {Hirschmann}, Michaela and {Kokorev}, Vasily and {Maseda}, Michael V. and {McConachie}, Ian and {Matthee}, Jorryt and {Miller}, Tim B. and {Naidu}, Rohan P. and {Oesch}, Pascal A. and {Pan}, Richard and {Price}, Sedona H. and {Suess}, Katherine A. and {Weaver}, John R. and {Xiao}, Mengyuan and {Zhang}, Yunchong and {Zitrin}, Adi},
        title = "{A Confirmed Deficit of Hot and Cold Dust Emission in the Most Luminous Little Red Dots}",
      journal = {\apjl},
         year = 2025,
        month = sep,
       volume = {991},
       number = {1},
          eid = {L10},
        pages = {L10},
          doi = {10.3847/2041-8213/ade78b},
archivePrefix = {arXiv},
       eprint = {2503.02059},
 primaryClass = {astro-ph.GA},
       adsurl = {https://ui.adsabs.harvard.edu/abs/2025ApJ...991L..10S}
}

@ARTICLE{Xiao2025,
       author = {{Xiao}, Mengyuan and {Oesch}, Pascal A. and {Bing}, Longji and {Elbaz}, David and {Matthee}, Jorryt and {Fudamoto}, Yoshinobu and {Fujimoto}, Seiji and {Marques-Chaves}, Rui and {Williams}, Christina C. and {Dessauges-Zavadsky}, Miroslava and {Valentino}, Francesco and {Brammer}, Gabriel and {Covelo-Paz}, Alba and {Daddi}, Emanuele and {Fynbo}, Johan P.~U. and {Gillman}, Steven and {Ginolfi}, Michele and {Giovinazzo}, Emma and {Greene}, Jenny E. and {Gu}, Qiusheng and {Illingworth}, Garth and {Inayoshi}, Kohei and {Kokorev}, Vasily and {Meyer}, Romain A. and {Naidu}, Rohan P. and {Reddy}, Naveen A. and {Schaerer}, Daniel and {Shapley}, Alice and {Stefanon}, Mauro and {Steinhardt}, Charles L. and {Setton}, David J. and {Vestergaard}, Marianne and {Wang}, Tao},
        title = "{No [C II] or dust detection in two Little Red Dots at z$_{spec}$> 7}",
      journal = {\aap},
         year = 2025,
        month = aug,
       volume = {700},
          eid = {A231},
        pages = {A231},
          doi = {10.1051/0004-6361/202554361},
archivePrefix = {arXiv},
       eprint = {2503.01945},
 primaryClass = {astro-ph.GA},
       adsurl = {https://ui.adsabs.harvard.edu/abs/2025A&A...700A.231X}
}

@article{Watanabe:WAIC,
  title={Asymptotic equivalence of Bayes cross validation and widely applicable information criterion in singular learning theory.},
  author={Watanabe, Sumio and Opper, Manfred},
  journal={Journal of machine learning research},
  volume={11},
  number={12},
  year={2010}
}

@ARTICLE{Begelman2006,
       author = {{Begelman}, Mitchell C. and {Volonteri}, Marta and {Rees}, Martin J.},
        title = "{Formation of supermassive black holes by direct collapse in pre-galactic haloes}",
      journal = {\mnras},
         year = 2006,
        month = jul,
       volume = {370},
       number = {1},
        pages = {289-298},
          doi = {10.1111/j.1365-2966.2006.10467.x},
archivePrefix = {arXiv},
       eprint = {astro-ph/0602363},
 primaryClass = {astro-ph},
       adsurl = {https://ui.adsabs.harvard.edu/abs/2006MNRAS.370..289B}
}

@ARTICLE{Begelman2008,
       author = {{Begelman}, Mitchell C. and {Rossi}, Elena M. and {Armitage}, Philip J.},
        title = "{Quasi-stars: accreting black holes inside massive envelopes}",
      journal = {\mnras},
         year = 2008,
        month = jul,
       volume = {387},
       number = {4},
        pages = {1649-1659},
          doi = {10.1111/j.1365-2966.2008.13344.x},
archivePrefix = {arXiv},
       eprint = {0711.4078},
 primaryClass = {astro-ph},
       adsurl = {https://ui.adsabs.harvard.edu/abs/2008MNRAS.387.1649B}
}

@ARTICLE{Coughlin2024,
       author = {{Coughlin}, Eric R. and {Begelman}, Mitchell C.},
        title = "{Quasi-stars as a Means of Rapid Black Hole Growth in the Early Universe}",
      journal = {\apj},
         year = 2024,
        month = aug,
       volume = {970},
       number = {2},
          eid = {158},
        pages = {158},
          doi = {10.3847/1538-4357/ad5723},
archivePrefix = {arXiv},
       eprint = {2405.00084},
 primaryClass = {astro-ph.GA},
       adsurl = {https://ui.adsabs.harvard.edu/abs/2024ApJ...970..158C}
}

@ARTICLE{DEugenio2025,
       author = {{D'Eugenio}, Francesco and {Maiolino}, Roberto and {Perna}, Michele and {{\"U}bler}, Hannah and {Ji}, Xihan and {McClymont}, William and {Koudmani}, Sophie and {Sijacki}, Debora and {Juod{\v{z}}balis}, Ignas and {Scholtz}, Jan and {Bennett}, Jake S. and {Bunker}, Andrew J. and {Carniani}, Stefano and {Charlot}, St{\'e}phane and {Cresci}, Giovanni and {Curtis-Lake}, Emma and {Bont{\`a}}, Elena Dalla and {Inayoshi}, Kohei and {Jones}, Gareth C. and {Lyu}, Jianwei and {Marconi}, Alessandro and {Mazzolari}, Giovanni and {Nelson}, Erica J. and {Parlanti}, Eleonora and {Robertson}, Brant E. and {Schneider}, Raffaella and {Simmonds}, Charlotte and {Tacchella}, Sandro and {Venturi}, Giacomo and {Willott}, Chris and {Witstok}, Joris and {Witten}, Callum},
        title = "{BlackTHUNDER strikes twice: Balmer-line absorption in an overmassive Little Red Dot at z = 7.04}",
      journal = {\mnras},
         year = 2026,
        month = apr,
       volume = {547},
       number = {4},
          eid = {stag401},
        pages = {stag401},
          doi = {10.1093/mnras/stag401},
archivePrefix = {arXiv},
       eprint = {2503.11752},
 primaryClass = {astro-ph.GA},
       adsurl = {https://ui.adsabs.harvard.edu/abs/2026MNRAS.547ag401D}
}

@ARTICLE{Rusakov2025,
       author = {{Rusakov}, V. and {Watson}, D. and {Nikopoulos}, G.~P. and {Brammer}, G. and {Gottumukkala}, R. and {Harvey}, T. and {Heintz}, K.~E. and {Damgaard}, R. and {Sim}, S.~A. and {Sneppen}, A. and {Vijayan}, A.~P. and {Adams}, N. and {Austin}, D. and {Conselice}, C.~J. and {Goolsby}, C.~M. and {Toft}, S. and {Witstok}, J.},
        title = "{Little red dots as young supermassive black holes in dense ionized cocoons}",
      journal = {\nat},
         year = 2026,
        month = jan,
       volume = {649},
       number = {8097},
        pages = {574-579},
          doi = {10.1038/s41586-025-09900-4},
archivePrefix = {arXiv},
       eprint = {2503.16595},
 primaryClass = {astro-ph.GA},
       adsurl = {https://ui.adsabs.harvard.edu/abs/2026Natur.649..574R}
}

@ARTICLE{Hviding2025,
       author = {{Hviding}, Raphael E. and {de Graaff}, Anna and {Miller}, Tim B. and {Setton}, David J. and {Greene}, Jenny E. and {Labb{\'e}}, Ivo and {Brammer}, Gabriel and {Bezanson}, Rachel and {Boogaard}, Leindert A. and {Cleri}, Nikko J. and {Leja}, Joel and {Maseda}, Michael V. and {McConachie}, Ian and {Matthee}, Jorryt and {Naidu}, Rohan P. and {Oesch}, Pascal A. and {Wang}, Bingjie and {Whitaker}, Katherine E. and {Williams}, Christina C.},
        title = "{RUBIES: A spectroscopic census of little red dots: All point sources with v-shaped continua have broad lines}",
      journal = {\aap},
         year = 2025,
        month = oct,
       volume = {702},
          eid = {A57},
        pages = {A57},
          doi = {10.1051/0004-6361/202555816},
archivePrefix = {arXiv},
       eprint = {2506.05459},
 primaryClass = {astro-ph.GA},
       adsurl = {https://ui.adsabs.harvard.edu/abs/2025A&A...702A..57H}
}

@ARTICLE{deGraaff2025,
       author = {{de Graaff}, Anna and {Rix}, Hans-Walter and {Naidu}, Rohan P. and {Labb{\'e}}, Ivo and {Wang}, Bingjie and {Leja}, Joel and {Matthee}, Jorryt and {Katz}, Harley and {Greene}, Jenny E. and {Hviding}, Raphael E. and {Baggen}, Josephine and {Bezanson}, Rachel and {Boogaard}, Leindert A. and {Brammer}, Gabriel and {Dayal}, Pratika and {van Dokkum}, Pieter and {Goulding}, Andy D. and {Hirschmann}, Michaela and {Maseda}, Michael V. and {McConachie}, Ian and {Miller}, Tim B. and {Nelson}, Erica and {Oesch}, Pascal A. and {Setton}, David J. and {Shivaei}, Irene and {Weibel}, Andrea and {Whitaker}, Katherine E. and {Williams}, Christina C.},
        title = "{A remarkable ruby: Absorption in dense gas, rather than evolved stars, drives the extreme Balmer break of a little red dot at z = 3.5}",
      journal = {\aap},
         year = 2025,
        month = sep,
       volume = {701},
          eid = {A168},
        pages = {A168},
          doi = {10.1051/0004-6361/202554681},
archivePrefix = {arXiv},
       eprint = {2503.16600},
 primaryClass = {astro-ph.GA},
       adsurl = {https://ui.adsabs.harvard.edu/abs/2025A&A...701A.168D}
}

@ARTICLE{Valentino2025,
       author = {{Valentino}, F. and {Heintz}, K.~E. and {Brammer}, G. and {Ito}, K. and {Kokorev}, V. and {Whitaker}, K.~E. and {Gallazzi}, A. and {de Graaff}, A. and {Weibel}, A. and {Frye}, B.~L. and {Kamieneski}, P.~S. and {Jin}, S. and {Ceverino}, D. and {Faisst}, A. and {Farcy}, M. and {Fujimoto}, S. and {Gillman}, S. and {Gottumukkala}, R. and {Hamadouche}, M. and {Harrington}, K.~C. and {Hirschmann}, M. and {Jespersen}, C.~K. and {Kakimoto}, T. and {Kubo}, M. and {Lagos}, C. d. P. and {Lee}, M. and {Magdis}, G.~E. and {Man}, A.~W.~S. and {Onodera}, M. and {Rizzo}, F. and {Shimakawa}, R. and {Setton}, D.~J. and {Tanaka}, M. and {Toft}, S. and {Wu}, P. -F. and {Zhu}, P.},
        title = "{Gas outflows in two recently quenched galaxies at z = 4 and 7}",
      journal = {\aap},
         year = 2025,
        month = jul,
       volume = {699},
          eid = {A358},
        pages = {A358},
          doi = {10.1051/0004-6361/202553908},
archivePrefix = {arXiv},
       eprint = {2503.01990},
 primaryClass = {astro-ph.GA},
       adsurl = {https://ui.adsabs.harvard.edu/abs/2025A&A...699A.358V}
}

@ARTICLE{Finkelstein2025,
       author = {{Finkelstein}, Steven L. and {Bagley}, Micaela B. and {Arrabal Haro}, Pablo and {Dickinson}, Mark and {Ferguson}, Henry C. and {Kartaltepe}, Jeyhan S. and {Kocevski}, Dale D. and {Koekemoer}, Anton M. and {Lotz}, Jennifer M. and {Papovich}, Casey and {P{\'e}rez-Gonz{\'a}lez}, Pablo G. and {Pirzkal}, Nor and {Somerville}, Rachel S. and {Trump}, Jonathan R. and {Yang}, Guang and {Yung}, L.~Y. Aaron and {Fontana}, Adriano and {Grazian}, Andrea and {Grogin}, Norman A. and {Kewley}, Lisa J. and {Kirkpatrick}, Allison and {Larson}, Rebecca L. and {Pentericci}, Laura and {Ravindranath}, Swara and {Wilkins}, Stephen M. and {Almaini}, Omar and {Amor{\'\i}n}, Ricardo O. and {Barro}, Guillermo and {Bhatawdekar}, Rachana and {Bisigello}, Laura and {Brooks}, Madisyn and {Buat}, V{\'e}ronique and {Buitrago}, Fernando and {Burgarella}, Denis and {Calabr{\`o}}, Antonello and {Castellano}, Marco and {Cheng}, Yingjie and {Cleri}, Nikko J. and {Cole}, Justin W. and {Cooper}, M.~C. and {Cooper}, Olivia R. and {Costantin}, Luca and {Cox}, Isa G. and {Croton}, Darren and {Daddi}, Emanuele and {Davis}, Kelcey and {Dekel}, Avishai and {Elbaz}, David and {Fern{\'a}ndez}, Vital and {Fujimoto}, Seiji and {Gandolfi}, Giovanni and {Gardner}, Jonathan P. and {Gawiser}, Eric and {Giavalisco}, Mauro and {G{\'o}mez-Guijarro}, Carlos and {Guo}, Yuchen and {Gupta}, Ansh R. and {Hathi}, Nimish P. and {Harish}, Santosh and {Henry}, Aur{\'e}lien and {Hirschmann}, Michaela and {Hu}, Weida and {Hutchison}, Taylor A. and {Iyer}, Kartheik G. and {Jaskot}, Anne E. and {Jha}, Saurabh W. and {Jung}, Intae and {Kassin}, Susan A. and {Kokorev}, Vasily and {Kurczynski}, Peter and {Leung}, Gene C.~K. and {Llerena}, Mario and {Long}, Arianna S. and {Lucas}, Ray A. and {Lu}, Shiying and {McGrath}, Elizabeth J. and {McIntosh}, Daniel H. and {Merlin}, Emiliano and {Mobasher}, Bahram and {Morales}, Alexa M. and {Napolitano}, Lorenzo and {Pacucci}, Fabio and {Pandya}, Viraj and {Rafelski}, Marc and {Rodighiero}, Giulia and {Rose}, Caitlin and {Santini}, Paola and {Seill{\'e}}, Lise-Marie and {Simons}, Raymond C. and {Shen}, Lu and {Straughn}, Amber N. and {Tacchella}, Sandro and {Taylor}, Anthony J. and {Vanderhoof}, Brittany N. and {Vega-Ferrero}, Jes{\'u}s and {Weiner}, Benjamin J. and {Willmer}, Christopher N.~A. and {Zhu}, Peixin and {Bell}, Eric F. and {Wuyts}, Stijn and {Holwerda}, Benne W. and {Wang}, Xin and {Wang}, Weichen and {Zavala}, Jorge A. and {CEERS Collaboration}},
        title = "{The Cosmic Evolution Early Release Science Survey (CEERS)}",
      journal = {\apjl},
         year = 2025,
        month = apr,
       volume = {983},
       number = {1},
          eid = {L4},
        pages = {L4},
          doi = {10.3847/2041-8213/adbbd3},
archivePrefix = {arXiv},
       eprint = {2501.04085},
 primaryClass = {astro-ph.GA},
       adsurl = {https://ui.adsabs.harvard.edu/abs/2025ApJ...983L...4F}
}

@ARTICLE{Bezanson2024,
       author = {{Bezanson}, Rachel and {Labbe}, Ivo and {Whitaker}, Katherine E. and {Leja}, Joel and {Price}, Sedona H. and {Franx}, Marijn and {Brammer}, Gabriel and {Marchesini}, Danilo and {Zitrin}, Adi and {Wang}, Bingjie and {Weaver}, John R. and {Furtak}, Lukas J. and {Atek}, Hakim and {Coe}, Dan and {Cutler}, Sam E. and {Dayal}, Pratika and {van Dokkum}, Pieter and {Feldmann}, Robert and {F{\"o}rster Schreiber}, Natascha M. and {Fujimoto}, Seiji and {Geha}, Marla and {Glazebrook}, Karl and {de Graaff}, Anna and {Greene}, Jenny E. and {Juneau}, St{\'e}phanie and {Kassin}, Susan and {Kriek}, Mariska and {Khullar}, Gourav and {Maseda}, Michael and {Mowla}, Lamiya A. and {Muzzin}, Adam and {Nanayakkara}, Themiya and {Nelson}, Erica J. and {Oesch}, Pascal A. and {Pacifici}, Camilla and {Pan}, Richard and {Papovich}, Casey and {Setton}, David J. and {Shapley}, Alice E. and {Smit}, Renske and {Stefanon}, Mauro and {Taylor}, Edward N. and {Williams}, Christina C.},
        title = "{The JWST UNCOVER Treasury Survey: Ultradeep NIRSpec and NIRCam Observations before the Epoch of Reionization}",
      journal = {\apj},
         year = 2024,
        month = oct,
       volume = {974},
       number = {1},
          eid = {92},
        pages = {92},
          doi = {10.3847/1538-4357/ad66cf},
archivePrefix = {arXiv},
       eprint = {2212.04026},
 primaryClass = {astro-ph.GA},
       adsurl = {https://ui.adsabs.harvard.edu/abs/2024ApJ...974...92B}
}

@ARTICLE{Price2025,
       author = {{Price}, Sedona H. and {Bezanson}, Rachel and {Labbe}, Ivo and {Furtak}, Lukas J. and {de Graaff}, Anna and {Greene}, Jenny E. and {Kokorev}, Vasily and {Setton}, David J. and {Suess}, Katherine A. and {Brammer}, Gabriel and {Cutler}, Sam E. and {Leja}, Joel and {Pan}, Richard and {Wang}, Bingjie and {Weaver}, John R. and {Whitaker}, Katherine E. and {Atek}, Hakim and {Burgasser}, Adam J. and {Chemerynska}, Iryna and {Dayal}, Pratika and {Feldmann}, Robert and {F{\"o}rster Schreiber}, Natascha M. and {Fudamoto}, Yoshinobu and {Fujimoto}, Seiji and {Glazebrook}, Karl and {Goulding}, Andy D. and {Khullar}, Gourav and {Kriek}, Mariska and {Marchesini}, Danilo and {Maseda}, Michael V. and {Miller}, Tim B. and {Muzzin}, Adam and {Nanayakkara}, Themiya and {Nelson}, Erica and {Oesch}, Pascal A. and {Shipley}, Heath and {Smit}, Renske and {Taylor}, Edward N. and {Dokkum}, Pieter van and {Williams}, Christina C. and {Zitrin}, Adi},
        title = "{The UNCOVER Survey: First Release of Ultradeep JWST/NIRSpec PRISM Spectra for {\ensuremath{\sim}}700 Galaxies from z {\ensuremath{\sim}} 0.3{\textendash}13 in A2744}",
      journal = {\apj},
         year = 2025,
        month = mar,
       volume = {982},
       number = {1},
          eid = {51},
        pages = {51},
          doi = {10.3847/1538-4357/adaec1},
archivePrefix = {arXiv},
       eprint = {2408.03920},
 primaryClass = {astro-ph.GA},
       adsurl = {https://ui.adsabs.harvard.edu/abs/2025ApJ...982...51P}
}

@ARTICLE{Shen2024,
       author = {{Shen}, Yue and {Zhuang}, Ming-Yang and {Li}, Junyao and {Burgasser}, Adam J. and {Fan}, Xiaohui and {Greene}, Jenny E. and {Narayan}, Gautham and {Shapley}, Alice E. and {Sun}, Fengwu and {Wang}, Feige and {Yang}, Qian},
        title = "{NEXUS: the North ecliptic pole EXtragalactic Unified Survey}",
      journal = {arXiv e-prints},
         year = 2024,
        month = aug,
          eid = {arXiv:2408.12713},
        pages = {arXiv:2408.12713},
          doi = {10.48550/arXiv.2408.12713},
archivePrefix = {arXiv},
       eprint = {2408.12713},
 primaryClass = {astro-ph.GA},
       adsurl = {https://ui.adsabs.harvard.edu/abs/2024arXiv240812713S}
}

@ARTICLE{Taylor2025,
       author = {{Taylor}, Anthony J. and {Kokorev}, Vasily and {Kocevski}, Dale D. and {Akins}, Hollis B. and {Cullen}, Fergus and {Dickinson}, Mark and {Finkelstein}, Steven L. and {Arrabal Haro}, Pablo and {Bromm}, Volker and {Giavalisco}, Mauro and {Inayoshi}, Kohei and {Juneau}, St{\'e}phanie and {Leung}, Gene C.~K. and {P{\'e}rez-Gonz{\'a}lez}, Pablo G. and {Somerville}, Rachel S. and {Trump}, Jonathan R. and {Amor{\'\i}n}, Ricardo O. and {Barro}, Guillermo and {Burgarella}, Denis and {Brooks}, Madisyn and {Carnall}, Adam C. and {Casey}, Caitlin M. and {Cheng}, Yingjie and {Chisholm}, John and {Chworowsky}, Katherine and {Davis}, Kelcey and {Donnan}, Callum T. and {Dunlop}, James S. and {Ellis}, Richard S. and {Fern{\'a}ndez}, Vital and {Fujimoto}, Seiji and {Grogin}, Norman A. and {Gupta}, Ansh R. and {Hathi}, Nimish P. and {Jung}, Intae and {Hirschmann}, Michaela and {Kartaltepe}, Jeyhan S. and {Koekemoer}, Anton M. and {Larson}, Rebecca L. and {Leung}, Ho-Hin and {Llerena}, Mario and {Lucas}, Ray A. and {McLeod}, Derek J. and {McLure}, Ross and {Napolitano}, Lorenzo and {Papovich}, Casey and {Stanton}, Thomas M. and {Tripodi}, Roberta and {Wang}, Xin and {Wilkins}, Stephen M. and {Yung}, L.~Y. Aaron and {Zavala}, Jorge A.},
        title = "{CAPERS-LRD-z9: A Gas-enshrouded Little Red Dot Hosting a Broad-line Active Galactic Nucleus at z = 9.288}",
      journal = {\apjl},
         year = 2025,
        month = aug,
       volume = {989},
       number = {1},
          eid = {L7},
        pages = {L7},
          doi = {10.3847/2041-8213/ade789},
archivePrefix = {arXiv},
       eprint = {2505.04609},
 primaryClass = {astro-ph.GA},
       adsurl = {https://ui.adsabs.harvard.edu/abs/2025ApJ...989L...7T}
}

@ARTICLE{Naidu2025,
       author = {{Naidu}, Rohan P. and {Matthee}, Jorryt and {Katz}, Harley and {de Graaff}, Anna and {Oesch}, Pascal and {Smith}, Aaron and {Greene}, Jenny E. and {Brammer}, Gabriel and {Weibel}, Andrea and {Hviding}, Raphael and {Chisholm}, John and {Labb\textbackslash'e}, Ivo and {Simcoe}, Robert A. and {Witten}, Callum and {Atek}, Hakim and {Baggen}, Josephine F.~W. and {Belli}, Sirio and {Bezanson}, Rachel and {Boogaard}, Leindert A. and {Bose}, Sownak and {Covelo-Paz}, Alba and {Dayal}, Pratika and {Fudamoto}, Yoshinobu and {Furtak}, Lukas J. and {Giovinazzo}, Emma and {Goulding}, Andy and {Gronke}, Max and {Heintz}, Kasper E. and {Hirschmann}, Michaela and {Illingworth}, Garth and {Inoue}, Akio K. and {Johnson}, Benjamin D. and {Leja}, Joel and {Leonova}, Ecaterina and {McConachie}, Ian and {Maseda}, Michael V. and {Natarajan}, Priyamvada and {Nelson}, Erica and {Setton}, David J. and {Shivaei}, Irene and {Sobral}, David and {Stefanon}, Mauro and {Tacchella}, Sandro and {Toft}, Sune and {Torralba}, Alberto and {van Dokkum}, Pieter and {van der Wel}, Arjen and {Volonteri}, Marta and {Walter}, Fabian and {Wang}, Bingjie and {Watson}, Darach},
        author = {Naidu, Rohan and Matthee, Jorryt and Katz, Harley and Graaff, Anna and Oesch, Pascal and Smith, Aaron and Greene, Jenny and Brammer, Gabriel and Weibel, Andrea and Hviding, Raphael and Chisholm, John and Labbe, Ivo and Simcoe, Robert and Witten, Callum and Sun, Wendy and Atek, Hakim and Baggen, Josephine and Belli, Sirio and Bezanson, Rachel and Whitaker, Katherine},
        year = {2026},
        month = {08},
        pages = {329-333},
        title = {A gas-enshrouded and gas-reddened black hole at cosmic dawn},
        volume = {656},
        journal = {Nature},
        doi = {10.1038/s41586-026-10846-4}
}

@ARTICLE{Casey2025,
       author = {{Casey}, Caitlin M. and {Akins}, Hollis B. and {Finkelstein}, Steven L. and {Franco}, Maximilien and {Fujimoto}, Seiji and {Liu}, Daizhong and {Long}, Arianna S. and {Magdis}, Georgios and {Manning}, Sinclaire M. and {McKinney}, Jed and {Shuntov}, Marko and {Tanaka}, Takumi S.},
        title = "{An Upper Limit of {}10$^{6}$ M$_{{\ensuremath{\odot}}}$ in Dust from ALMA Observations in 60 Little Red Dots}",
      journal = {\apjl},
         year = 2025,
        month = sep,
       volume = {990},
       number = {2},
          eid = {L61},
        pages = {L61},
          doi = {10.3847/2041-8213/adfa91},
archivePrefix = {arXiv},
       eprint = {2505.18873},
 primaryClass = {astro-ph.GA},
       adsurl = {https://ui.adsabs.harvard.edu/abs/2025ApJ...990L..61C}
}

@ARTICLE{Begelman2025,
       author = {{Begelman}, Mitchell C. and {Dexter}, Jason},
        title = "{Little Red Dots as Late-stage Quasi-stars}",
      journal = {\apj},
         year = 2026,
        month = jan,
       volume = {996},
       number = {1},
          eid = {48},
        pages = {48},
          doi = {10.3847/1538-4357/ae274a},
archivePrefix = {arXiv},
       eprint = {2507.09085},
 primaryClass = {astro-ph.GA},
       adsurl = {https://ui.adsabs.harvard.edu/abs/2026ApJ...996...48B}
}

@ARTICLE{Kido2025,
       author = {{Kido}, Daisaburo and {Ioka}, Kunihito and {Hotokezaka}, Kenta and {Inayoshi}, Kohei and {Irwin}, Christopher M.},
        title = "{Black hole envelopes in Little Red Dots}",
      journal = {\mnras},
         year = 2025,
        month = dec,
       volume = {544},
       number = {4},
        pages = {3407-3416},
          doi = {10.1093/mnras/staf1898},
archivePrefix = {arXiv},
       eprint = {2505.06965},
 primaryClass = {astro-ph.HE},
       adsurl = {https://ui.adsabs.harvard.edu/abs/2025MNRAS.544.3407K}
}

@ARTICLE{Liu2025,
       author = {{Liu}, Hanpu and {Jiang}, Yan-Fei and {Quataert}, Eliot and {Greene}, Jenny E. and {Ma}, Yilun},
        title = "{The Balmer Break and Optical Continuum of Little Red Dots from Super-Eddington Accretion}",
      journal = {\apj},
         year = 2025,
        month = nov,
       volume = {994},
       number = {1},
          eid = {113},
        pages = {113},
          doi = {10.3847/1538-4357/ae0c19},
archivePrefix = {arXiv},
       eprint = {2507.07190},
 primaryClass = {astro-ph.GA},
       adsurl = {https://ui.adsabs.harvard.edu/abs/2025ApJ...994..113L}
}

@ARTICLE{Lin2025,
       author = {{Lin}, Xiaojing and {Fan}, Xiaohui and {Cai}, Zheng and {Bian}, Fuyan and {Liu}, Hanpu and {Sun}, Fengwu and {Ma}, Yilun and {Greene}, Jenny E. and {Strauss}, Michael A. and {Green}, Richard and {Lyu}, Jianwei and {Champagne}, Jaclyn B. and {Goulding}, Andy D. and {Inayoshi}, Kohei and {Jin}, Xiangyu and {Leung}, Gene C.~K. and {Li}, Mingyu and {Liu}, Weizhe and {Liu}, Yichen and {Mao}, Junjie and {Pudoka}, Maria Anne and {Tee}, Wei Leong and {Wang}, Ben and {Wang}, Feige and {Wu}, Yunjing and {Yang}, Jinyi and {Zhang}, Haowen and {Zhu}, Yongda},
        title = "{The Discovery of Little Red Dots in the Local Universe: Signatures of Cool Gas Envelopes}",
      journal = {\apj},
         year = 2026,
        month = feb,
       volume = {997},
       number = {2},
          eid = {364},
        pages = {364},
          doi = {10.3847/1538-4357/ae2bdf},
archivePrefix = {arXiv},
       eprint = {2507.10659},
 primaryClass = {astro-ph.GA},
       adsurl = {https://ui.adsabs.harvard.edu/abs/2026ApJ...997..364L}
}

@ARTICLE{Pacucci2024,
       author = {{Pacucci}, Fabio and {Narayan}, Ramesh},
        title = "{Mildly Super-Eddington Accretion onto Slowly Spinning Black Holes Explains the X-Ray Weakness of the Little Red Dots}",
      journal = {\apj},
         year = 2024,
        month = nov,
       volume = {976},
       number = {1},
          eid = {96},
        pages = {96},
          doi = {10.3847/1538-4357/ad84f7},
archivePrefix = {arXiv},
       eprint = {2407.15915},
 primaryClass = {astro-ph.HE},
       adsurl = {https://ui.adsabs.harvard.edu/abs/2024ApJ...976...96P}
}

@ARTICLE{abmag,
       author = {{Oke}, J.~B. and {Gunn}, J.~E.},
        title = "{Secondary standard stars for absolute spectrophotometry.}",
      journal = {\apj},
         year = 1983,
        month = mar,
       volume = {266},
        pages = {713-717},
          doi = {10.1086/160817},
       adsurl = {https://ui.adsabs.harvard.edu/abs/1983ApJ...266..713O}
}

@ARTICLE{Chang2025,
       author = {{Chang}, Seok-Jun and {Gronke}, Max and {Matthee}, Jorryt and {Mason}, Charlotte},
        title = "{Impact of resonance, Raman, and Thomson scattering on hydrogen line formation in Little Red Dots}",
      journal = {\mnras},
         year = 2026,
        month = feb,
       volume = {545},
       number = {4},
          eid = {staf2131},
        pages = {staf2131},
          doi = {10.1093/mnras/staf2131},
archivePrefix = {arXiv},
       eprint = {2508.08768},
 primaryClass = {astro-ph.GA},
       adsurl = {https://ui.adsabs.harvard.edu/abs/2026MNRAS.545f2131C}
}

@ARTICLE{Brazzini2025,
       author = {{Brazzini}, Matilde and {D'Eugenio}, Francesco and {Maiolino}, Roberto and {Juod{\v{z}}balis}, Ignas and {Ji}, Xihan and {Scholtz}, Jan and {Chang}, Seok-Jun},
        title = "{Ruling out dominant electron scattering in Little Red Dots' Rosetta Stone using multiple hydrogen lines}",
      journal = {\mnras},
         year = 2025,
        month = nov,
       volume = {544},
       number = {1},
        pages = {L167-L173},
          doi = {10.1093/mnrasl/slaf116},
archivePrefix = {arXiv},
       eprint = {2507.08929},
 primaryClass = {astro-ph.GA},
       adsurl = {https://ui.adsabs.harvard.edu/abs/2025MNRAS.544L.167B}
}

@ARTICLE{Lin2024,
       author = {{Lin}, Xiaojing and {Wang}, Feige and {Fan}, Xiaohui and {Cai}, Zheng and {Champagne}, Jaclyn B. and {Sun}, Fengwu and {Volonteri}, Marta and {Yang}, Jinyi and {Hennawi}, Joseph F. and {Ba{\~n}ados}, Eduardo and {Barth}, Aaron and {Eilers}, Anna-Christina and {Farina}, Emanuele Paolo and {Liu}, Weizhe and {Jin}, Xiangyu and {Jun}, Hyunsung D. and {Lupi}, Alessandro and {Kakiichi}, Koki and {Mazzucchelli}, Chiara and {Onoue}, Masafusa and {Pan}, Zhiwei and {Pizzati}, Elia and {Rojas-Ruiz}, Sof{\'\i}a and {Schindler}, Jan-Torge and {Trakhtenbrot}, Benny and {Shen}, Yue and {Trebitsch}, Maxime and {Zhuang}, Ming-Yang and {Endsley}, Ryan and {Meyer}, Romain A. and {Li}, Zihao and {Li}, Mingyu and {Pudoka}, Maria and {Tee}, Wei Leong and {Wu}, Yunjing and {Zhang}, Haowen},
        title = "{A SPectroscopic Survey of Biased Halos In the Reionization Era (ASPIRE): Broad-line AGN at z = 4‑5 Revealed by JWST/NIRCam WFSS}",
      journal = {\apj},
         year = 2024,
        month = oct,
       volume = {974},
       number = {1},
          eid = {147},
        pages = {147},
          doi = {10.3847/1538-4357/ad6565},
archivePrefix = {arXiv},
       eprint = {2407.17570},
 primaryClass = {astro-ph.GA},
       adsurl = {https://ui.adsabs.harvard.edu/abs/2024ApJ...974..147L}
}

@ARTICLE{Shetty2009a,
       author = {{Shetty}, Rahul and {Kauffmann}, Jens and {Schnee}, Scott and {Goodman}, Alyssa A. and {Ercolano}, Barbara},
        title = "{The Effect of Line-of-Sight Temperature Variation and Noise on Dust Continuum Observations}",
      journal = {\apj},
         year = 2009,
        month = may,
       volume = {696},
       number = {2},
        pages = {2234-2251},
          doi = {10.1088/0004-637X/696/2/2234},
archivePrefix = {arXiv},
       eprint = {0902.3477},
 primaryClass = {astro-ph.GA},
       adsurl = {https://ui.adsabs.harvard.edu/abs/2009ApJ...696.2234S}
}

@ARTICLE{Shetty2009b,
       author = {{Shetty}, Rahul and {Kauffmann}, Jens and {Schnee}, Scott and {Goodman}, Alyssa A.},
        title = "{The Effect of Noise on the Dust Temperature-Spectral Index Correlation}",
      journal = {\apj},
         year = 2009,
        month = may,
       volume = {696},
       number = {1},
        pages = {676-680},
          doi = {10.1088/0004-637X/696/1/676},
archivePrefix = {arXiv},
       eprint = {0902.0636},
 primaryClass = {astro-ph.GA},
       adsurl = {https://ui.adsabs.harvard.edu/abs/2009ApJ...696..676S}
}

@ARTICLE{Hildebrand1983,
       author = {{Hildebrand}, R.~H.},
        title = "{The determination of cloud masses and dust characteristics from submillimetre thermal emission.}",
      journal = {\qjras},
         year = 1983,
        month = sep,
       volume = {24},
        pages = {267-282},
       adsurl = {https://ui.adsabs.harvard.edu/abs/1983QJRAS..24..267H}
}

@ARTICLE{Furtak2023:lensing,
       author = {{Furtak}, Lukas J. and {Zitrin}, Adi and {Weaver}, John R. and {Atek}, Hakim and {Bezanson}, Rachel and {Labb{\'e}}, Ivo and {Whitaker}, Katherine E. and {Leja}, Joel and {Price}, Sedona H. and {Brammer}, Gabriel B. and {Wang}, Bingjie and {Marchesini}, Danilo and {Pan}, Richard and {Dayal}, Pratika and {van Dokkum}, Pieter and {Feldmann}, Robert and {Fujimoto}, Seiji and {Franx}, Marijn and {Khullar}, Gourav and {Nelson}, Erica J. and {Mowla}, Lamiya A.},
        title = "{UNCOVERing the extended strong lensing structures of Abell 2744 with the deepest JWST imaging}",
      journal = {\mnras},
         year = 2023,
        month = aug,
       volume = {523},
       number = {3},
        pages = {4568-4582},
          doi = {10.1093/mnras/stad1627},
archivePrefix = {arXiv},
       eprint = {2212.04381},
 primaryClass = {astro-ph.GA},
       adsurl = {https://ui.adsabs.harvard.edu/abs/2023MNRAS.523.4568F}
}

@ARTICLE{Tripodi2025,
       author = {{Tripodi}, Roberta and {Brada{\v{c}}}, Maru{\v{s}}a and {D'Eugenio}, Francesco and {Martis}, Nicholas and {Rihtar{\v{s}}i{\v{c}}}, Gregor and {Willott}, Chris and {Pentericci}, Laura and {Moreschini}, Bianca and {Markevitch}, Maxim and {Asada}, Yoshihisa and {Calabr{\'o}}, Antonello and {Desprez}, Guillaume and {Felicioni}, Giordano and {Gaspar}, Gaia and {Gonzalez}, Anthony H. and {Harshan}, Anishya and {Ji}, Xihan and {Jude{\v{z}}}, Jon and {Lemaux}, Brian C. and {Marconi}, Alessandro and {Markov}, Vladan and {Merida}, Rosa M. and {Napolitano}, Lorenzo and {Noirot}, Ga{\"e}l and {Parente}, Massimiliano and {Peter}, Annika H.~G. and {Robbins}, Luke and {Robertson}, Andrew and {Sarrouh}, Ghassan T.~E. and {Sawicki}, Marcin},
        title = "{A Deep Dive down the Broad-line Region: Permitted O I, Ca II, and Fe II Emission in an Active Galactic Nucleus Little Red Dot at z = 5.3}",
      journal = {\apjl},
         year = 2025,
        month = nov,
       volume = {994},
       number = {1},
          eid = {L6},
        pages = {L6},
          doi = {10.3847/2041-8213/ae13a9},
archivePrefix = {arXiv},
       eprint = {2507.20684},
 primaryClass = {astro-ph.GA},
       adsurl = {https://ui.adsabs.harvard.edu/abs/2025ApJ...994L...6T}
}

@ARTICLE{Mathew2018,
       author = {{Mathew}, Blesson and {Manoj}, P. and {Narang}, Mayank and {Banerjee}, D.~P.~K. and {Nayak}, Pratheeksha and {Muneer}, S. and {Vig}, S. and {Pramod Kumar}, S. and {Paul}, K.~T. and {Maheswar}, G.},
        title = "{Excitation Mechanism of O I Lines in Herbig Ae/Be Stars}",
      journal = {\apj},
         year = 2018,
        month = apr,
       volume = {857},
       number = {1},
          eid = {30},
        pages = {30},
          doi = {10.3847/1538-4357/aab3d8},
archivePrefix = {arXiv},
       eprint = {1803.02025},
 primaryClass = {astro-ph.SR},
       adsurl = {https://ui.adsabs.harvard.edu/abs/2018ApJ...857...30M}
}

@ARTICLE{Strittmatter1977,
       author = {{Strittmatter}, P.~A. and {Woolf}, N.~J. and {Thompson}, R.~I. and {Wilkerson}, S. and {Angel}, J.~R.~P. and {Stockman}, H.~S. and {Gilbert}, G. and {Grandi}, S.~A. and {Larson}, H. and {Fink}, U.},
        title = "{The spectral development of Nova Cygni 1975.}",
      journal = {\apj},
         year = 1977,
        month = aug,
       volume = {216},
        pages = {23-32},
          doi = {10.1086/155438},
       adsurl = {https://ui.adsabs.harvard.edu/abs/1977ApJ...216...23S}
}

@ARTICLE{Zwick2025,
       author = {{Zwick}, Lorenz and {Tiede}, Christopher and {Mayer}, Lucio},
        title = "{Little Red Dots as Self-gravitating Disks Accreting on Supermassive Stars: Spectral Appearance and Formation Pathway of the Progenitors to Direct Collapse Black Holes}",
      journal = {\apj},
         year = 2026,
        month = may,
       volume = {1002},
       number = {1},
          eid = {7},
        pages = {7},
          doi = {10.3847/1538-4357/ae4d47},
archivePrefix = {arXiv},
       eprint = {2507.22014},
 primaryClass = {astro-ph.GA},
       adsurl = {https://ui.adsabs.harvard.edu/abs/2026ApJ..1002....7Z}
}

@ARTICLE{Hayashi1961,
       author = {{Hayashi}, Chushiro},
        title = "{Stellar evolution in early phases of gravitational contraction.}",
      journal = {\pasj},
         year = 1961,
        month = dec,
       volume = {13},
        pages = {450-452},
       adsurl = {https://ui.adsabs.harvard.edu/abs/1961PASJ...13..450H}
}

@ARTICLE{Wang2025,
       author = {{Wang}, Bingjie and {Leja}, Joel and {Katz}, Harley and {Inayoshi}, Kohei and {Cleri}, Nikko J. and {de Graaff}, Anna and {Hviding}, Raphael E. and {van Dokkum}, Pieter and {Greene}, Jenny E. and {Labb{\'e}}, Ivo and {Matthee}, Jorryt and {McConachie}, Ian and {Naidu}, Rohan P. and {Nelson}, Erica J.},
        title = "{The Missing Hard Photons of Little Red Dots: Their Incident Ionizing Spectra Resemble Massive Stars}",
      journal = {\apj},
         year = 2026,
        month = may,
       volume = {1003},
       number = {1},
          eid = {10},
        pages = {10},
          doi = {10.3847/1538-4357/ae5bab},
archivePrefix = {arXiv},
       eprint = {2508.18358},
 primaryClass = {astro-ph.GA},
       adsurl = {https://ui.adsabs.harvard.edu/abs/2026ApJ..1003...10W}
}

@ARTICLE{Nandal2025,
       author = {{Nandal}, Devesh and {Loeb}, Abraham},
        title = "{Supermassive Stars Match the Spectral Signatures of JWST's Little Red Dots}",
      journal = {\apj},
         year = 2026,
        month = feb,
       volume = {998},
       number = {1},
          eid = {124},
        pages = {124},
          doi = {10.3847/1538-4357/ae32f3},
archivePrefix = {arXiv},
       eprint = {2507.12618},
 primaryClass = {astro-ph.GA},
       adsurl = {https://ui.adsabs.harvard.edu/abs/2026ApJ...998..124N}
}

@dataset{brammer_djav44,
  author       = {Brammer, Gabriel and
                  Valentino, Francesco},
  title        = {The DAWN JWST Archive: Compilation of Public
                   NIRSpec Spectra
                  },
  month        = may,
  year         = 2025,
  publisher    = {Zenodo},
  version      = {4.4},
  doi          = {10.5281/zenodo.15472354},
  url          = {https://doi.org/10.5281/zenodo.15472354},
}

@ARTICLE{Pollock2025,
       author = {{Pollock}, Clara L. and {Gottumukkala}, Rashmi and {Heintz}, Kasper E. and {Brammer}, Gabriel B. and {Roberts-Borsani}, Guido and {Oesch}, Pascal A. and {Witstok}, Joris and {Arellano-C{\'o}rdova}, Karla Z. and {Cullen}, Fergus and {Scholte}, Dirk and {Terp}, Chamilla and {Rowland}, Lucie and {Sneppen}, Albert and {Ito}, Kei and {Valentino}, Francesco and {Matthee}, Jorryt and {Watson}, Darach and {Toft}, Sune},
        title = "{Novel z {\ensuremath{\sim}} 10 auroral line measurements extend the gradual offset of the fundamental metallicity relation deep into the first gigayear of cosmic time}",
      journal = {\aap},
         year = 2026,
        month = apr,
       volume = {708},
          eid = {A203},
        pages = {A203},
          doi = {10.1051/0004-6361/202556032},
archivePrefix = {arXiv},
       eprint = {2506.15779},
 primaryClass = {astro-ph.GA},
       adsurl = {https://ui.adsabs.harvard.edu/abs/2026A&A...708A.203P}
}

@ARTICLE{canucsdr1,
       author = {{Sarrouh}, Ghassan T.~E. and {Asada}, Yoshihisa and {Martis}, Nicholas S. and {Willott}, Chris J. and {Iyer}, Kartheik G. and {Noirot}, Ga{\"e}l and {Muzzin}, Adam and {Sawicki}, Marcin and {Brammer}, Gabriel and {Desprez}, Guillaume and {Rihtar{\v{s}}i{\v{c}}}, Gregor and {Zabl}, Johannes and {Abraham}, Roberto and {Brada{\v{c}}}, Maru{\v{s}}a and {Doyon}, Ren{\'e} and {Antwi-Danso}, Jacqueline and {Berek}, Samantha and {Brown}, Westley and {Estrada-Carpenter}, Vince and {Favaro}, Jeremy and {Felicioni}, Giordano and {Forrest}, Ben and {Gaspar}, Gaia and {Gould}, Katriona M.~L. and {Gledhill}, Rachel and {Harshan}, Anishya and {Jahan}, Nusrath and {Jagga}, Naadiyah and {Jude{\v{z}}}, Jon and {Marchesini}, Danilo and {Markov}, Vladan and {Matharu}, Jasleen and {MacFarland}, Shannon and {Merchant}, Maya and {M{\'e}rida}, Rosa M. and {Mowla}, Lamiya and {Myers}, Katherine and {Omori}, Kiyoaki C. and {Pacifici}, Camilla and {Ravindranath}, Swara and {Robbins}, Luke and {Strait}, Victoria and {Sok}, Visal and {Tan}, Vivian Yun Yan and {Tripodi}, Roberta and {Wilson}, Gillian and {Withers}, Sunna},
        title = "{CANUCS/Technicolor Data Release 1: Imaging, Photometry, Slit Spectroscopy, and Stellar Population Parameters}",
      journal = {\apjs},
         year = 2026,
        month = jan,
       volume = {282},
       number = {1},
          eid = {3},
        pages = {3},
          doi = {10.3847/1538-4365/ae1611},
archivePrefix = {arXiv},
       eprint = {2506.21685},
 primaryClass = {astro-ph.GA},
       adsurl = {https://ui.adsabs.harvard.edu/abs/2026ApJS..282....3S}
}

@ARTICLE{Torralba2025,
       author = {{Torralba}, Alberto and {Matthee}, Jorryt and {Pezzulli}, Gabriele and {Naidu}, Rohan P. and {Ishikawa}, Yuzo and {Brammer}, Gabriel B. and {Chang}, Seok-Jun and {Chisholm}, John and {de Graaff}, Anna and {D'Eugenio}, Francesco and {Di Cesare}, Claudia and {Eilers}, Anna-Christina and {Greene}, Jenny E. and {Gronke}, Max and {Iani}, Edoardo and {Kokorev}, Vasily and {Kotiwale}, Gauri and {Kramarenko}, Ivan and {Ma}, Yilun and {Mascia}, Sara and {Navarrete}, Benjam{\'\i}n and {Nelson}, Erica and {Oesch}, Pascal and {Simcoe}, Robert A. and {Wuyts}, Stijn},
        title = "{The warm outer layer of a little red dot as the source of [Fe II] and collisional Balmer lines with scattering wings}",
      journal = {\aap},
         year = 2026,
        month = feb,
       volume = {707},
          eid = {A75},
        pages = {A75},
          doi = {10.1051/0004-6361/202557537},
archivePrefix = {arXiv},
       eprint = {2510.00103},
 primaryClass = {astro-ph.GA},
       adsurl = {https://ui.adsabs.harvard.edu/abs/2026A&A...707A..75T}
}

@ARTICLE{Lambrides2025,
       author = {{Lambrides}, Erini and {Larson}, Rebecca and {Hutchison}, Taylor and {Arrabal Haro}, Pablo and {Wang}, Bingjie and {Welch}, Brian and {Kocevski}, Dale D. and {Richardson}, Chris T. and {Papovich}, Casey and {Trump}, Jonathan R. and {Bosman}, Sarah E.~I. and {Rigby}, Jane R. and {Finkelstein}, Steven L. and {Barro}, Guillermo and {Antwi-Danso}, Jacqueline and {Long}, Arianna and {Taylor}, Anthony J. and {Cann}, Jenna and {McKaig}, Jeffrey and {Koekemoer}, Anton M. and {Cleri}, Nikko J. and {Akins}, Hollis B. and {Bagley}, Mic B. and {Berg}, Danielle A. and {Bromm}, Volker and {Chisholm}, John and {Chworowsky}, Katherine and {Coffin}, Sadie and {Cooper}, M.~C. and {Cooper}, Olivia and {Cox}, Isa and {Dickinson}, Mark and {Ferguson}, Henry C. and {Franco}, Maximilien and {Gardner}, Jonathan P. and {Grogin}, Norman A. and {Hirschmann}, Michaela and {Huertas-Company}, Marc and {Jung}, Intae and {Kartaltepe}, Jeyhan S. and {Khullar}, Gourav P. and {Lucas}, Ray A. and {McGrath}, Elizabeth J. and {Morales}, Alexa M. and {Olivier}, Grace M. and {Ch{\'a}vez Ortiz}, {\'O}scar A. and {P{\'e}rez-Gonz{\'a}lez}, Pablo G. and {Pirzkal}, Norbert and {Somerville}, Rachel S. and {Vanderhoof}, Brittany and {Weiner}, Benjamin J. and {Yung}, L.~Y. Aaron and {Zavala}, Jorge A.},
        title = "{Discovery of Multiply Ionized Iron Emission Powered by an Active Galactic Nucleus in a z\raisebox{-0.5ex}\textasciitilde7 Little Red Dot}",
      journal = {arXiv e-prints},
         year = 2025,
        month = sep,
          eid = {arXiv:2509.09607},
        pages = {arXiv:2509.09607},
          doi = {10.48550/arXiv.2509.09607},
archivePrefix = {arXiv},
       eprint = {2509.09607},
 primaryClass = {astro-ph.GA},
       adsurl = {https://ui.adsabs.harvard.edu/abs/2025arXiv250909607L}
}

@ARTICLE{DEugenio2025:iron,
       author = {{D'Eugenio}, Francesco and {Nelson}, Erica and {Ji}, Xihan and {Baggen}, Josephine and {Greene}, Jenny and {Labb{\'e}}, Ivo and {Pezzulli}, Gabriele and {Brown}, Vanessa and {Maiolino}, Roberto and {Matthee}, Jorryt and {Terlevich}, Elena and {Terlevich}, Roberto and {Torralba}, Alberto and {Carniani}, Stefano},
        title = "{Irony at z=6.68: a bright AGN with forbidden Fe emission and multi-component Balmer absorption}",
      journal = {arXiv e-prints},
         year = 2025,
        month = sep,
          eid = {arXiv:2510.00101},
        pages = {arXiv:2510.00101},
          doi = {10.48550/arXiv.2510.00101},
archivePrefix = {arXiv},
       eprint = {2510.00101},
 primaryClass = {astro-ph.GA},
       adsurl = {https://ui.adsabs.harvard.edu/abs/2025arXiv251000101D}
}

@ARTICLE{Planck2020LCDM,
       author = {{Planck Collaboration} and {Aghanim}, N. and {Akrami}, Y. and {Ashdown}, M. and {Aumont}, J. and {Baccigalupi}, C. and {Ballardini}, M. and {Banday}, A.~J. and {Barreiro}, R.~B. and {Bartolo}, N. and {Basak}, S. and {Battye}, R. and {Benabed}, K. and {Bernard}, J.-P. and {Bersanelli}, M. and {Bielewicz}, P. and {Bock}, J.~J. and {Bond}, J.~R. and {Borrill}, J. and {Bouchet}, F.~R. and {Boulanger}, F. and {Bucher}, M. and {Burigana}, C. and {Butler}, R.~C. and {Calabrese}, E. and {Cardoso}, J.-F. and {Carron}, J. and {Challinor}, A. and {Chiang}, H.~C. and {Chluba}, J. and {Colombo}, L.~P.~L. and {Combet}, C. and {Contreras}, D. and {Crill}, B.~P. and {Cuttaia}, F. and {de Bernardis}, P. and {de Zotti}, G. and {Delabrouille}, J. and {Delouis}, J.-M. and {Di Valentino}, E. and {Diego}, J.~M. and {Dor{\'e}}, O. and {Douspis}, M. and {Ducout}, A. and {Dupac}, X. and {Dusini}, S. and {Efstathiou}, G. and {Elsner}, F. and {En{\ss}lin}, T.~A. and {Eriksen}, H.~K. and {Fantaye}, Y. and {Farhang}, M. and {Fergusson}, J. and {Fernandez-Cobos}, R. and {Finelli}, F. and {Forastieri}, F. and {Frailis}, M. and {Fraisse}, A.~A. and {Franceschi}, E. and {Frolov}, A. and {Galeotta}, S. and {Galli}, S. and {Ganga}, K. and {G{\'e}nova-Santos}, R.~T. and {Gerbino}, M. and {Ghosh}, T. and {Gonz{\'a}lez-Nuevo}, J. and {G{\'o}rski}, K.~M. and {Gratton}, S. and {Gruppuso}, A. and {Gudmundsson}, J.~E. and {Hamann}, J. and {Handley}, W. and {Hansen}, F.~K. and {Herranz}, D. and {Hildebrandt}, S.~R. and {Hivon}, E. and {Huang}, Z. and {Jaffe}, A.~H. and {Jones}, W.~C. and {Karakci}, A. and {Keih{\"a}nen}, E. and {Keskitalo}, R. and {Kiiveri}, K. and {Kim}, J. and {Kisner}, T.~S. and {Knox}, L. and {Krachmalnicoff}, N. and {Kunz}, M. and {Kurki-Suonio}, H. and {Lagache}, G. and {Lamarre}, J.-M. and {Lasenby}, A. and {Lattanzi}, M. and {Lawrence}, C.~R. and {Le Jeune}, M. and {Lemos}, P. and {Lesgourgues}, J. and {Levrier}, F. and {Lewis}, A. and {Liguori}, M. and {Lilje}, P.~B. and {Lilley}, M. and {Lindholm}, V. and {L{\'o}pez-Caniego}, M. and {Lubin}, P.~M. and {Ma}, Y.-Z. and {Mac{\'\i}as-P{\'e}rez}, J.~F. and {Maggio}, G. and {Maino}, D. and {Mandolesi}, N. and {Mangilli}, A. and {Marcos-Caballero}, A. and {Maris}, M. and {Martin}, P.~G. and {Martinelli}, M. and {Mart{\'\i}nez-Gonz{\'a}lez}, E. and {Matarrese}, S. and {Mauri}, N. and {McEwen}, J.~D. and {Meinhold}, P.~R. and {Melchiorri}, A. and {Mennella}, A. and {Migliaccio}, M. and {Millea}, M. and {Mitra}, S. and {Miville-Desch{\^e}nes}, M.-A. and {Molinari}, D. and {Montier}, L. and {Morgante}, G. and {Moss}, A. and {Natoli}, P. and {N{\o}rgaard-Nielsen}, H.~U. and {Pagano}, L. and {Paoletti}, D. and {Partridge}, B. and {Patanchon}, G. and {Peiris}, H.~V. and {Perrotta}, F. and {Pettorino}, V. and {Piacentini}, F. and {Polastri}, L. and {Polenta}, G. and {Puget}, J.-L. and {Rachen}, J.~P. and {Reinecke}, M. and {Remazeilles}, M. and {Renzi}, A. and {Rocha}, G. and {Rosset}, C. and {Roudier}, G. and {Rubi{\~n}o-Mart{\'\i}n}, J.~A. and {Ruiz-Granados}, B. and {Salvati}, L. and {Sandri}, M. and {Savelainen}, M. and {Scott}, D. and {Shellard}, E.~P.~S. and {Sirignano}, C. and {Sirri}, G. and {Spencer}, L.~D. and {Sunyaev}, R. and {Suur-Uski}, A.-S. and {Tauber}, J.~A. and {Tavagnacco}, D. and {Tenti}, M. and {Toffolatti}, L. and {Tomasi}, M. and {Trombetti}, T. and {Valenziano}, L. and {Valiviita}, J. and {Van Tent}, B. and {Vibert}, L. and {Vielva}, P. and {Villa}, F. and {Vittorio}, N. and {Wandelt}, B.~D. and {Wehus}, I.~K. and {White}, M. and {White}, S.~D.~M. and {Zacchei}, A. and {Zonca}, A.},
        title = "{Planck 2018 results. VI. Cosmological parameters}",
      journal = {\aap},
         year = 2020,
        month = sep,
       volume = {641},
          eid = {A6},
        pages = {A6},
          doi = {10.1051/0004-6361/201833910},
archivePrefix = {arXiv},
       eprint = {1807.06209},
 primaryClass = {astro-ph.CO},
       adsurl = {https://ui.adsabs.harvard.edu/abs/2020A&A...641A...6P}
}

@ARTICLE{Torralba2025a,
       author = {{Torralba}, Alberto and {Matthee}, Jorryt and {Pezzulli}, Gabriele and {Urrutia}, Tanya and {Gronke}, Max and {Mascia}, Sara and {D'Eugenio}, Francesco and {Di Cesare}, Claudia and {Eilers}, Anna-Christina and {Greene}, Jenny E. and {Iani}, Edoardo and {Ishikawa}, Yuzo and {Mackenzie}, Ruari and {Naidu}, Rohan P. and {Navarrete}, Benjam{\'\i}n and {Kotiwale}, Gauri},
        title = "{A weak Ly{\ensuremath{\alpha}} halo for an extremely bright little red dot: Indications of enshrouded supermassive black hole growth}",
      journal = {\aap},
         year = 2026,
        month = jan,
       volume = {705},
          eid = {A147},
        pages = {A147},
          doi = {10.1051/0004-6361/202555596},
archivePrefix = {arXiv},
       eprint = {2505.09542},
 primaryClass = {astro-ph.GA},
       adsurl = {https://ui.adsabs.harvard.edu/abs/2026A&A...705A.147T}
}

@ARTICLE{Kokorev2025,
       author = {{Kokorev}, Vasily and {Chisholm}, John and {Naidu}, Rohan P. and {Fujimoto}, Seiji and {Atek}, Hakim and {Brammer}, Gabriel and {Finkelstein}, Steven L. and {Akins}, Hollis B. and {Berg}, Danielle A. and {Furtak}, Lukas J. and {Fei}, Qinyue and {Hsiao}, Tiger Yu-Yang and {Labb{\'e}}, Ivo and {Matthee}, Jorryt and {Mu{\~n}oz}, Julian B. and {Oesch}, Pascal A. and {Pan}, Richard and {Rinaldi}, Pierluigi and {Saldana-Lopez}, Alberto and {Schaerer}, Daniel and {Volonteri}, Marta and {Zitrin}, Adi},
        title = "{The Deepest GLIMPSE of a Dense Gas Cocoon Enshrouding a Little Red Dot}",
      journal = {\apj},
         year = 2026,
        month = jun,
       volume = {1004},
       number = {2},
          eid = {153},
        pages = {153},
          doi = {10.3847/1538-4357/ae4ed7},
archivePrefix = {arXiv},
       eprint = {2511.07515},
 primaryClass = {astro-ph.GA},
       adsurl = {https://ui.adsabs.harvard.edu/abs/2026ApJ..1004..153K}
}

@ARTICLE{Rodriguez2002,
       author = {{Rodr{\'\i}guez-Ardila}, A. and {Viegas}, S.~M. and {Pastoriza}, M.~G. and {Prato}, L. and {Donzelli}, Carlos J.},
        title = "{The O I Line Emission in Active Galactic Nuclei Revisited}",
      journal = {\apj},
         year = 2002,
        month = jun,
       volume = {572},
       number = {1},
        pages = {94-104},
          doi = {10.1086/340192},
archivePrefix = {arXiv},
       eprint = {astro-ph/0202252},
 primaryClass = {astro-ph},
       adsurl = {https://ui.adsabs.harvard.edu/abs/2002ApJ...572...94R}
}

@BOOK{Kippenhahn1994,
       author = {{Kippenhahn}, Rudolf and {Weigert}, Alfred},
        title = "{Stellar Structure and Evolution}",
         year = 1994,
       adsurl = {https://ui.adsabs.harvard.edu/abs/1994sse..book.....K}
}

@ARTICLE{Zhang2025,
       author = {{Zhang}, Chenxuan and {Wu}, Qingwen and {Fan}, Xiao and {Ho}, Luis C. and {Wu}, Jiancheng and {Zhang}, Huanian and {Lyu}, Bing and {Cao}, Xinwu and {Wang}, Jianmin},
        title = "{The composite spectrum of little red dots from a standard inner disk and an unstable outer disk}",
      journal = {Nature Astronomy},
         year = 2026,
        month = may,
       volume = {10},
        pages = {753-761},
          doi = {10.1038/s41550-026-02785-x},
archivePrefix = {arXiv},
       eprint = {2505.12719},
 primaryClass = {astro-ph.HE},
       adsurl = {https://ui.adsabs.harvard.edu/abs/2026NatAs..10..753Z}
}

@ARTICLE{Nikopoulos2025,
       author = {{Nikopoulos}, G.~P. and {Watson}, D. and {Sneppen}, A. and {Rusakov}, V. and {Heintz}, K.~E. and {Witstok}, J. and {Brammer}, G.},
        title = "{Evidence of violation of case B recombination in little red dots}",
      journal = {\aap},
         year = 2026,
        month = jun,
       volume = {710},
          eid = {A136},
        pages = {A136},
          doi = {10.1051/0004-6361/202557604},
archivePrefix = {arXiv},
       eprint = {2510.06362},
 primaryClass = {astro-ph.GA},
       adsurl = {https://ui.adsabs.harvard.edu/abs/2026A&A...710A.136N}
}

@ARTICLE{Juodzbalis2025,
       author = {{Juod{\v{z}}balis}, Ignas and {Maiolino}, Roberto and {Baker}, William M. and {Lake}, Emma Curtis and {Scholtz}, Jan and {D'Eugenio}, Francesco and {Trefoloni}, Bartolomeo and {Isobe}, Yuki and {Tacchella}, Sandro and {Bunker}, Andrew J. and {Carniani}, Stefano and {Charlot}, St{\'e}phane and {Jones}, Gareth C. and {Parlanti}, Eleonora and {Perna}, Michele and {Rinaldi}, Pierluigi and {Robertson}, Brant and {{\"U}bler}, Hannah and {Venturi}, Giacomo and {Willott}, Chris},
        title = "{JADES: comprehensive census of broad-line AGN from reionization to cosmic noon revealed by JWST}",
      journal = {\mnras},
         year = 2026,
        month = mar,
       volume = {546},
       number = {3},
          eid = {stag086},
        pages = {stag086},
          doi = {10.1093/mnras/stag086},
archivePrefix = {arXiv},
       eprint = {2504.03551},
 primaryClass = {astro-ph.GA},
       adsurl = {https://ui.adsabs.harvard.edu/abs/2026MNRAS.546ag086J}
}

@ARTICLE{Casey2024,
       author = {{Casey}, Caitlin M. and {Akins}, Hollis B. and {Kokorev}, Vasily and {McKinney}, Jed and {Cooper}, Olivia R. and {Long}, Arianna S. and {Franco}, Maximilien and {Manning}, Sinclaire M.},
        title = "{Dust in Little Red Dots}",
      journal = {\apjl},
         year = 2024,
        month = nov,
       volume = {975},
       number = {1},
          eid = {L4},
        pages = {L4},
          doi = {10.3847/2041-8213/ad7ba7},
archivePrefix = {arXiv},
       eprint = {2407.05094},
 primaryClass = {astro-ph.GA},
       adsurl = {https://ui.adsabs.harvard.edu/abs/2024ApJ...975L...4C}
}

@ARTICLE{Schindler2025,
       author = {{Schindler}, Jan-Torge and {Hennawi}, Joseph F. and {Davies}, Frederick B. and {Bosman}, Sarah E.~I. and {Endsley}, Ryan and {Wang}, Feige and {Yang}, Jinyi and {Barth}, Aaron J. and {Eilers}, Anna-Christina and {Fan}, Xiaohui and {Kakiichi}, Koki and {Maseda}, Michael and {Pizzati}, Elia and {Nanni}, Riccardo},
        title = "{A little red dot at z = 7.3 within a large galaxy overdensity}",
      journal = {Nature Astronomy},
         year = 2025,
        month = sep,
          doi = {10.1038/s41550-025-02660-1},
archivePrefix = {arXiv},
       eprint = {2411.11534},
 primaryClass = {astro-ph.GA},
       adsurl = {https://ui.adsabs.harvard.edu/abs/2025NatAs.tmp..191S}
}

@ARTICLE{Maiolino2025,
       author = {{Maiolino}, Roberto and {{\"U}bler}, Hannah and {D'Eugenio}, Francesco and {Scholtz}, Jan and {Juod{\v{z}}balis}, Ignas and {Ji}, Xihan and {Perna}, Michele and {Bromm}, Volker and {Dayal}, Pratika and {Koudmani}, Sophie and {Liu}, Boyuan and {Schneider}, Raffaella and {Sijacki}, Debora and {Valiante}, Rosa and {Trinca}, Alessandro and {Zhang}, Saiyang and {Volonteri}, Marta and {Inayoshi}, Kohei and {Carniani}, Stefano and {Nakajima}, Kimihiko and {Isobe}, Yuki and {Witstok}, Joris and {Jones}, Gareth C. and {Tacchella}, Sandro and {Arribas}, Santiago and {Bunker}, Andrew and {Cataldi}, Elisa and {Charlot}, Stephane and {Curti}, Giovanni Cresci Mirko and {Fabian}, Andrew C. and {Katz}, Harley and {Kumari}, Nimisha and {Laporte}, Nicolas and {Mazzolari}, Giovanni and {Robertson}, Brant and {Sun}, Fengwu and {Rodriguez Del Pino}, Bruno and {Venturi}, Giacomo},
        title = "{A black hole in a near pristine galaxy 700 Myr after the big bang}",
      journal = {\mnras},
         year = 2026,
        month = may,
       volume = {548},
       number = {1},
          eid = {staf2109},
        pages = {staf2109},
          doi = {10.1093/mnras/staf2109},
archivePrefix = {arXiv},
       eprint = {2505.22567},
 primaryClass = {astro-ph.GA},
       adsurl = {https://ui.adsabs.harvard.edu/abs/2026MNRAS.548f2109M}
}

@ARTICLE{Scholtz2025,
       author = {{Scholtz}, J. and {Carniani}, S. and {Parlanti}, E. and {D'Eugenio}, F. and {Curtis-Lake}, E. and {Jakobsen}, P. and {Bunker}, A.~J. and {Cameron}, A.~J. and {Arribas}, S. and {Baker}, W.~M. and {Charlot}, S. and {Chevellard}, J. and {Circosta}, C. and {Curti}, M. and {Duan}, Q. and {Eisenstein}, D.~J. and {Hainline}, K. and {Ji}, Z. and {Johnson}, B.~D. and {Jones}, G.~C. and {Kumari}, N. and {Maiolino}, R. and {Maseda}, M.~V. and {Perna}, M. and {P{\'e}rez-Gonz{\'a}lez}, P.~G. and {Rawle}, T. and {Rieke}, M. and {Rinaldi}, P. and {Robertson}, B. and {Saxena}, A. and {Shivaei}, I. and {Silcock}, M.~S. and {Sun}, Y. and {Rodr{\'\i}guez Del Pino}, B. and {Tacchella}, S. and {{\"U}bler}, H. and {Venturi}, G. and {Williams}, C.~C. and {Willmer}, C.~N.~A. and {Willott}, C. and {Witstok}, J.},
        title = "{JADES Data Release 4 ─ Paper II. Data reduction, analysis, and emission-line fluxes of the complete spectroscopic sample}",
      journal = {\mnras},
         year = 2026,
        month = jul,
       volume = {549},
       number = {4},
          eid = {stag939},
        pages = {stag939},
          doi = {10.1093/mnras/stag939},
archivePrefix = {arXiv},
       eprint = {2510.01034},
 primaryClass = {astro-ph.GA},
       adsurl = {https://ui.adsabs.harvard.edu/abs/2026MNRAS.549ag939S}
}

@ARTICLE{CurtisLake2025,
       author = {{Curtis-Lake}, Emma and {Cameron}, Alex J. and {Bunker}, Andrew J. and {Scholtz}, Jan and {Carniani}, Stefano and {Parlanti}, Eleonora and {D'Eugenio}, Francesco and {Jakobsen}, Peter and {Willmer}, Christopher N.~A. and {Arribas}, Santiago and {Baker}, William M. and {Charlot}, St{\'e}phane and {Chevallard}, Jacopo and {Circosta}, Chiara and {Curti}, Mirko and {Duan}, Qiao and {Eisenstein}, Daniel J. and {Hainline}, Kevin and {Ji}, Zhiyuan and {Johnson}, Benjamin D. and {Jones}, Gareth C. and {Maiolino}, Roberto and {Maseda}, Michael V. and {Perna}, Michele and {P{\'e}rez-Gonz{\'a}lez}, Pablo G. and {Rawle}, Tim and {Rieke}, Marcia and {Rinaldi}, Pierluigi and {Robertson}, Brant and {Rodr{\'\i}guez Del Pino}, Bruno and {Saxena}, Aayush and {Shivaei}, Irene and {Smit}, Renske and {Tacchella}, Sandro and {{\"U}bler}, Hannah and {Venturi}, Giacomo and {Williams}, Christina C. and {Willott}, Chris},
        title = "{JADES data release 4 ─ Paper I. Sample selection, observing strategy and redshifts of the complete spectroscopic sample}",
      journal = {\mnras},
         year = 2026,
        month = jul,
       volume = {549},
       number = {4},
          eid = {stag836},
        pages = {stag836},
          doi = {10.1093/mnras/stag836},
archivePrefix = {arXiv},
       eprint = {2510.01033},
 primaryClass = {astro-ph.GA},
       adsurl = {https://ui.adsabs.harvard.edu/abs/2026MNRAS.549ag836C}
}

@ARTICLE{Naidu2025z14,
       author = {{Naidu}, Rohan P. and {Oesch}, Pascal A. and {Brammer}, Gabriel and {Weibel}, Andrea and {Li}, Yijia and {Matthee}, Jorryt and {Chisolm}, John and {Pollock}, Clara L. and {Heintz}, Kasper E. and {Johnson}, Benjamin D. and {Shen}, Xuejian and {Hviding}, Raphael E. and {Leja}, Joel and {Tacchella}, Sandro and {Ganguly}, Arpita and {Witten}, Callum and {Atek}, Hakim and {Belli}, Siro and {Bose}, Sownak and {Bouwens}, Rychard and {Dayal}, Pratika and {Decarli}, Roberto and {de Graaff}, Anna and {Fudamoto}, Yoshinobu and {Giovinazzo}, Emma and {Greene}, Jenny E. and {Illingworth}, Garth and {Inoue}, Akio K. and {Kane}, Sarah G. and {Labbe}, Ivo and {Leonova}, Ecaterina and {Marques-Chaves}, Rui and {Meyer}, Roman A. and {Nelson}, Erica J. and {Roberts-Borsani}, Guido and {Schaerer}, Daniel and {Simcoe}, Robert A. and {Stefanon}, Mauro and {Sugahara}, Yuma and {Toft}, Sune and {van der Wel}, Arjen and {van Dokkum}, Pieter and {Walter}, Fabian and {Watson}, Darrach and {Weaver}, John R. and {Whitaker}, Katherine E.},
        title = "{A Cosmic Miracle: A Remarkably Luminous Galaxy at zspec = 14.44 Confirmed with JWST}",
      journal = {The Open Journal of Astrophysics},
         year = 2026,
        month = jan,
       volume = {9},
        pages = {56033},
          doi = {10.33232/001c.156033},
archivePrefix = {arXiv},
       eprint = {2505.11263},
 primaryClass = {astro-ph.GA},
       adsurl = {https://ui.adsabs.harvard.edu/abs/2026OJAp....956033N}
}

@ARTICLE{Tal2014,
       author = {{Alexander}, Tal and {Natarajan}, Priyamvada},
        title = "{Rapid growth of seed black holes in the early universe by supra-exponential accretion}",
      journal = {Science},
         year = 2014,
        month = sep,
       volume = {345},
       number = {6202},
        pages = {1330-1333},
          doi = {10.1126/science.1251053},
archivePrefix = {arXiv},
       eprint = {1408.1718},
 primaryClass = {astro-ph.GA},
       adsurl = {https://ui.adsabs.harvard.edu/abs/2014Sci...345.1330A}
}

@ARTICLE{Glikman2006,
       author = {{Glikman}, Eilat and {Helfand}, David J. and {White}, Richard L.},
        title = "{A Near-Infrared Spectral Template for Quasars}",
      journal = {\apj},
         year = 2006,
        month = apr,
       volume = {640},
       number = {2},
        pages = {579-591},
          doi = {10.1086/500098},
archivePrefix = {arXiv},
       eprint = {astro-ph/0511640},
 primaryClass = {astro-ph},
       adsurl = {https://ui.adsabs.harvard.edu/abs/2006ApJ...640..579G}
}

@ARTICLE{VandenBerk2001,
       author = {{Vanden Berk}, Daniel E. and {Richards}, Gordon T. and {Bauer}, Amanda and {Strauss}, Michael A. and {Schneider}, Donald P. and {Heckman}, Timothy M. and {York}, Donald G. and {Hall}, Patrick B. and {Fan}, Xiaohui and {Knapp}, G.~R. and {Anderson}, Scott F. and {Annis}, James and {Bahcall}, Neta A. and {Bernardi}, Mariangela and {Briggs}, John W. and {Brinkmann}, J. and {Brunner}, Robert and {Burles}, Scott and {Carey}, Larry and {Castander}, Francisco J. and {Connolly}, A.~J. and {Crocker}, J.~H. and {Csabai}, Istv{\'a}n and {Doi}, Mamoru and {Finkbeiner}, Douglas and {Friedman}, Scott and {Frieman}, Joshua A. and {Fukugita}, Masataka and {Gunn}, James E. and {Hennessy}, G.~S. and {Ivezi{\'c}}, {\v{Z}}eljko and {Kent}, Stephen and {Kunszt}, Peter Z. and {Lamb}, D.~Q. and {Leger}, R. French and {Long}, Daniel C. and {Loveday}, Jon and {Lupton}, Robert H. and {Meiksin}, Avery and {Merelli}, Aronne and {Munn}, Jeffrey A. and {Newberg}, Heidi Jo and {Newcomb}, Matt and {Nichol}, R.~C. and {Owen}, Russell and {Pier}, Jeffrey R. and {Pope}, Adrian and {Rockosi}, Constance M. and {Schlegel}, David J. and {Siegmund}, Walter A. and {Smee}, Stephen and {Snir}, Yehuda and {Stoughton}, Chris and {Stubbs}, Christopher and {SubbaRao}, Mark and {Szalay}, Alexander S. and {Szokoly}, Gyula P. and {Tremonti}, Christy and {Uomoto}, Alan and {Waddell}, Patrick and {Yanny}, Brian and {Zheng}, Wei},
        title = "{Composite Quasar Spectra from the Sloan Digital Sky Survey}",
      journal = {\aj},
         year = 2001,
        month = aug,
       volume = {122},
       number = {2},
        pages = {549-564},
          doi = {10.1086/321167},
archivePrefix = {arXiv},
       eprint = {astro-ph/0105231},
 primaryClass = {astro-ph},
       adsurl = {https://ui.adsabs.harvard.edu/abs/2001AJ....122..549V}
}

@ARTICLE{Chen2026,
       author = {{Chen}, Yi-Xian and {Liu}, Hanpu and {Li}, Ruancun and {Wang}, Bingjie and {Ma}, Yilun and {Jiang}, Yan-Fei and {Greene}, Jenny E. and {Quataert}, Eliot and {Goodman}, Jeremy},
        title = "{Spectral Appearance of Self-gravitating Disks Powered by Stellar Objects: Universal Effective Temperature in the Optical Continuum and Application to Little Red Dots}",
      journal = {\apjl},
         year = 2026,
        month = jun,
       volume = {1003},
       number = {2},
          eid = {L46},
        pages = {L46},
          doi = {10.3847/2041-8213/ae6b69},
archivePrefix = {arXiv},
       eprint = {2602.06954},
 primaryClass = {astro-ph.HE},
       adsurl = {https://ui.adsabs.harvard.edu/abs/2026ApJ..1003L..46C}
}

@ARTICLE{Baggen2026,
       author = {{Baggen}, Josephine F.~W. and {Scoggins}, Matthew T. and {van Dokkum}, Pieter and {Haiman}, Zolt{\'a}n and {Torralba}, Alberto and {Matthee}, Jorryt},
        title = "{Connecting the Dots: UV-bright Companions of Little Red Dots as Lyman─Werner Sources Enabling Direct-collapse Black Hole Formation}",
      journal = {\apjl},
         year = 2026,
        month = may,
       volume = {1002},
       number = {1},
          eid = {L4},
        pages = {L4},
          doi = {10.3847/2041-8213/ae58a5},
archivePrefix = {arXiv},
       eprint = {2602.02702},
 primaryClass = {astro-ph.GA},
       adsurl = {https://ui.adsabs.harvard.edu/abs/2026ApJ..1002L...4B}
}

@ARTICLE{Cloonan2026,
       author = {{Cloonan}, Aidan P. and {Whitaker}, Katherine E. and {Manning}, Sinclaire M. and {Williams}, Christina C. and {Greene}, Jenny E. and {Oesch}, Pascal A. and {Weibel}, Andrea and {Brammer}, Gabriel and {de Graaff}, Anna and {Hviding}, Raphael E. and {Dayal}, Pratika and {Jespersen}, Christian Kragh and {Ji}, Zhiyuan and {Labbe}, Ivo and {Xiao}, Mengyuan and {Zhang}, Yunchong},
        title = "{A PANORAMIC of UV-optical morphologies of ``Little Red Dots'': Two groups of LRDs distinguished by UV half-light radius}",
      journal = {arXiv e-prints},
         year = 2026,
        month = mar,
          eid = {arXiv:2603.24700},
        pages = {arXiv:2603.24700},
          doi = {10.48550/arXiv.2603.24700},
archivePrefix = {arXiv},
       eprint = {2603.24700},
 primaryClass = {astro-ph.GA},
       adsurl = {https://ui.adsabs.harvard.edu/abs/2026arXiv260324700C}
}

\section*{Affiliations}
\noindent
{\it $^{1}$Max-Planck-Institut f\"ur Astronomie, K\"onigstuhl 17, D-69117, Heidelberg, Germany \\
$^{2}$Center for Astrophysics $|$ Harvard \& Smithsonian, 60 Garden St., Cambridge MA 02138 USA \\
$^{3}$MIT Kavli Institute for Astrophysics and Space Research, 70 Vassar Street, Cambridge, MA 02139, USA \\
$^{4}$Department of Astrophysical Sciences, Princeton University, 4 Ivy Lane, Princeton, NJ 08544, USA      \\
$^{5}$Center for Interdisciplinary Exploration and Research in Astrophysics (CIERA), Northwestern University, 1800 Sherman Ave, Evanston, IL 60201, USA      \\
$^{6}$Department of Astronomy \& Astrophysics, The Pennsylvania State University, University Park, PA 16802, USA      \\
$^{7}$Institute for Computational \& Data Sciences, The Pennsylvania State University, University Park, PA 16802, USA      \\
$^{8}$Institute for Gravitation and the Cosmos, The Pennsylvania State University, University Park, PA 16802, USA      \\
$^{9}$Institute of Science and Technology Austria (ISTA), Am Campus 1, 3400 Klosterneuburg, Austria      \\
$^{10}$Cosmic Dawn Center (DAWN), Niels Bohr Institute, University of Copenhagen, Jagtvej 128, K{\o}benhavn N, DK-2200, Denmark      \\
$^{11}$Department of Astronomy \& Astrophysics, University of Chicago, 5640 S. Ellis Avenue, Chicago, IL 60637, USA      \\
$^{12}$Kavli Institute for Cosmological Physics, University of Chicago, Chicago IL 60637, USA      \\
$^{13}$Department of Physics and Astronomy and PITT PACC, University of Pittsburgh, Pittsburgh, PA 15260, USA      \\
$^{14}$Leiden Observatory, Leiden University, PO Box 9513, NL-2300 RA Leiden, The Netherlands      \\
$^{15}$Institute for Computational Cosmology, Department of Physics, Durham University, South Road, Durham DH1 3LE, UK      \\
$^{16}$Department of Astronomy, The University of Texas at Austin, 2515 Speedway, Stop C1400, Austin, TX 78712, USA      \\
$^{17}$Cosmic Frontier Center, The University of Texas at Austin, Austin, TX 78712, USA      \\
$^{18}$Canadian Institute for Theoretical Astrophysics, 60 St George St, University of Toronto, Toronto, ON M5S 3H8, Canada      \\
$^{19}$David A. Dunlap Department of Astronomy and Astrophysics, University of Toronto, 50 St George St, Toronto ON M5S 3H4, Canada      \\
$^{20}$Department of Physics, 60 St George St, University of Toronto, Toronto, ON M5S 3H8, Canada      \\
$^{21}$Department of Astrophysics, University of Zurich, Zurich CH-8057, Switzerland      \\
$^{22}$Center for Frontier Science, Chiba University, 1-33 Yayoi-cho, Inage-ku, Chiba 263-8522, Japan      \\
$^{23}$Dunlap Institute for Astronomy and Astrophysics, 50 St. George Street, Toronto, Ontario, M5S 3H4, Canada      \\
$^{24}$Centre for Astrophysics and Supercomputing, Swinburne University of Technology, Melbourne, VIC 3122, Australia      \\
$^{25}$Department of Astronomy, University of Geneva, Chemin Pegasi 51, 1290 Versoix, Switzerland      \\
$^{26}$Department of Astronomy, University of Wisconsin-Madison, 475 N. Charter St., Madison, WI 53706 USA      \\
$^{27}$Department for Astrophysical and Planetary Science, University of Colorado, Boulder, CO 80309, USA      \\
$^{28}$Astronomical Observatory Institute, Faculty of Physics and Astronomy, Adam Mickiewicz University, ul. S{\l}oneczna 36, 60-286, Pozna{\'n}, Poland      \\
$^{29}$California Institute of Technology, Pasadena, CA 91125, USA      \\
$^{30}$Sterrenkundig Observatorium, Universiteit Gent, Krijgslaan 281 S9, 9000 Gent, Belgium      \\
}


\appendix

\section{{Slit losses}}\label{apdx:slitloss}

\begin{figure}
    \centering
    \includegraphics[width=\linewidth]{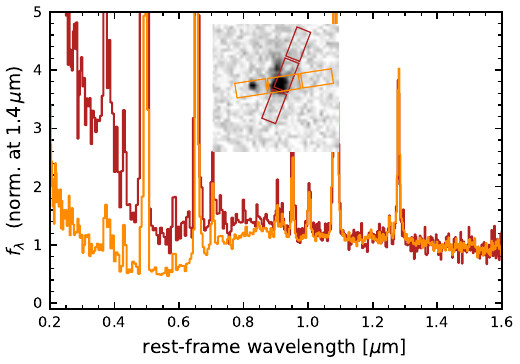}
    \caption{{Duplicate spectra of NEXUS-10687 reveal that the precise slit placement can have a significant effect on the observed spectral shape due to the presence of extended blue emission in the northern vicinity of the LRD.}}
    \label{fig:nexus}
\end{figure}

{As described in Section~\ref{sec:spec_data}, the spectroscopic reduction pipeline performs a morphology-dependent and wavelength-dependent correction for the slit losses based on empirical modelling and calibration. However, this pipeline cannot account for colour gradients within a spatially extended source if part of the source fall outside the slit. That is, if a LRD has extended UV emission (as is commonly observed among LRDs, e.g. \citealt{Cloonan2026}), the slit placement can have a significant effect on the spatially-integrated spectrum. We show an example of the most extreme case found in our sample of LRDs with duplicate spectra in Figure~\ref{fig:nexus}. One spectrum is well centred on the LRD engine itself (orange), whereas the other includes flux from a nearby blue clump north of the LRD (red). The inclusion of this UV flux leads to a substantial change in the (flux-normalised) spectral shape. In turn, this affects the precise inferred value of, for instance, $\betaBB$ and $\lpeak$. This indicates that individual LRDs need to be interpreted with caution, although our broader analysis of LRDs with duplicate spectra indicates that population aggregate results (the primary focus of this paper) are robust. }

\section{Data tables}\label{sec:apdx_tables}

\begin{table*}
\caption{Program information for the LRD spectra used in this work, including duplicate spectra and multiple images of strongly lensed LRDs ({181 spectra of 146} unique objects). \label{tab:surveys}}
\begin{tabular}{l|c}
{Program} & {\# LRD spectra}  \\
\hline
CANUCS \citep[GTO-1208;][]{canucsdr1} & 5 \\
CAPERS (GO-6368; PI: Dickinson) & {20} \\
CEERS \citep[GO-1345;][]{Finkelstein2025} & 2 \\
DDT-2750 (PI: Arrabal Haro) & 2 \\
DDT-2767 (PI: Kelly) & 1 \\
DDT-6585 (PI: Coulter) & 3 \\
GO-1433 (PI: Coe) & 1 \\
{GO-2198 (PI: Barrufet; \citealt{Barrufet2025})} & 3 \\
GO-4106 (PIs: Nelson \& Labbe) & {9} \\ 
{GO-5545 (PI: Barrufet)} & {7} \\
{GO-5997 (PI: Looser)} & {1} \\
{GO-8018 (PI: Lin)} & {8} \\
{GO-8060 (PI: Egami)} & {3} \\
GTO-WIDE \citep[GTO-1212, GTO-1213, GTO-1215;][]{Maseda2024} & 5  \\
{JADES \citep[GTO-1180, GTO-1181, GTO-1286; GTO-1287][]{CurtisLake2025,Scholtz2025}} & {16} \\
MoM (GO-5224; PIs: Oesch \& Naidu) & 13 \\
NEXUS \citep[GO-5105;][]{Shen2024} & {27} \\
RUBIES \citep[GO-4233;][]{deGraaff2024d} & {38} \\
UNCOVER \citep[GO-2561][]{Bezanson2024,Price2025} & {17} \\
\end{tabular}
\end{table*}

\begin{table*}
\caption{Description of data table released together with this work. \label{tab:data_release}}
\begin{tabular}{l|l|l}
{Column name} & {Unit}  & {Description} \\
\hline
pid & & JWST program ID \\ 
srcid & & MSA ID number \\ 
root & & DJA root name\\
file && DJA filename \\ 
ra & deg & Right ascension \\ 
dec& deg & Declination \\ 
zspec&& Spectroscopic redshift \\ 
f444w\_aper0.1 & nJy & NIRCam/F444W circular aperture flux (radius $0.1\arcsec$)  \\ 
mu & & Lensing magnification factor (obtained from \citealt{Furtak2023:lensing} and \citealt{canucsdr1}) \\ 
Ndup & & Number of duplicate PRISM spectra in LRD table \\ 
dup\_filenames&& DJA filenames of duplicate spectra \\ 
use\_dG26 &&  Flag to filter the 146 unique sources (with their highest S/N PRISM spectra) used in this work \\
lambda\_v & $\micron$ & Inflection wavelength of v-shape ($\lambda_{\rm v}$, [5, 16, 50, 84, 95] percentiles) \\ 
beta\_UV && UV slope (measured from PRISM spectrum; [5, 16, 50, 84, 95] percentiles)\\ 
M\_UV & mag & Absolute magnitude at 1500\,\AA\xspace (measured from PRISM spectrum, [5, 16, 50, 84, 95] percentiles)\\ 
break\_strength && Balmer break strength ([5, 16, 50, 84, 95] percentiles) \\ 
beta\_MBB && Power-law slope $\betaBB$ of modified blackbody ([5, 16, 50, 84, 95] percentiles)\\  
Teff & K & Effective temperature of modified blackbody ([5, 16, 50, 84, 95] percentiles)\\ 
peak\_wave & $\micron$ & Peak wavelength $\lpeak$ of modified blackbody ([5, 16, 50, 84, 95] percentiles)\\ 
logL\_MBB & $\log \ergs$ & Integrated luminosity of modified blackbody ($L_{\rm blackbody}$; [5, 16, 50, 84, 95] percentiles) \\ 
logL\_5100 & $\log \ergs$ & Optical luminosity $\lambda L_\lambda$ at $\lambda=5100\,$\AA\xspace ([5, 16, 50, 84, 95] percentiles) \\ 
L5100\_LMBB && Ratio of $L_{5100}$ to $L_{\rm blackbody}$ ([5, 16, 50, 84, 95] percentiles) \\ 
Ha\_total\_ew & \AA\xspace (rest) & Total (broad+narrow) \Ha EW  ([5, 16, 50, 84, 95] percentiles) \\ 
Hb\_total\_ew & \AA\xspace (rest) & Total (broad+narrow) \Hb EW  ([5, 16, 50, 84, 95] percentiles)  \\ 
Balmer\_dec\_total & & Total (broad+narrow) Balmer decrement $\rm H\alpha/H\beta$  ([5, 16, 50, 84, 95] percentiles)  \\ 
LHa\_total & $\ergs$ & Total (broad+narrow) \Ha luminosity ([5, 16, 50, 84, 95] percentiles) \\ 
logLHa\_total & $\log\ergs$ & Total (broad+narrow) \Ha luminosity ([5, 16, 50, 84, 95] percentiles) \\ 
OIII\_5007\_ew &  \AA\xspace (rest) & \Oiii$_{\lambda 5007}$  EW ([5, 16, 50, 84, 95] percentiles) \\ 
LOIII\_5007 &  $\ergs$ & \Oiii$_{\lambda 5007}$ luminosity ([5, 16, 50, 84, 95] percentiles) \\ 
logLOIII\_5007 & $\log\ergs$& \Oiii$_{\lambda 5007}$  luminosity ([5, 16, 50, 84, 95] percentiles) \\ 
LOI\_8446 &  $\ergs$ & \Oi$_{\lambda 8446}$ luminosity ([5, 16, 50, 84, 95] percentiles) \\ 
logLOI\_8446& $\log\ergs$& \Oi$_{\lambda 8446}$ luminosity ([5, 16, 50, 84, 95] percentiles)  \\ 
\end{tabular}
\end{table*}

\clearpage
\section{Supplementary figures of the modified blackbody models}\label{apdx:Teff}

\begin{figure}
    \centering
    \includegraphics[width=\linewidth]{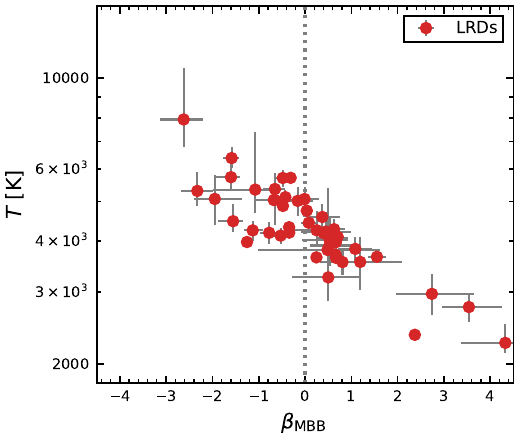}
    \caption{Distribution of the power law slopes and temperatures from modified blackbody fits to the LRD sample restricted to $z<4.5$ (red). For reference, the results of fits to the complete RUBIES sample at $2<z<5$ and $\rm F444W<26$ are shown in grey. A slope of $\beta_{\rm MBB}=0$ corresponds to a perfect single-temperature blackbody. LRDs are well approximated by single-temperature blackbodies of characteristic temperature $T\sim5000\,$K, although the precise inferred temperature is strongly anti-correlated with $\betaBB$, and therefore more sensitive to the details of the data reduction and fitting than the measured $\lpeak$ of Figure~\ref{fig:BB_beta}.  }
    \label{fig:beta_Teff_corr}
\end{figure}

\begin{figure}
    \centering
    \includegraphics[width=\linewidth]{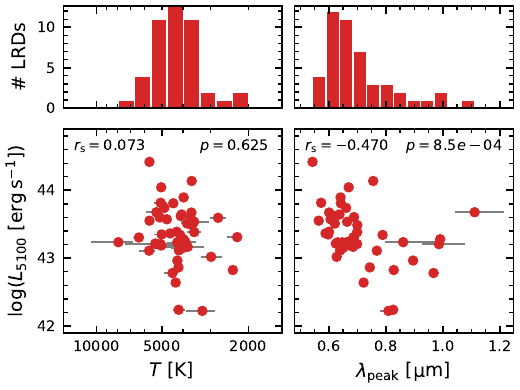}
    \caption{HR diagram, but with $L_{5100}$ rather than the integrated luminosity ($L_{\rm blackbody}$, Fig.~\ref{fig:HR}). Because the shape of the SED differs systematically for sources that peak at longer wavelengths, a significant correlation arises between $L_{5100}$ and $\lpeak$ that is not apparent for $L_{\rm blackbody}$ in Fig.~\ref{fig:HR}. }
    \label{fig:HR5100}
\end{figure}

\begin{figure*}
    \centering
    \includegraphics[width=0.89\linewidth]{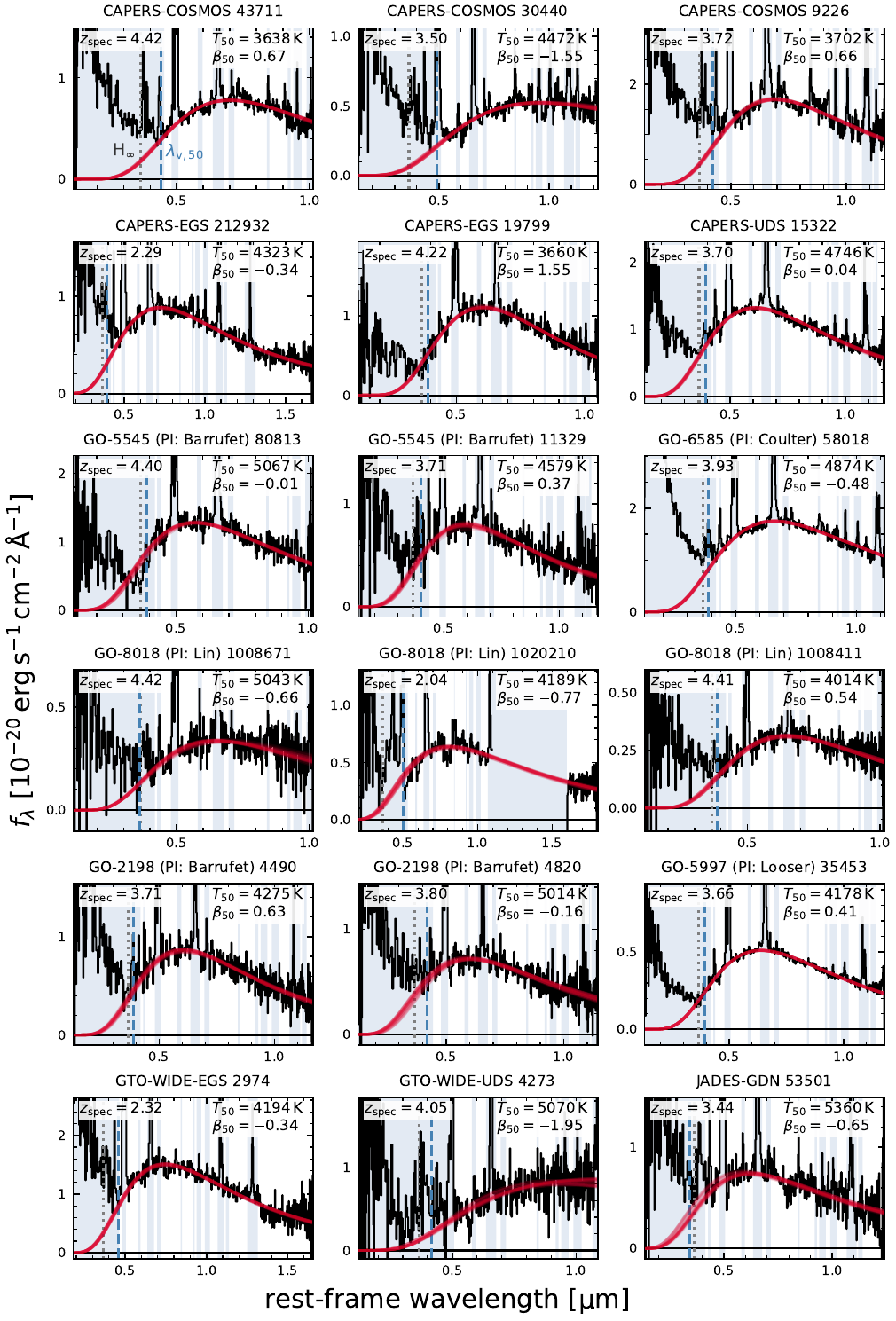}
    \caption{Modified blackbody fits (red lines, 100 random draws from posterior) to the $z<4.5$ LRD spectra (see Section~\ref{sec:selection}). Shaded regions indicate masked areas, and blue dashed lines the posterior median of the v-shape wavelength, $\lambda_{\rm v}$.  }
    \label{fig:all_BBs}
\end{figure*}

\begin{figure*}
    \centering
    \includegraphics[width=0.89\linewidth]{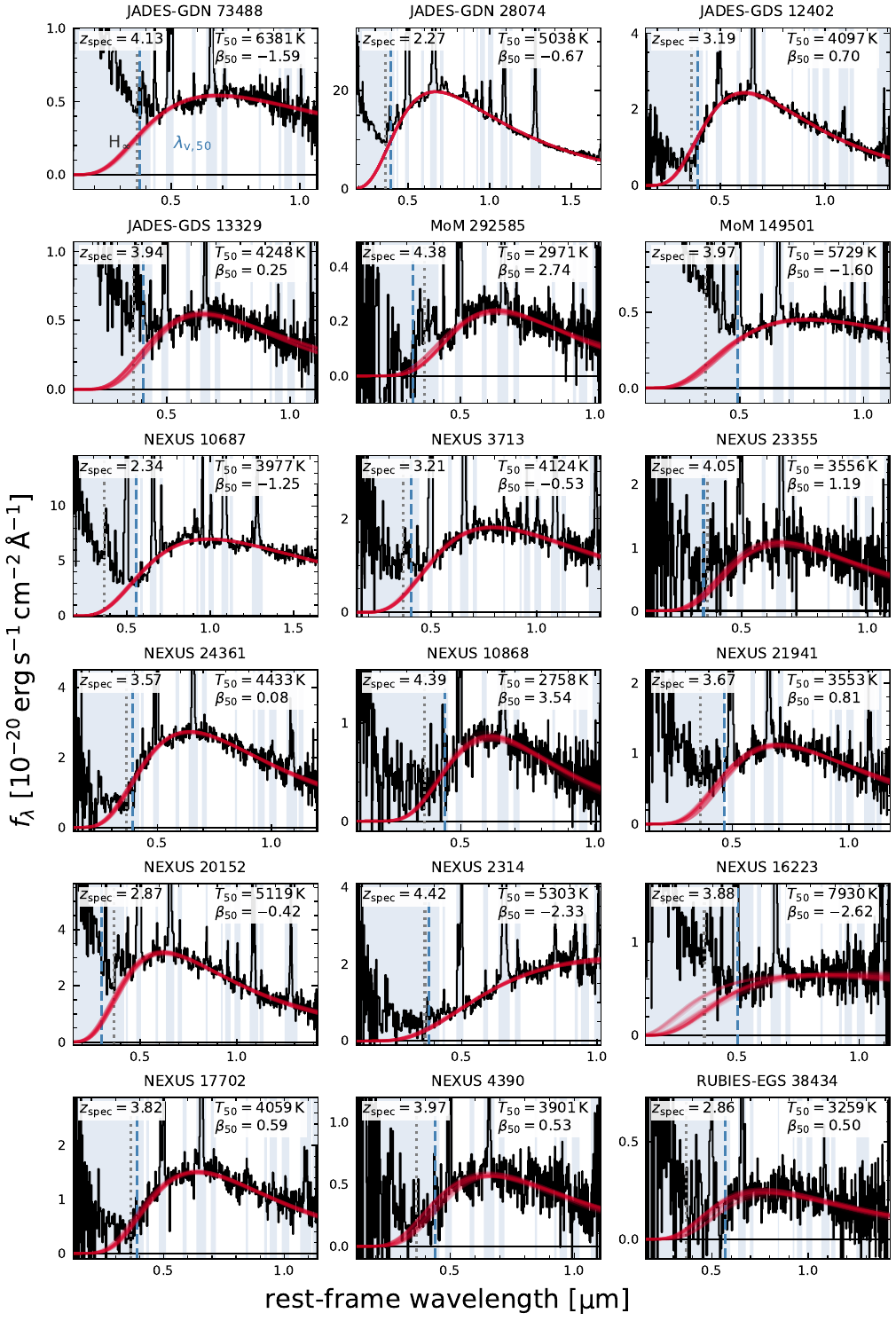}
    \caption{As Figure~\ref{fig:all_BBs}, showing {the second part} of the $z<4.5$ LRD sample.}
\end{figure*}

\begin{figure*}
    \centering
    \includegraphics[width=0.89\linewidth]{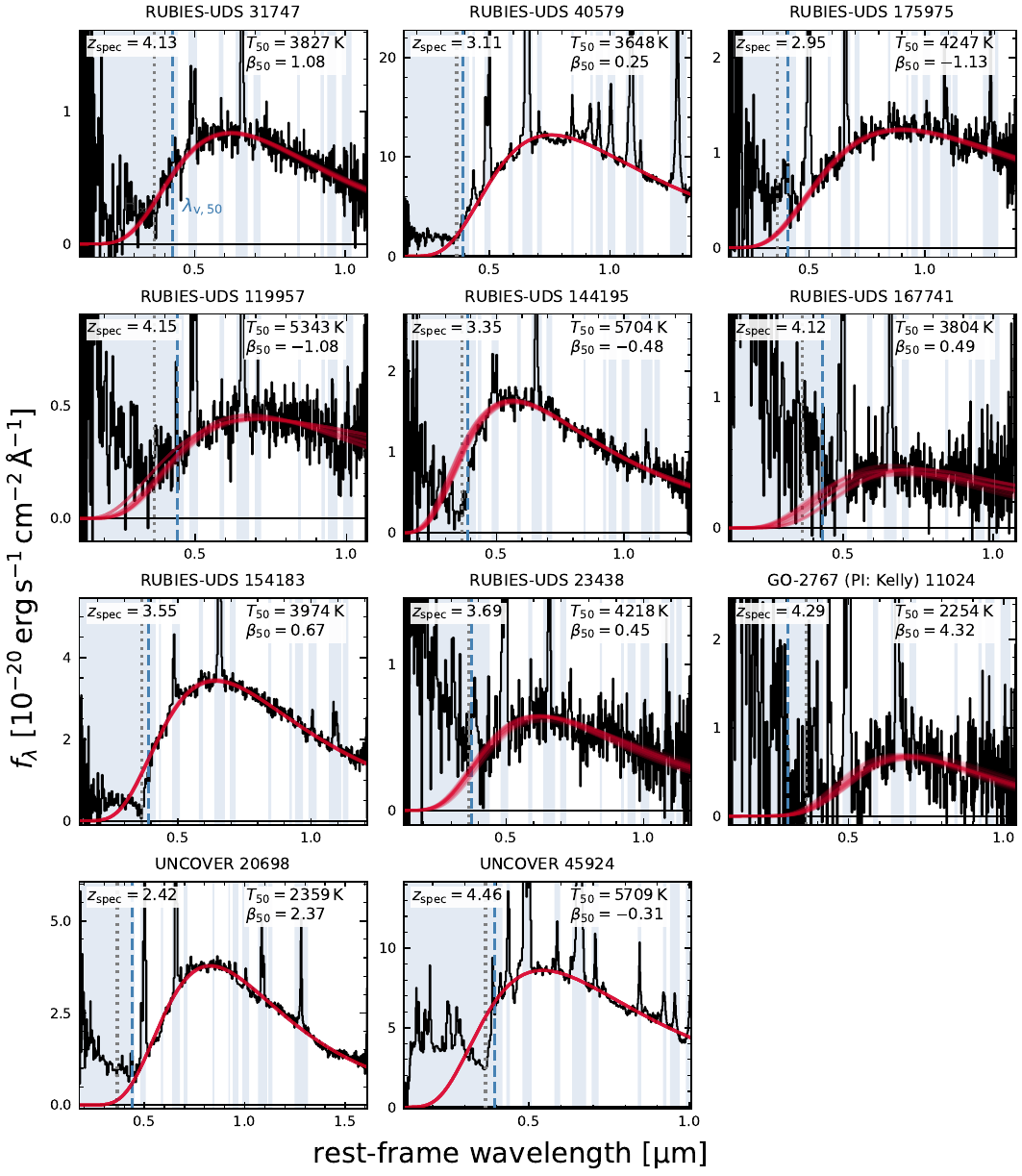}
    \caption{{As Figure~\ref{fig:all_BBs}, showing the third part of the $z<4.5$ LRD sample.}}
\end{figure*}

\bsp	
\label{lastpage}
\end{document}